\documentclass[useAMS,usenatbib]{mnras}

\usepackage{graphicx}
\usepackage[outdir=./]{epstopdf}
\DeclareGraphicsExtensions{.eps}
\usepackage{pdflscape}
\usepackage{multirow}
\usepackage{numprint}
\usepackage{float}
\usepackage{booktabs}
\usepackage{natbib}
\usepackage{amsmath}
\usepackage{comment}
\usepackage{verbatim}
\usepackage[labelfont=bf]{caption}
\bibpunct{(}{)}{;}{a}{,}{;}
\usepackage{placeins}
\usepackage{tikz}
\def\checkmark{\tikz\fill[scale=0.4](0,.35) -- (.25,0) -- (1,.7) -- (.25,.15) -- cycle;}

\makeatletter
\let\jnl@style=\rmfamily
\def\ref@jnl#1{{\jnl@style#1}}

\newcommand{\magphys}{{\sc{magphys}}}

\newcommand{\high}{H{\sc i}GH}
\newcommand{\sfrmb}{${\rm SFR}/M_{\rm bary}$}
\newcommand{\mdmb}{$M_d/M_{\rm bary}$}
\newcommand{\hi}{H{\sc i}}
\newcommand{\MHI}{M_{\text{H{\sc i}}}}

\newcommand{\citeb}[1]{\citeauthor{#1}, \citeyear{#1}}
\makeatother

\usepackage{color}

\title[Dust sources and sinks in low metallicity galaxies]{Using dust, gas and stellar mass selected samples to probe dust sources and sinks in low metallicity galaxies}
\author[P. De Vis, H. L. Gomez et al. ]{P. De Vis$^{1,2}$\thanks{E-mail:
pieter.devis@pg.canterbury.ac.nz}, H. L. Gomez$^{3}$, S. P. Schofield$^{3}$, S. Maddox$^{3,4}$, L. Dunne$^{3,4}$, \and M. Baes$^{2}$, P. Cigan$^3$, C.J.R. Clark$^{3}$, E. L. Gomez$^{5,3}$, M. Lara-L{\'o}pez$^6$, M. Owers$^{7,8}$
\\
$^{1}$ Department of Physics \& Astronomy, University of Canterbury, Private Bag 4800, Christchurch, New Zealand\\
$^{2}$ Sterrenkundig Observatorium, Universiteit Gent, Krijgslaan 281, B-9000 Gent, Belgium \\
$^{3}$ School of Physics \& Astronomy, Cardiff University, Queen's Buildings, The Parade, Cardiff CF24 3AA, United Kingdom \\
$^{4}$ Institute for Astronomy, University of Edinburgh, Royal Observatory, Blackford Hill, Edinburgh, EH9 3HJ, United Kingdom\\
$^{5}$ Las Cumbres Observatory Global Telescope Network, 6740 Cortona Dr, Suite 102, Goleta, CA9117, Unites States \\
$^{6}$ Instituto de Astronom{\'i}a, Universidad Nacional Aut{\'o}mana de M{\'e}xico, A.P. 70-264, 04510 M{\'e}xico, D.F., M{\'e}xico \\
$^{7}$ Australian Astronomical Observatory, PO Box 915, North Ryde, NSW 1670, Australia\\
$^{8}$ Department of Physics and Astronomy, Macquarie University, NSW 2109, Australia \\
}
\begin{document}

\date{Accepted version 05/05/2017 }

\pagerange{\pageref{firstpage}--\pageref{lastpage}} \pubyear{2017}

\maketitle

\label{firstpage}
\begin{abstract}
We combine samples of nearby galaxies with \textit{Herschel} photometry selected on their dust, metal, \hi, and stellar mass content, and compare these to chemical evolution models in order to discriminate between different dust sources. In a companion paper, we used a H{\sc i}-selected sample of nearby galaxies to reveal a sub-sample of very gas rich (gas fraction $>80$\,per\,cent) sources with dust masses significantly below predictions from simple chemical evolution models, and well below $M_{d}/M_{*}$ and $M_{d}/M_{\rm gas}$ scaling relations seen in dust and stellar-selected samples of local galaxies. We use a chemical evolution model to explain these dust-poor, but gas-rich, sources as well as the observed star formation rates (SFRs) and dust-to-gas ratios. We find that (i) a delayed star formation history is required to model the observed SFRs; (ii) inflows and outflows are required to model the observed metallicities at low gas fractions; (iii) a reduced contribution of dust from supernovae (SNe) is needed to explain the dust-poor sources with high gas fractions. These dust-poor, low stellar mass galaxies require a typical core-collapse SN to produce $\rm 0.01-0.16\ M_{\odot}$ of dust. To match the observed dust masses at lower gas fractions, significant grain growth is required to counteract the reduced contribution from dust in SNe and dust destruction from SN shocks. These findings are statistically robust, though due to intrinsic scatter it is not always possible to find one single model that successfully describes all the data. We also show that the dust-to-metals ratio decreases towards lower metallicity.
\end{abstract}


\defcitealias{Clark2015}{C15} \defcitealias{DeVis2017}{DV17}

\section{Introduction}
Interstellar dust is formed in the winds of evolved
low-to-intermediate mass stars (LIMS,
\citealp{Ferrarotti2006,Sargent2010}) and in core-collapse supernovae (SNe)
as massive, short-lived stars end their lives
{\citep[e.g.][]{Dunne2003,Rho2008,Dunne2009,Barlow2010,Matsuura2011,Gomez2012b,Indebetouw2014,Gall2014}.
There are also strong indications for grain growth in the Interstellar Medium (ISM) as
dust grains acquire additional mass from the gas phase by accretion
in both high and low redshift galaxies
\citep{Dwek2007,Mattsson2012,Asano2013,Zhukovska2014,Rowlands2014b}.
Recent surveys with the \textit{Herschel Space Observatory} (hereafter \textit{Herschel}, \citeb{Pilbratt2010}) of nearby galaxies have produced dust mass scaling relations with stellar mass, gas mass and star formation rate in both targeted samples of nearby galaxies, such as the \textit{Herschel} Reference Survey (HRS, e.g. \citealt{Boselli2010,Cortese2012,SmithD2012,Cortese2014}), and in wide-area blind surveys \citep{Dunne2011,Bourne2012,Clark2015} including the \textit{Herschel}-Astrophysical Terahertz Large area Survey (H-ATLAS, \citealp{Eales2010}). The dust properties of the blind, volume-limited, dust-selected HAPLESS sample of 42 galaxies over the equatorial H-ATLAS fields were presented by \citeauthor{Clark2015} (2015, hereafter \citetalias{Clark2015}).

\citetalias{Clark2015} attempted to model the HRS and HAPLESS galaxies using a simple closed box chemical evolution model and suggested the following: as galaxies evolve, their dust content first rises steeply, then levels off and reaches its peak about half way through its evolution, and finally declines towards lower gas fractions. In \citet[][hereafter \citetalias{DeVis2017}]{DeVis2017}, a local H{\sc i}-selected sample taken from the same H-ATLAS fields (dubbed `\high') was used to complement the stellar mass selected HRS and the dust-selected HAPLESS sample. They showed that the H{\sc i} selection recovered similar dust- and gas-rich galaxies as were seen in HAPLESS, but also revealed gas-rich sources {\it with much lower dust content}. \citetalias{DeVis2017} showed that these dust-poor sources are offset from the simple evolutionary scenario put forward in \citetalias{Clark2015}.

\citet{Zhukovska2014} compared the sample from \citet{Remy-Ruyer2014} (including the Dwarf Galaxy Survey \citep[DGS,][]{Madden2013}, the largest sample of low metallicity sources observed with {\it Herschel}) with a chemical evolution model to show that the observed variation in dust-to-gas ratio and metallicity in local star-forming dwarfs can be explained using models with bursty star formation histories, low dust yields from core-collapse SNe and additional grain growth in the ISM. \citet{Feldmann2015} took the sample from \cite{Remy-Ruyer2014} and used both an analytic approximation and dynamic one-zone chemical evolution models to fit the observed relationships in the 126 local galaxies. These models require very rapid grain growth, which activates at a critical metallicity, to match the observed dust-to-metal ratio in the galaxies. \citet{Feldmann2015} also argues that there is a balance between metal-poor inflows and enriched outflows which regulates the dust-to-metal ratio. \citet{Popping2016} study the dust content of galaxies from z = 0 to z = 9 using chemical evolution models including stellar dust, dust grain growth, destruction of dust by supernovae and in the hot halo, and dusty winds and inflows.

In this companion paper to \citetalias{DeVis2017}, we add additional metallicity information to \citetalias{DeVis2017}'s compilation of dust, stellar mass and H{\sc i}-selected samples of nearby galaxies, and add the metal-selected DGS, in order to investigate the dust-to-gas and dust-to-metal properties for a total of 382 sources (44 DGS sources, 58 HAPLESS+\high\ sources and 280 HRS sources have dust masses from \textit{Herschel} photometry). The combined sample here allows one to sample a wider range of gas fractions than possible before (from $0.05<f_g<0.97$). We apply a chemical evolution model to interpret the data by relaxing the closed box model assumption from \citetalias{Clark2015} and \citetalias{DeVis2017} and adding inflows and outflows, using different star formation histories (SFH), allowing for dust grain growth in the ISM and dust destruction. Section~\ref{sec:sample} reminds the reader of the samples from \citetalias{DeVis2017} and introduces the DGS. Section~\ref{sec:metals} describes the new metallicity data and the used calibrations. Section~\ref{sec:chemmodel} briefly describes the chemical model and the combination of parameters modelled in this work. The results are discussed in Section \ref{sec:results}, where we attempt to determine the contribution from different dust sources and to explain the dust-poor, gas-rich sample first shown in \citetalias{DeVis2017}. Our conclusions are listed in Section~\ref{sec:conclusions}.

\section{Nearby galaxy samples}
\label{sec:sample}

A detailed description of the HRS, HAPLESS and \high\ samples used in this work is provided in detail in \citetalias{DeVis2017}. Here we briefly remind the reader of the different datasets and parameters, introduce the new metallicity measurements and introduce the Dwarf Galaxy Survey \citep{Madden2013,Remy-Ruyer2013} which we add to our sample of local \textit{Herschel} galaxies in Section~\ref{sec:chemmodel}. The average properties for the samples used in this work are shown in Table~\ref{tab:medians}. By compiling the different nearby galaxy samples, we can model the dust properties for a total of 382 sources, compared to 126 sources in \citet{Zhukovska2014} and \citet{Feldmann2015}.
We also increase the number of low metallicity sources (additional 67 sources with $Z<1/3 \, Z_\odot$), which lie off the scaling relations for more evolved sources. This is particularly important given the relevance of immature, unevolved low metallicity sources as analogues for the first galaxies. When comparing all samples, \mdmb\ first rises steeply, then levels off and then drops again as galaxies evolve from high to low gas fractions.

\subsection{H{\sc i}, dust and stellar mass selected samples}
The dust-selected HAPLESS (\citetalias{Clark2015}) is a blind, volume-limited sample of 42 local ($z<0.01$) galaxies detected at 250\,$\mu$m from the H-ATLAS Phase 1 Version-3 internal data release, covering 160\,sq. degrees of the sky \citep{Valiante2016,Bourne2016}. The H{\sc i}-selected sample (\high, \citetalias{DeVis2017}) is extracted from the same H-ATLAS area and includes 40 unconfused H{\sc i} sources identified in the H{\sc i} Parkes All Sky Survey (HIPASS, \citealt{Barnes1992,Meyer2004}) and the Arecibo Legacy Fast ALFA Survey (ALFALFA, \citealt{Giovanelli2005,Haynes2011}, Haynes et al. {\it priv comm.}); 24 of these sources overlap with the HAPLESS sample. \citetalias{DeVis2017} further split their sample by stellar mass into \high-high and \high-low categories based on whether the sources were above or below $M_{*} = 10^9\,\rm M_{\odot}$. To supplement the dust and \hi\ selected samples taken from H-ATLAS, we follow \citetalias{Clark2015} and \citetalias{DeVis2017} and use the HRS \citep{Boselli2010} which provides a (quasi) stellar mass selected sample of nearby sources. The HRS targeted 323 galaxies ranging from late-type to early type sources.  H{\sc i} masses were taken from \citet{Boselli2014}.
\citetalias{DeVis2017} compiled FUV-submm photometry for each of these samples, and subsequently derived dust masses, stellar masses and star formation rates consistently using {\sc magphys} \citep{daCunha2008}. The {\sc magphys} properties for the \high\, and HAPLESS samples are provided in Table~\ref{tab:properties}.

\begin{table*}
\caption{The average properties for the samples used in this work quoted as the mean $\pm$ standard deviation.  Where data is not available for all the sample we quote the number of sources in the brackets. We only show the late type galaxies (LTGs) in the HRS.}
\centering
\begin{tabular}{cccccc} \hline\hline
Galaxy Sample & log SFR & log $M_{\rm HI}$  & log $M_*$  & log $M_d$ & $f_g$ \\
& ($\rm M_{\odot}\,yr^{-1}$)& ($\rm M_{\odot}$) &  ($\rm M_{\odot}$) & ($\rm M_{\odot}$) & \\ \hline
DGS & $-0.68 \pm 0.85$ (45) & $8.57 \pm 0.78 $ (35) & $8.10 \pm 0.99 $&$5.12 \pm 1.77$ & $ 0.74\pm 0.23 $ (35) \\
\high-low  &$-1.19 \pm 0.52$ & $9.02 \pm 0.46$ &$ 8.17 \pm 0.56$ &$5.21 \pm 0.97$ & $ 0.87 \pm 0.09$  \\
\high-high & $-0.07 \pm 0.40 $& $9.76 \pm 0.39 $&$ 9.89 \pm 0.65$&$7.12 \pm 0.43$ & $ 0.50 \pm 0.24 $ \\
HRS (LTGs) &  $-0.70 \pm 0.67$& $8.94 \pm 0.56$ (231) &$ 9.64 \pm 0.57$ & $6.70 \pm 0.54$ (239)&  $ 0.28 \pm 0.22$ (231) \\ \hline
\end{tabular}
\label{tab:medians}
\end{table*}

\subsection{The Dwarf Galaxy Survey}
In this work, we also include results from the Dwarf Galaxy Survey (DGS; \citeb{Madden2013}) to improve our sampling of galaxies at low stellar masses and metallicities.  The DGS sources were selected from several other surveys in order to make a broad sample of 50 galaxies ranging from very low ($\sim 1/50 Z_{\odot}$) to moderate metallicity ($\sim 1/3 Z_{\odot}$). In order to compare the samples, we need consistent methods to calculate the different galactic properties.  Unfortunately we do not have the same complete UV-submm coverage for DGS sources as we have available for the H-ATLAS and HRS.  Consequently, we redetermined the dust masses for the DGS galaxies using a two-component modified blackbody (MBB) fit to the 70-500\,$\mu$m photometry provided in \citet{Remy-Ruyer2013,Remy-Ruyer2015}. This method produces consistent results with the dust masses output by {\sc magphys} for the HAPLESS and \high\ sources with dust temperatures $>$15\,K and both methods assume the same dust absorption coefficient of $\kappa_{850} = 0.07\,\rm{ m^2\,kg^{-1}}$ \citep{James2002}.
However, for some sources fitted by a MBB with $\rm T_c < 15K$, {\sc magphys} results in warmer temperatures (by $\sim$3\,K) and therefore smaller dust masses (see also \citetalias{DeVis2017}). After scaling the \citet{Remy-Ruyer2015} masses for graphite grains by the difference in $\kappa$ used in their work and {\sc magphys}, these dust masses are entirely consistent with the MBB results for all sources with $\rm T_c > 15K$. For some of the colder sources there is an offset, yet this is of the same magnitude as the offset between {\sc magphys} and the MBB method. Therefore in what follows, we simply \emph{scale the \citet{Remy-Ruyer2015} dust masses for the difference in $\kappa$ to be consistent across samples.}

\citet{Remy-Ruyer2015} derived stellar masses based on the \citet{Eskew2012} method. To check consistency with the {\sc magphys} stellar masses, we re-derived stellar masses for our \high\ sample using their calibration and found that the DGS stellar masses were a factor of $\sim 3.2$ larger than {\sc magphys}. This difference stems mainly from \citet{Eskew2012} adopting a Salpeter initial mass function (IMF), whereas in this work we assume a Chabrier IMF. After scaling the DGS values to be consistent with the other samples here, the \high-low and DGS contain galaxies with similar average stellar masses.
SFRs for DGS were taken from \citet{Remy-Ruyer2015} and were estimated using a combination of $L_{\rm TIR}$ and the observed H$\alpha$ luminosity \citep{Kennicutt2009}. As there are no integrated H$\alpha$ luminosities available for the \high~and HAPLESS samples, we compared this SFR method with the SFRs output by {\sc magphys} for the HRS galaxies. We found that these methods were compatible for all but the most quiescent sources in the HRS (specific star formation rate $< 10^{-11} \rm{\,M_\odot\,yr^{-1}}$) which are not discussed here. H{\sc i} masses are available for 35 DGS galaxies from \citet{Madden2013}.

\section{Metallicities and calibrations}
\label{sec:metals}
To calculate metallicities for the \high\ and HAPLESS samples, we use fibre optical spectroscopy from SDSS \citep{Thomas2013}, supplemented by GAMA \citep{Hopkins2013} (v17). Although GAMA is an extragalactic survey of thousands of galaxies, we have used GAMA fibre spectra that, for our nearby galaxies, correspond to H{\sc ii} regions within the galaxies. Emission lines were measured by running each spectrum through a modified version of the Gas AND Absorption Line Fitting algorithm (GANDALF; \citeb{Sarzi2006}). Results were cross checked with GAMA's GaussFitComplexv05 (GFC; \citeb{Gordon2016}) and both techniques gave comparable results for all but 15 H{\sc ii} regions.  The results for these 15 H{\sc ii} regions were checked against the Fit3D \citep{Sanchez2016} and the GAMA SpecLines (v4) catalogues \citep{Hopkins2013} and found to be consistent with the GFC method.

For many of the \high\ and HAPLESS sources, we found multiple fibres within the same galaxy.  Star forming (H{\sc ii}) regions were selected using the criteria in \citet{Kauffmann2003} by placing sources on the BPT diagram \citep{Baldwin1981}.  This resulted in a sample of 95 H{\sc ii} regions for the 40 \high\ galaxies and 85 H{\sc ii} regions for the 42 HAPLESS galaxies (67 H{\sc ii} regions overlap as their galaxies are in both samples). The emission line fluxes for each H{\sc ii} region were corrected for stellar absorption, and for internal and galactic extinction using the Balmer decrement $C(H_\beta)$ and the \citet{Cardelli1989} dust obscuration curve. Errors on the line measurements were provided by GANDALF or GFC.  We then bootstrapped the measurements by generating 1000 new emission line fluxes assuming a normal distribution with the extinction-corrected emission line fluxes as mean and the measured error as the standard deviation of the distribution. The \high\ and HAPLESS emission lines and their errors are presented in Table~\ref{tab:HIGHlines}. For the HRS, emission lines from  integrated spectroscopy are available from \citet{Boselli2013} for 170 LTGs. DGS line measurements are taken from the literature (Table~\ref{tab:DGSlines}; S. Madden, \emph{priv. comm.}\footnote{We include the DGS emission lines and metallicities for the community in Table~\ref{tab:DGSlines}, and we correct some of the references.}).

To derive metallicities from the emission line spectra, we compared the results from different empirical and theoretical methods in order to understand any systematic differences that may result from our methods. Here, our aim is to compare the different calibrations to determine which one fits our model best. Empirical calibrations are only valid for the same range of excitation and metallicity as the H{\sc ii} regions that were used to build the calibration. Since they are determined assuming an electron temperature, these methods may systematically underestimate the true metallicity if there are temperature inhomogeneities in a galaxy.  This is thought to be more severe in metal-rich H{\sc ii} regions because the higher efficiency of metal-line cooling leads to stronger temperature gradients \citep{Garnett1992,Stasinska2005,Moustakas2010}.  On the other hand, theoretical calibrations require inputs including stellar population synthesis and photoionization models; often the theoretical metallicities are higher than those found with the empirical calibrations.

We therefore chose to compare three common empirical methods including O3N2 and (third order polynomial) N2 from \citet{Pettini2004}, and the $S$ calibration from \citet[][hereafter PG16S]{Pilyugin2016}. All three methods produce metallicities that correlate well with stellar mass and gas fraction (e.g. \citealp{Kewley2008}).
However, the O3N2 calibration is only calibrated for metallicities $12+\mathrm{log(O/H)} > 8$ \citep{Pettini2004,Marino2013}, and the N2 method also runs into difficulties at the lowest metallicities due to the large scatter observed in $\mathrm{N/O}$ ratios \citep{Morales-Luis2014} and instead provides upper limits to the true metallicity for galaxies when $12+\mathrm{log(O/H)_{N2}} < 8$. We also derived metallicities consistent with the theoretical calculations from \citet[][hereafter T04]{Tremonti2004}. As we do not have access to their code we used the scaling relation between O3N2-T04 from \citet{Kewley2008}, calibrated against 27,730 SDSS star-forming galaxies (hereafter KE08/T04). We note that this conversion is only valid for $8.05<12+\mathrm{log(O/H)_{O3N2}}<8.9$. Additionally, we have determined metallicities using the Bayesian-based IZI tool \citep{Blanc2015} which also provides a theoretical calibration based on photo-ionisation models. The IZI results were found to be entirely consistent with the KE08/T04 calibration and are thus not further listed in this work.
When the different calibrations are compared, we find the PG16S calibration produces lower metallicities, followed by N2, 03N2 and then KE08/T04 (see also Table~\ref{table:dust_metals}).
Because the limited validity of the other calibrations below $12+\mathrm{log(O/H)} = 8$, PG16S is the most reliable calibration for the low metallicity sources. Given the importance of low metallicity sources in this work, we will plot metallicities using the PG16S calibration throughout.
However, for high metallicity sources, there is a remaining uncertainty in the metallicity relations in this work with respect to which calibration is used, and we therefore highlight the main differences throughout the text and include results derived from all three calibrations where appropriate.

Metallicities for each \high\ and HAPLESS galaxy were derived from a weighted average of the multiple H{\sc ii} regions within the same source (Table~\ref{tab:HIGHmetals}). Errors were derived by adding in quadrature the bootstrapped values on the extinction-corrected emission lines and the intrinsic scatter observed between the different H{\sc ii} regions within the same galaxy. The latter amounts to an uncertainty of 0.06\,dex (see also \citealp{Bresolin2015}). Note that these metallicities are based on an average of multiple small (2$^{\prime\prime}$ for GAMA, 3$^{\prime\prime}$ for SDSS) fibres per source and not on integrated measurements. They are thus not ideally suited for extended sources, yet they at least provide a good first estimate for the metallicity of the galaxies in our sample. The calibrations used in this work are derived using the electron temperature method. The uncertainty in the absolute metallicity determination by this method is $\sim 0.1$ dex \citep{Kewley2008}. We thus add 0.1 dex in quadrature to the uncertainty on the averaged metallicity for each galaxy and each calibration. The resulting metallicities and uncertainties for \high\ and HAPLESS are listed in Table~\ref{tab:HIGHmetals}.

\begin{figure}
  \includegraphics[trim=11mm 11mm 0mm 0mm clip=true,width=0.48\textwidth]{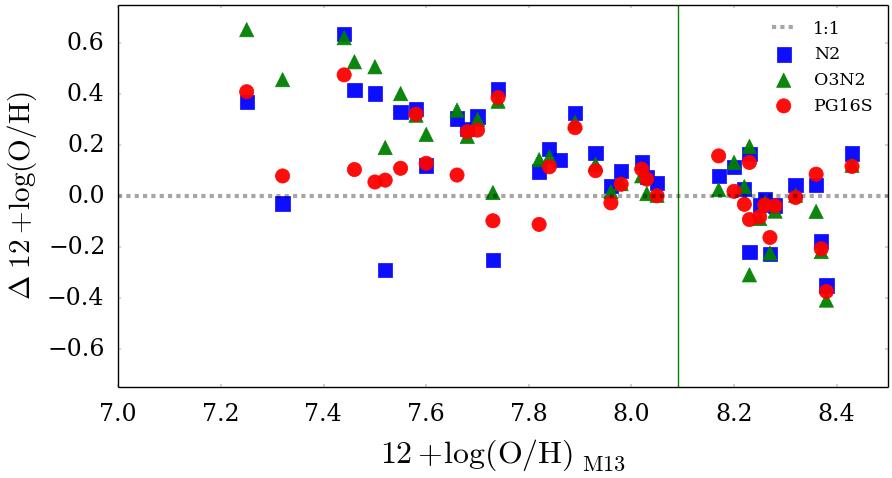}
  \caption{A comparison of the differences in metallicity calibrations $\Delta (12+\mathrm{log(O/H))}$ using the N2 (blue), O3N2 (green) and PG16S (red) methods with the published DGS metallicities derived using PT05 (M13, \citealp{Madden2013}). The significant offset between the O3N2 and PT05 at the lowest metallicities may be because this calibration is known to break down here (indicated by the vertical line; \citealp{Pettini2004}), though the N2 and PG16S results also suggest PT05 tends to underestimate metallicities in this regime.}
  \label{fig:DGSmetals}
\end{figure}

For the HRS and DGS, we simply derive metallicities based on the integrated spectroscopy for each of the three calibrations. The DGS metallicities were originally estimated using the \citet[][hereafter PT05]{Pilyugin2005} calibration \citep{Madden2013}. PT05 is calibrated over a similar range of metallicity to N2 and PG16S, but there are a number of reasons we choose not to use this as a method to determine metallicities in this study. First, PT05 is not a good estimator for metal-rich galaxies that have low excitation parameters $P$ and high values of $R_{23}$ \citep{Moustakas2010}, such as the HRS and \high-high galaxies.   PT05 is therefore more suited for the DGS and \high-low sources that have appropriate $P$ and $R_{23}$ values but this means we cannot apply a consistent method to derive metallicities across the different samples of nearby galaxies.  Second, PT05 metallicities have been shown to have a lot of scatter with stellar mass compared to other calibrators \citep{Kewley2008}, suggesting it is not a good tracer of metallicity across a wide range of galaxy properties.  Third, the PT05 method has two `branches' of metallicity values versus the $R_{23}$ emission line ratio with a transition region between the two branches.  Because of this, a large difference in $Z$ can be derived for galaxies with very small changes in emission line ratio. The PG16S calibration also uses different relations for high and low metallicities. However the appropriate ranges where the high- and low-metallicity relations can be used, overlap for adjacent metallicities, and the transition zone thus disappears. Fig.~\ref{fig:DGSmetals} compares the DGS metallicities derived here with the published values from \citet[][M13]{Madden2013}, which are based on the PT05 calibration. We find, on average, lower metallicities for galaxies with $12+\mathrm{log(O/H)_{M13}} < 8$, and higher metallicities for high metallicity souces (though with less scatter).
We see that the differences between PT05 and PG16S are less pronounced than for the other calibrations, though using PG16S still suggest higher metallicities (0.09 dex on average) than previously published, especially for low metallicity sources. We thus use PG16S metallicities throughout this work.

\section{The chemical evolution model}
\label{sec:chemmodel}
A chemical and dust evolution model can be used to build a consistent picture of how the metals, dust and gas content change as galaxies evolve \citep{Tinsley1980}. The simple chemical evolution model used in \citetalias{Clark2015} to interpret the different scaling relations for dust, gas and stellar mass selected samples neglected dust destruction and grain growth, and assumed that the system was a closed box (no inflows or outflows). Following \citet{Zhukovska2014} and \citet{Feldmann2015}, we relax all of these assumptions. The chemical model is presented in full in Rowlands et al. (\citeyear{Rowlands2014b}; see also \citeb{Morgan2003}) and the Python code used is freely available on {\sc github}\footnote{https://github.com/zemogle/chemevol}. The equation for the dust mass evolution is given in Appendix~\ref{app:dusteq}. In short, the model uses a prescription for the Star Formation History (SFH) and a \citet{Chabrier2003} IMF to calculate how much of the initial gas is converted into stars at any given time. The model also tracks the continuous build-up of metals as stars end their lives, though metals can be removed in outflows of material. For dust, the picture is more complex. Dust is produced by supernovae and evolved low-intermediate mass stars, and additional mass is gained from the ISM by dust grain growth. Dust is primarily destroyed by SN shocks and astration (the removal of gas and dust due to star formation). We use a one-zone chemical evolution model, i.e. we study the integrated properties of galaxies without spatial resolution, and assume instant mixing of dust, gas and metals.

In this model we include simple analytical prescriptions for grain growth and dust destruction via shocks as described in \citet{Rowlands2014b}. The timescale for dust destruction ($\tau_{\rm dest}$, following \citealt{Dwek2007}) is described as a function of the rate of SN ($R_{\rm SN}$):
\begin{equation}
\tau_{\rm dest} = \frac{M_g}{m_{\rm ISM}R_{\rm SN}(t)},
\end{equation}
where $M_g$ is the gas mass of the galaxy and $m_{\rm ISM}$ is the mass of ISM that is swept up by each individual SN event.  In some models (Table~\ref{table:models}) we set this to $m_{\rm ISM} = 100\rm \,M_{\odot}$, indicative of SN shocks ploughing into typical interstellar densities of $10^3\,\rm cm^{-3}$ (\citealt{Gall2011,Dwek2011}), although we also explore models with $1000\,\rm M_{\odot}$ (\citealp{Dwek2007}), consistent with dust destruction in the diffuse ISM and possibly more appropriate for a low metallicity ISM.

The grain growth prescription is taken from \cite{Mattsson2012} where the timescale for dust growth is given by:
\begin{equation}
\tau_{\rm grow} = \frac{M_g}{\epsilon Z \psi(t)}\  \left(1 - \frac{\eta_d}{Z}\right)^{-1},
\label{eq:tau_gg}
\end{equation}
and $\eta_{d}$ is the dust-to-gas ratio, $Z$ is the metallicity ($Z=M_{\rm metals}/M_{\rm gas}$) and $\epsilon$ is a free parameter, which is set to $\epsilon = 500$ in \citet{Mattsson2012}, appropriate for accretion timescales of $10^7\,\rm yr$ for a galaxy similar to the Milky Way.

We also test four `representative' star formation histories, including a Milky Way-type exponentially declining SFR \citep{Yin2009}, and two versions of a delayed SFH as parameterised by \citet{Lee2010}:
\begin{equation}
{\rm SFR}(t)\propto \frac{t}{\tau^2} \, e^{-t/\tau}
\label{eq:SFH}
\end{equation}
where $t$ is the age of the galaxy and $\tau$ is the star formation timescale.  First, we assume a SFH with $\tau=6.9\rm \,Gyr$ with peak SFR $4.4\,\rm M_\odot\,yr^{-1}$ in order to produce the same stellar mass as the Milky Way-type SFH. The second delayed SFH is reduced by a factor of 3, and has $\tau = 15 \rm \, Gyr$ (see Section~\ref{sec:SFR}). Finally, a model including a bursty SFH, similar to that used in \citet{Zhukovska2014} to explain the SFR properties of the DGS sources, is also included. In Section~\ref{sec:results}, we test various parameter combinations, including changing SFHs, IMFs, inflows, outflows and including different dust sources, in order to interpret the observed dust, metal, gas and star formation rates of the nearby galaxy samples. The parameters for Models I-VII, which are good representations for the sampled parameter-space, are listed in Table~\ref{table:models}.

We note that this model differs from \citet{Rowlands2014b} in the following ways: (i) the initial remnant mass function is updated. (ii) We now take into account the formation of a black hole for stars with initial mass $m_i>40\,\rm M_{\odot}$ when accounting for gas and metals released into the ISM.  Stars with progenitor mass above this cut-off mass only contribute gas and metals lost via stellar winds before the collapse. (iii) We add an additional term $f_c$ to account for the fraction of gas that is cold enough for grain growth in the ISM.  We follow \citet{Mancini2015} and \citet{Inoue2003} by setting this equal to 0.5. This parameter is likely to be higher at earlier times (e.g. \citealt{Popping2014,Nozawa2015}) though we choose to keep it constant here. (iv) We no longer interpolate the yields from stars of a given mass but just choose the nearest neighbour value, this has a small effect on the resulting stellar yields.  (v) We directly input the dust masses for core collapse SN for stars with initial mass $8.5 < M_i \le 40$ from \citet{Todini2001}.  \citet{Rowlands2014b} used the \citet{Todini2001} dust masses to estimate a condensation efficiency for SN dust ($\delta_{\rm SN}$) and applied that to the metal yields from \citet{Maeder1992}.  Using the former technique reduces the dust mass by a factor of $\sim$1.8 for a MW-like galaxy at early times ($\rm <0.8\,Gyr$) compared to the latter.
\begin{table*}
\caption{Parameters for the different chemical evolution models used.}
\centering
\begin{tabular}{lccccccc} \hline\hline
Name & IMF & SFH & Reduced SN dust &  Destruction &  Grain Growth &Inflow &Outflow \\ \hline
Model I & Chabrier & Milky Way & N & N &  N  & N  & N \\
Model II & Chabrier & Delayed & N & N & N & N  & N \\
Model III & Chabrier & Delayed & N & N  & N & N &  $1.5\times$ SFR \\
Model IV & Chabrier & Delayed & $\times 6$ & $m_{\rm ISM}=150$ &  $\epsilon=700$ &  $1.7\times$ SFR &  $1.7\times$ SFR \\
Model V & Chabrier & Delayed & $\times 12$ & $m_{\rm ISM}=1500$ &  $\epsilon=5000$ &  $2.5\times$ SFR &  $2.5\times$ SFR \\
Model VI & Chabrier & Delayed/3 & $\times 100$ & $m_{\rm ISM}=150$ & $\epsilon=8000$ &  $2.5\times$ SFR &  $2.5\times$ SFR \\ \hline
Model VII & Chabrier & Bursty & $\times 12$ &  $m_{\rm ISM} = 150$ &  $\epsilon = 12000$ &  $4.0\times$ SFR &  $4.0\times$  SFR \\ \hline
\end{tabular}
\label{table:models}
\end{table*}

\section{Results}
\label{sec:results}
\subsection{A simple model fit to dust in nearby galaxies}
\label{sec:simple}
In Fig.~\ref{fig:baryonic}, we follow	 \citetalias{Clark2015} and \citetalias{DeVis2017} and compare the evolution of the dust-to-baryonic mass ratio (\mdmb) with gas fraction for the different nearby galaxy samples compiled here. This plot is an excellent starting point as it tracks the relative changes in dust mass in terms of the evolutionary state.
We define the baryon mass and gas fraction as $M_{\rm bary}=M_g+M_*$ and $f_g=\frac{M_g}{M_*+M_g}$ respectively, where $M_g = 1.32\ \MHI$ to take into account the mass of neutral helium. Due to the difficulty in obtaining CO detections for all the different samples considered here, particularly for low stellar mass sources, we do not take into account any molecular component. We refer to Section~\ref{sec:caveats} for further discussion, though we note here that $\rm{H_2}$ does not dominate the total gas mass for our samples and thus including $\rm{H_2}$ would not affect the conclusions reached in this work.

In Fig.~\ref{fig:baryonic}, we find \mdmb\ follows a tight relation at low gas fractions. However at high gas fraction there is more scatter, at least in part due to differences in the contributions from the different dust sources. We also show how the observations from the different samples compare with a chemical evolution track similar to \citetalias{Clark2015} and \citetalias{DeVis2017}. This model uses a SFH consistent with the Milky Way \citep{Yin2009}, though here we use our updated code (Model I, Table~\ref{table:models}). Model I overlaps with Model II in Fig.~\ref{fig:baryonic} (see also Section~\ref{sec:baryon}). Although galaxies are more complex than this simple model, Model I does explain the overall trend in these samples, yet not all sources at gas fractions $< 50$\,per\,cent are well-matched. We note that our model peaks at a lower gas fraction ($\sim 0.3$) than in \citetalias{Clark2015} and \citetalias{DeVis2017} due to the changes made to the assumptions and dust inputs described in Section~\ref{sec:chemmodel}. Indeed, as our model has less dust injection from supernovae but the same dust injection from low-intermediate mass stars (LIMS) compared to \citet{Rowlands2014b}, this shifts the peak \mdmb\ towards lower gas fractions. In this work we assume a dust condensation efficiency for LIMS of 0.16, consistent with predictions of \cite{Morgan2003}. This value is somewhat lower than the high condensation efficiencies from theoretical models of dust formation in stellar winds \citep{Zhukovska2008,Ventura2012}. By choosing an even lower value for the dust condensation efficiency in LIMS we could obtain a better fit to \mdmb\ at low gas fractions for the closed box model of \citetalias{Clark2015}. However, as we will show in Section~\ref{sec:baryon}, an equally good fit to \mdmb\ can be obtained through the introduction of inflows and outflows when we relax the assumptions of the closed box model.

In Fig.~\ref{fig:baryonic}, we see that a large fraction of the highest gas fraction galaxies ($f_g > 85$\,per\,cent, \high-low) have significantly lower \mdmb\ than expected from Model I. We note, however, that the dust mass of these galaxies have large error bars due to poorly constrained dust temperatures from the \magphys\ fitting.  In order to ensure the offset in \mdmb\ for these sources is not due to this, we stacked the MIR-submm fluxes for the 8 \high-low sources with poorly constrained temperatures. The resulting stacked SED is well fitted by a single modified black body curve with dust temperature $T\sim 35$\,K. The lower dust masses for these sources are therefore consistent with them having warmer dust temperatures than the HAPLESS and HRS sources (on average). We conclude that a different set of chemical model properties are necessary to explain this slower build-up of dust for these high gas fraction sources compared to the dust-rich HAPLESS galaxies at the same \mdmb.

Note that in Fig.~\ref{fig:baryonic} we also highlight the well-studied galaxy I Zw 18 (part of the DGS sample) thought to be a local analogue of low-metallicity, high-redshift systems (e.g. \citealt{HerreraCamus2012,Fisher2014}).   The location of this source on this \mdmb\ `scaling relation' (and in later Sections) is indicated by the black diamond using the measured properties from \citet{Fisher2014}.  As we have re-evaluated the DGS measurements to be consistent across all samples (Section~\ref{sec:sample}), we have indicated where this galaxy moves with our revised measurements (dashed line). We will see in later sections that the dust properties of I Zw 18 are entirely consistent with its gas fraction and metallicity.

\begin{figure}
  \includegraphics[width=1.0\columnwidth]{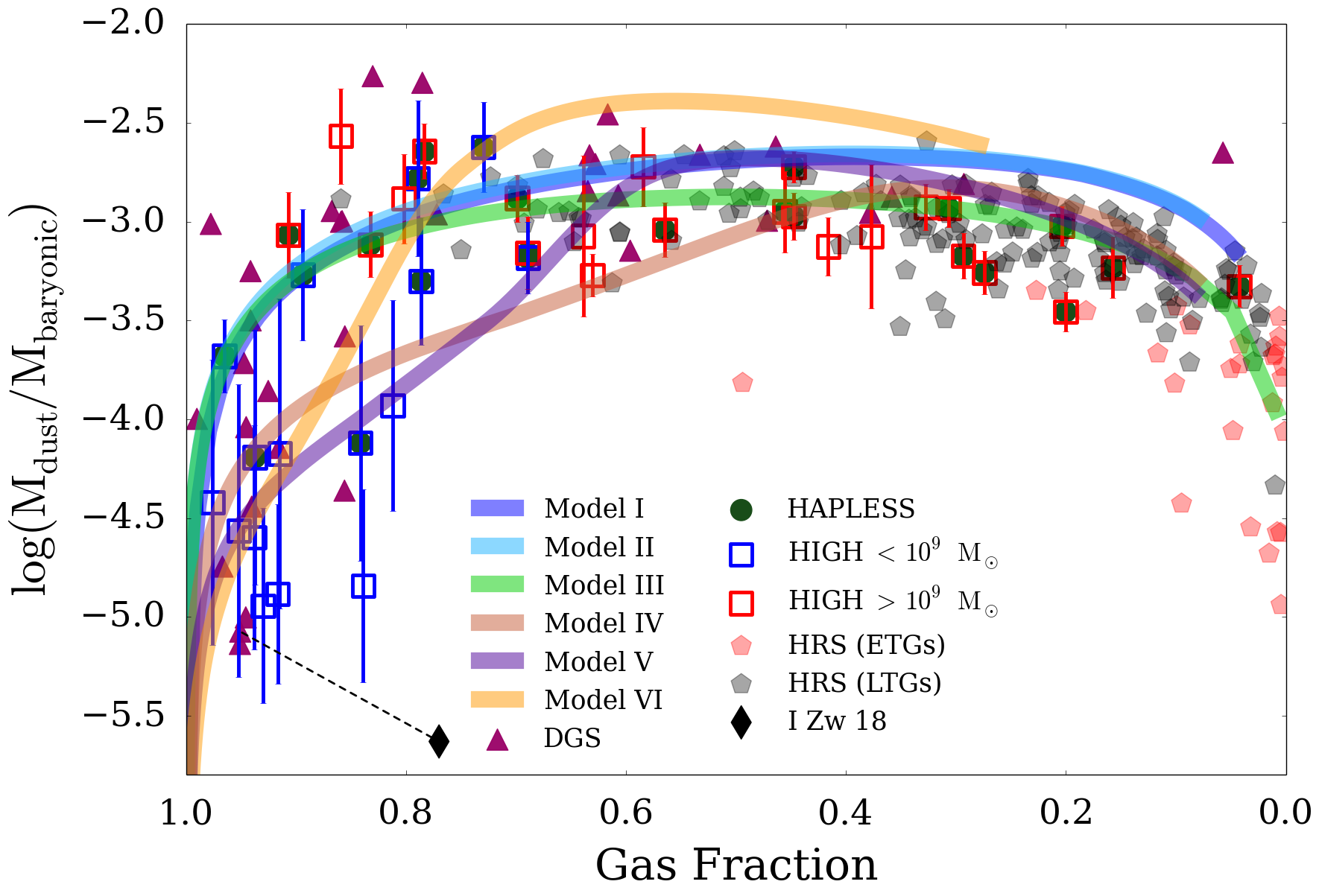}
  \caption{Variation of $M_{d}/M_{\rm bary}$ with gas fraction for the different nearby galaxy samples. The solid lines show how galaxies with the same initial gas mass but different combinations of SFHs, inflows, outflows and dust sources evolve as the gas is consumed into stars (Models I-VI; Table~\ref{table:models}). Models I and II overlap in this plot. The observed properties of dust-poor local galaxy I Zw 18 (black diamond) are also added for comparison \citep{Fisher2014}, with dashed line indicating where this source `moves' using the methods in this work.}
  \label{fig:baryonic}
\end{figure}

\subsection{Relaxing the closed box assumption}
\label{sec:baryon}
Fig.~\ref{fig:baryonic} also compares the \mdmb\ of these samples with different chemical evolution tracks including different SFHs and/or relaxing the closed box assumption from Model I (Models II-VI, Table~\ref{table:models}). There are significant differences between some of the models and the data, especially at $f_g \sim 80$\,per\,cent. Even for the same gas fraction, nearby low $M_*$ galaxies split into two categories: dust-rich and dust-poor and require {\it different chemical evolution models} to explain their dust-to-baryonic mass properties. Briefly, we see that the dust-rich low $M_*$ sources are matched relatively well by Models I-III, which show a steep rise in \mdmb\ at the highest gas fractions ($f_g > 95$\,per\,cent) and correspond to core-collapse SNe producing $0.17-1.0\,\rm M_{\odot}$ of dust per explosion when there is \textit{no dust destruction or grain growth} (or the net interstellar grain growth is matched by equal dust destruction). For the sources with high \mdmb\ for their gas fraction, an increased SN dust yield results in a better fit. In this case the dust contribution from LIMS needs to be reduced. Even though the SFH for models I and II are very different, their chemical evolution tracks in Fig.~\ref{fig:baryonic} nearly overlap, indicating the \mdmb\ evolution is not dependent on the SFH for models without dust grain growth.

To model the dust-poor \high-low and DGS sources, we follow \citet{Zhukovska2014} and \citet{Feldmann2015} and relax the closed box assumption, reduce the contribution from supernova dust, and include dust grain growth in our model.
Models IV-VI therefore require a reduction in the dust production in SNe by a factor of 6-100 compared to the models required to fit the HRS, \high-high and HAPLESS.
In contrast to our approach (and \citeb{Zhukovska2014}), \citet{Feldmann2015} even uses reduced supernova dust yields for sources that are not dust poor given their metallicity and instead uses extremely fast (timescale of $\sim 5$ Myr) grain growth  to achieve high dust masses at high gas fractions. There is thus a degeneracy between using a significant contribution from supernova dust, and using very fast dust grain growth. Their grain growth timescales of $\sim5$ Myr are much faster than typically found in nearby galaxies \citep{Mattsson2012,Mattsson2014} or from basic theoretical estimates of the underlying growth rate \citep{Draine2009}.

At late times (low gas fractions), Models I and II overestimate the amount of \mdmb\ and require (stronger) inflows and dust-rich outflows of gas or a reduced dust contribution from LIMS to explain the observed properties. The choice of Models I-VI will be motivated in Sections~\ref{sec:SFR}-\ref{sec:dust_metals}.

\subsection{Star formation histories}
\label{sec:SFR}
\begin{figure}
  \includegraphics[width=1.0\columnwidth]{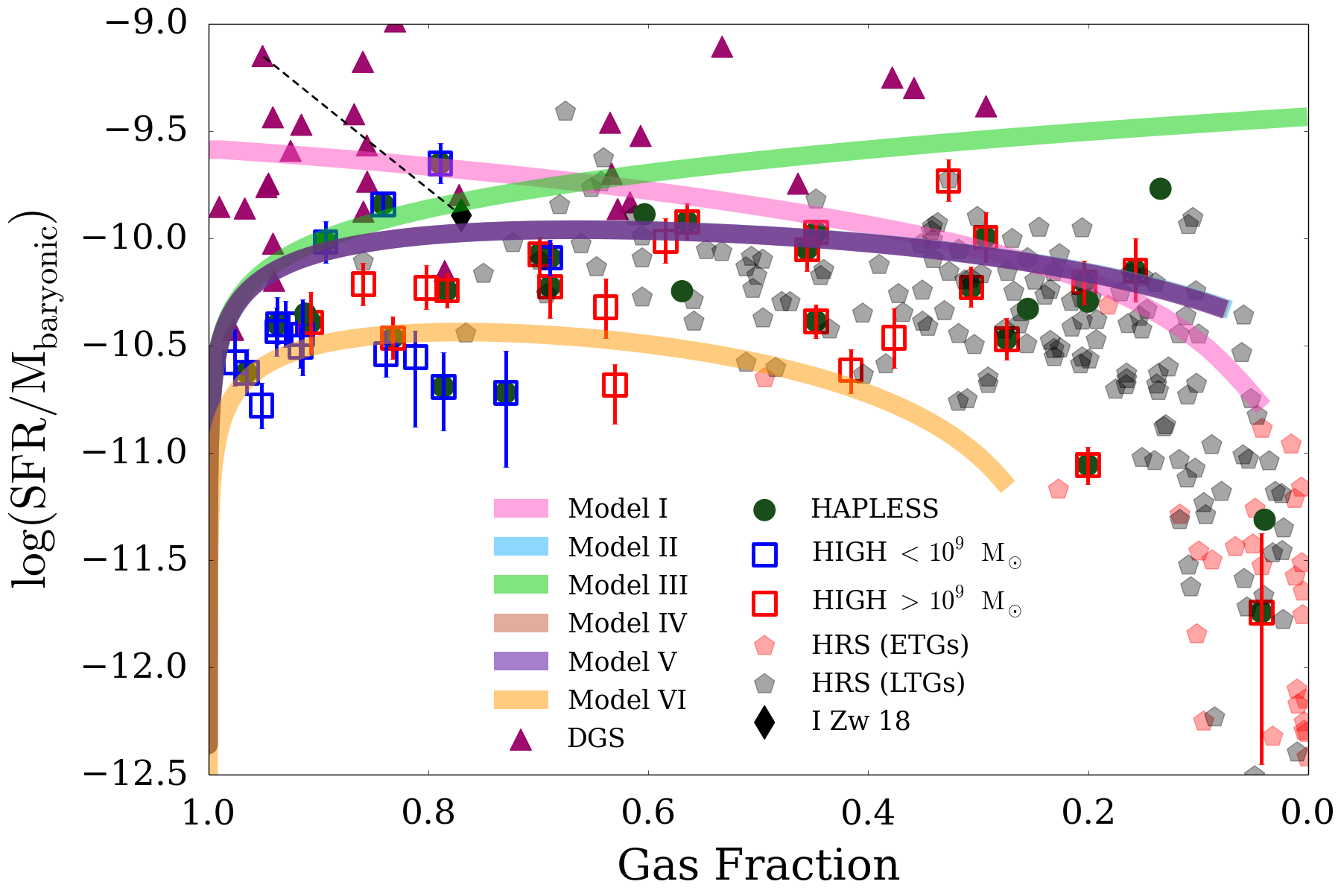}
  \caption{\sfrmb\ against the gas fraction reveals the need for a delayed SFH (Models II-VI) to explain the \high-low and HAPLESS sources at high $f_g$. In this parameter space, Models II, IV and V overlap as they have the same SFH and their inflows and outflows are balanced. Bursty SFHs are needed for the DGS (Appendix~\ref{app:dgs}).}
  \label{fig:SFRMbary}
\end{figure}
Next we attempt to explain the observed SFR properties with these models by comparing the change in \sfrmb\ with gas fraction. Fig.~\ref{fig:SFRMbary} compares the HAPLESS, HRS, \high, and DGS samples. In the high gas fraction regime ($f_g>80$\,per\,cent), we see that Model I overpredicts the \sfrmb, particularly in comparison to the \high-low sources. Delayed SFH models provide a closer match to this sample (as used in Models II-VI) by reducing the SFR per unit baryonic mass at early evolutionary stages. The values of the delayed SFHs in Eq.~\ref{eq:SFH} were chosen to match the data in Fig.~\ref{fig:SFRMbary}, with Model VI providing a good fit to most\footnote{Three \high-low sources actually have a higher SFR, more in line with the actively star-forming DGS sources rather than the other normal star-forming \high\ galaxies.} of the \high-low sources. In models with strong outflows but no inflows (Model III), the baryonic mass is significantly reduced at low gas fractions, and therefore \sfrmb\ increases as the gas fraction decreases. Model III thus poorly matches the observed \sfrmb\ at low gas fractions and can be discarded as an unrealistic model. However, when the outflow is matched by an equal inflow as in Model V, $M_{\rm bary}$ stays constant and we find the same \sfrmb\ track as for the same model without inflows and outflows (i.e. Models II, IV and V overlap in Fig.~\ref{fig:SFRMbary}).

The DGS sources lie significantly above the HRS, \high\ and HAPLESS samples in Fig.~\ref{fig:SFRMbary}, with higher \sfrmb\ for the same gas fraction.
DGS tends to contain more actively star-forming galaxies (average SFR $0.21\,\rm M_{\odot}\,yr^{-1}$, Table~\ref{tab:medians}) than is typical of nearby dwarfs (e.g. \citealt{Hunter2012}). Their selection towards more star-forming, low-stellar mass systems could be a consequence of their original selection of galaxies with moderate to very low PT05 metallicities. We return to this in the next section. The intensely SF nature of the DGS was highlighted in \citet{Zhukovska2014}, where they found they required bursty SFRs
 to fit the gas and dust properties of these dwarf galaxies.  Even with the revised dust masses and metallicities and the different model assumptions in this work, we also require a bursty SFH to fit the DGS properties (Appendix~\ref{app:dgs}).  This demonstrates that despite having similar stellar masses, dust temperatures and gas fractions as the \high-low sources, the DGS are more actively star forming than the \high\ galaxies and \emph{do not appear to be the same sources at a different evolutionary stage}. However, we cannot rule out that DGS and \high-low are both part of the same evolutionary sequence, with DGS sources undergoing a burst and \high-low sources in a quiescent period between bursts. The \high-low and HAPLESS samples therefore complement the DGS and provide additional, new, information of more normal star-forming systems at low metallicities, high $f_g$, and potentially different dust properties.

\subsection{Metallicity build-up}
\label{sec:metalsbuild}
We next wish to compare how the metallicity of galaxies changes as they evolve from high to low gas fractions. The chemical evolution code traces both the total metal mass fraction $Z$ and the oxygen mass, we can directly convert the models to oxygen abundance using:
\begin{equation}
  12 + {\rm log \left({O \over{H}} \right)} = 12 + {\rm log}\left( \frac{{\rm oxygen \, mass}/16}{{\rm gas \,  mass}/1.32} \right).
\end{equation}
\noindent In Fig.~\ref{fig:Z_gasfraction} we see in both the model behaviour and the observations that, in general, the metallicity increases monotonically as galaxies evolve from high to low gas fractions. The models are almost indistinguishable at gas fractions $>80$\,per\,cent in this parameter space.
At low gas fractions, Models I and II clearly overestimate the amount of metals; we find models with moderate ($2.5\times$ SFR) outflows of enriched gas and metal-poor inflows are necessary (Models V-VI).
Here we note that it is possible that empirical calibrations such as PG16S are underestimated (particularly at low gas fractions) due to temperature inhomogeneities. If this is the case, a theoretical calibration such as KE08/T04 would be more applicable and less strong inflows and outflows would be sufficient.

\begin{figure}
  \includegraphics[width=1.0\columnwidth]{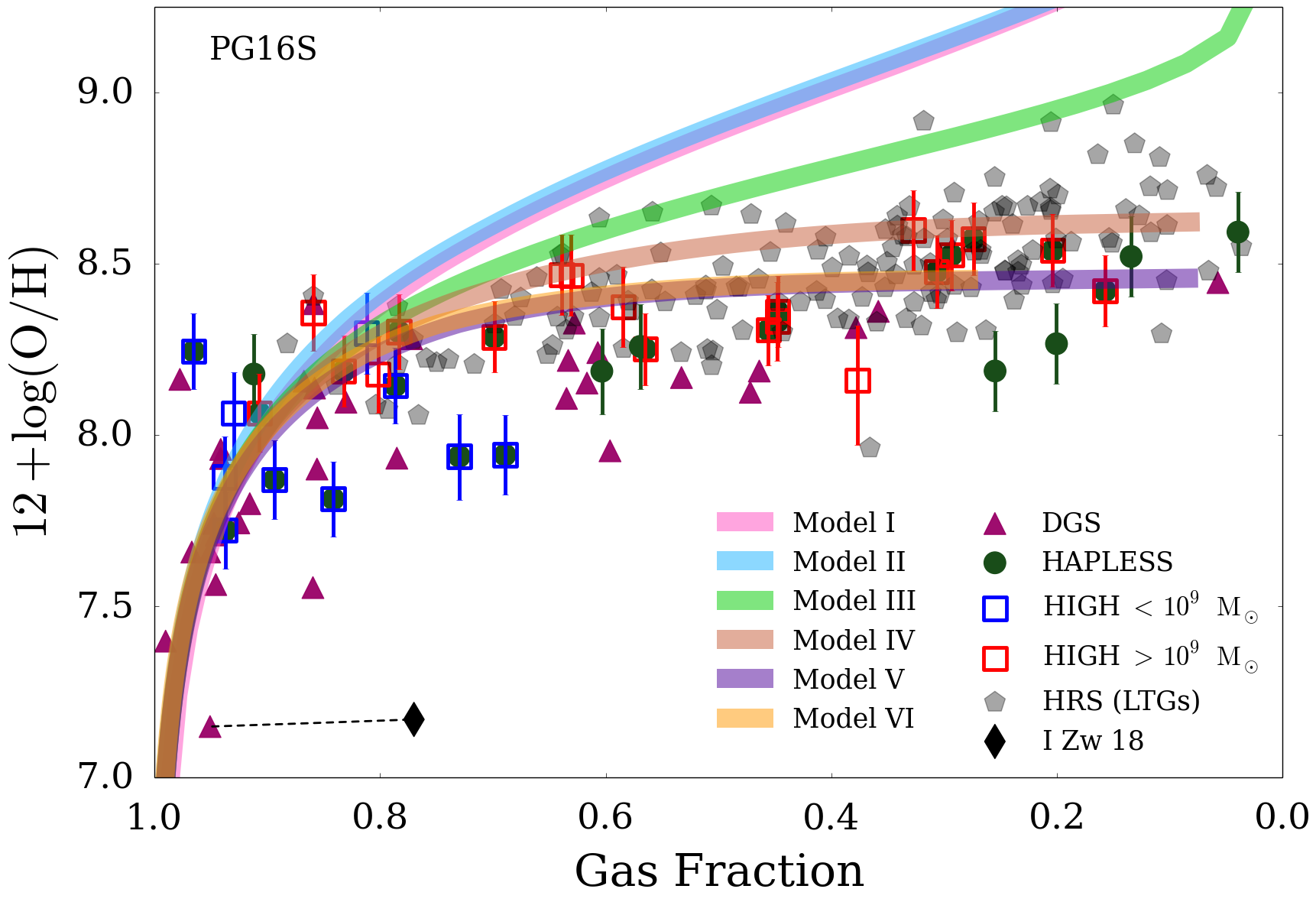}
\caption{Metallicity variation with gas fraction for the different samples using the PG16S metallicity calibration. The different chemical evolution models (see text and Table~\ref{table:models}) are also included. }
  \label{fig:Z_gasfraction}
\end{figure}

Fig.~\ref{fig:Z_gasfraction} also shows that the DGS appears to have lower metallicities than the HRS at low gas fractions and, to a lesser extent the \high-low sources at high gas fractions. In other words, \emph{the DGS galaxies are, on average, more metal-poor given their evolutionary state}. In the previous section we also found DGS are, on average, more actively forming stars. Selecting galaxies ranging from low to moderate metallicity at a given gas fraction appears to result in a sample selection biased towards galaxies with very high SFRs due to the mass-metallicity-SFR relation \citep{Mannucci2010,LaraLopez2010}. Additionally, higher SFR (and thus brighter) sources are easier to observe with \textit{Herschel}, and it is easier to obtain high signal-to-noise spectra (in order to determine metallicities). This is another reason DGS consists mainly of high SFR sources. As suggested in \citet{Feldmann2015}, the low metallicities at a given gas fraction for DGS sources, requires the addition of strong inflows and outflows to regulate the build-up of metals in the DGS galaxies. Only Model VII (Fig.~\ref{fig:bursty_results}), with bursty SFH and stronger inflows/outflows ($4\times$ SFR) than the models (V and VI) used to match the other nearby galaxy samples in this work, can be used to explain the $Z-f_g$ properties of the DGS sample.

\subsection{Dust-to-gas ratio}
Next we compare metallicity with the dust-to-gas ratio for the 253 galaxies in our combined sample that have metallicities available (Fig.~\ref{fig:HIDust}). The $M_{d}/M_{g}$ ratio correlates with the gas-phase metallicity over a wide range $7.3<12 + {\rm log(O/H)_{PG16S}} < 9.0$, yet we again identify two regimes. If dust traces the metals or a constant fraction of metals remains in dust grains, we expect a linear $M_d/M_g - Z$ relationship, with a slope as for Models I-III. Models I-III are consistent with the slope of those galaxies with highest $M_{d}/M_{g}$ in the metallicity range $7.3<12 + {\rm log(O/H)_{PG16S}} < 8.2$, but the HRS sources do lie offset from these models\footnote{Here we do not aim to find a single model that explains the HRS galaxies, rather we simply show how models without grain growth evolve in this plot.}. This could be explained either by increasing the dust produced by stars by a factor of 5 or more which would move Models I-III up the y-axis whilst keeping the slope constant. However we note that the amount of dust formed in LIMS stars and SNe is already substantial, and these models have no dust destruction suggesting that adding more dust produced from stars in this way is unrealistic. Alternatively, one could add interstellar grain growth which would act to steepen the ‘tips’ of Models I-III at metallicites $12 + {\rm log(O/H)_{PG16S}} > 8.2$ (see for example the tips of Models IV-VI). We return to how this may be a result of different dust-to-metal ratios in Section~\ref{sec:dust_metals}.

\begin{figure}
  \includegraphics[width=1.0\columnwidth]{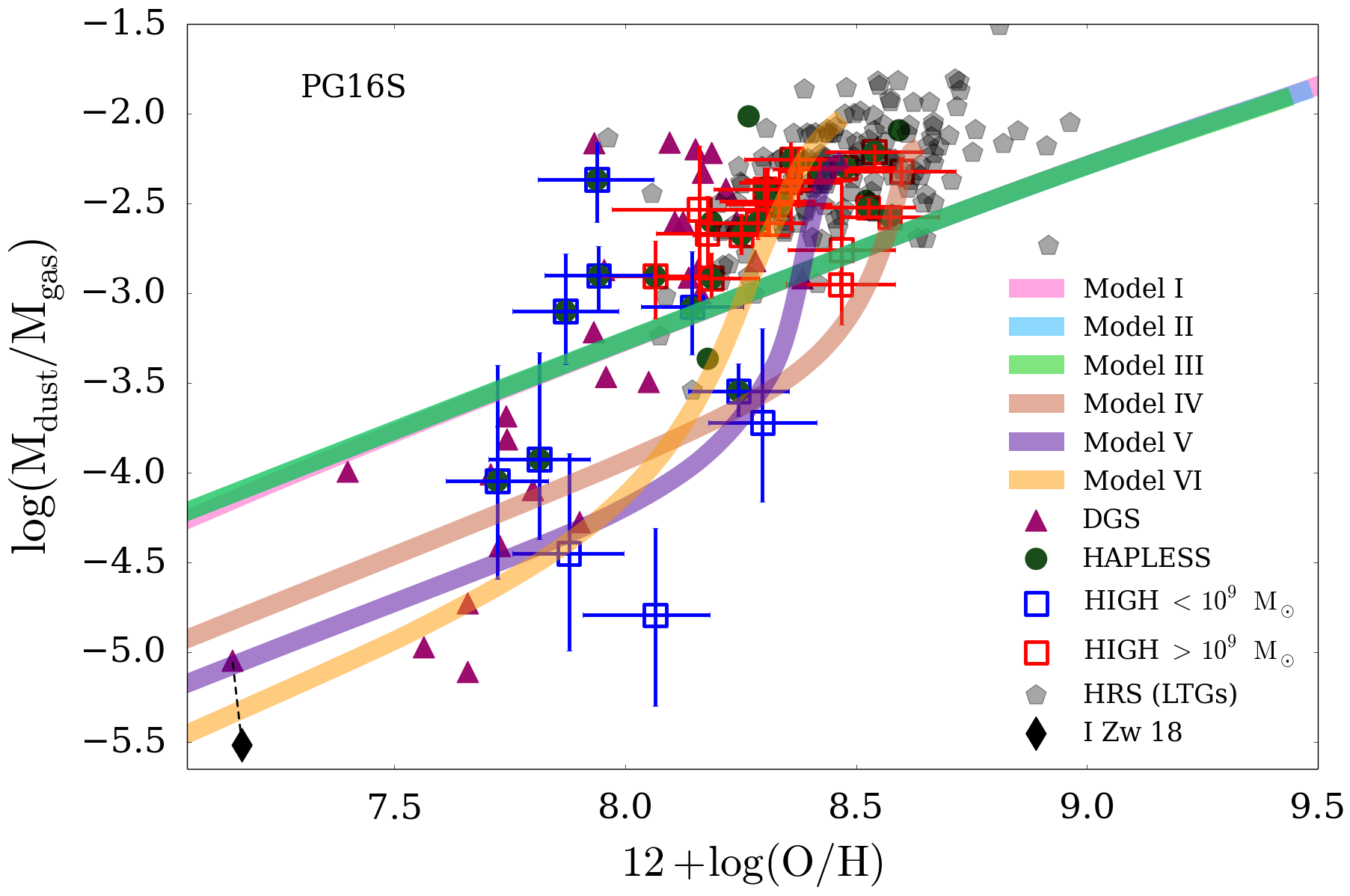}
\caption{Metallicity variation with gas-to-dust ratio $M_{d}/M_{g}$.
Models IV, V and VI provide a better match between metallicity and $M_{d}/M_{g}$ for the \high-low and many DGS sources than Models I-III.
}
  \label{fig:HIDust}
\end{figure}

The dust-poor high gas fraction sources (\high-low and some of the DGS galaxies) lie well below the linear trend from Model I-III. This offset was already discussed in \citet{Remy-Ruyer2013,Remy-Ruyer2015,Zhukovska2014,Feldmann2015}, who explained this by suggesting the supernova contribution to the dust budget needs be reduced and a dust grain growth term added. The argument is such: for the highest gas-fraction galaxies in Fig.~\ref{fig:HIDust} the dust mass needs to be significantly suppressed without reducing the metals. The only way\footnote{Note the dust destruction caveat in Section \ref{sec:caveats}.} to do this is to reduce the amount of dust formed by stars (including SN) in each stellar population.  As the dust-to-gas ratio is already lower than expected from a linear trend at high gas fractions, this suggests the SN dust production must be suppressed. The observed dust-to-gas ratio in the lowest metallicity \high-low and DGS galaxies requires models with a maximum of $0.01-0.16\,\rm M_\odot$ (Models VI-IV) of dust per core collapse SNe, which corresponds to a condensation efficiency of 0.2-3.2\,per\,cent for a 25$\rm M_{\odot}$ progenitor (assuming 5$\rm M_{\odot}$ of metals ejected and that all of this mass can be condensed into dust).

Therefore Models IV, V and VI include a reduced SN dust component (by a factor of 6-100 in mass, Table~\ref{table:models}) compared to the MW model.  Since there is less stardust in these models, if we require galaxies to ultimately reach the typical dust-to-gas ratios observed at low $f_g$ (Fig.~\ref{fig:HIDust}), we need to also include interstellar grain growth.  This dust source is strongly metal-dependent and only becomes important once the galaxy reaches a critical metallicity \citep{Asano2013}. This means that different values of the grain growth parameters $\epsilon$, and consequently $\tau_{\rm grow}$, move the model tracks. An increase of $\epsilon$ steepens the slope of $M_{d}/M_{g}$ (shown by Models IV-VI as they reach the end of their tracks); any offset from the linear trend in Fig.~\ref{fig:HIDust} can therefore be mitigated by changing $\epsilon$ such that grain growth starts at a lower metallicity (thereby increasing the dust-to-gas ratio). Alternatively, offsets in Fig.~\ref{fig:HIDust} can also be explained through the use of different bursty SFHs, because long quiescent phases allow accretion of existing metals
after short active enrichment episodes \citep{Zhukovska2014}.

The relative contributions to the dust mass budget for Models IV-VI are displayed in Fig.~\ref{fig:dust_sources}. At high gas fractions, stellar sources dominate (mostly SN dust, \citealp{Rowlands2014b}, \citetalias{Clark2015}), yet dust grain growth becomes the largest source of dust mass at gas fraction below 0.88, 0.79 and 0.53 for Model VI, V and IV respectively. The metallicity at which dust grain growth exceeds dust production from stars in our model is reached between $0.003 < Z <0.012$ (or $\rm 7.97 < 12 + log(O/H) < 8.63$, or $0.88>f_g>0.53$), though low values in this range result in the best match with observations. High values of $\epsilon$ (Table \ref{table:models}) lead to short grain growth timescales and low critical metallicities. By the time Models IV-VI reach the lowest gas fractions, dust grain growth produced 70-93\,per\,cent of the total dust mass created over the galaxy's lifetime. 

The ($Z$-dependent) dust grain growth timescales for Models IV, V and VI are shown in Fig.~\ref{fig:dust_sources} (bottom). For Model IV and V, the dust grain growth timescale $\tau_{\rm grow}$ decreases steeply at high gas fractions (when the critical metallicity is reached), and decreases gradually after that. 
For Model VI, we again find an initial steep decrease in $\tau_{\rm grow}$ at high gas fractions, yet afterwards we find some fluctuation and then an increase towards the lowest gas fractions. $\tau_{\rm grow}$ is inversely proportional to the SFR (Eqn. \ref{eq:tau_gg}) and the Model VI SFR decreases steeply towards low gas fractions (and is low in general). This explains the higher $\tau_{\rm grow}$ of Model VI compared to Model V, even though $\epsilon$ is higher. The fluctuation in the $\tau_{\rm grow}$ for Model VI is thus a result of the balance of the growing efficiency of grain growth as the metallicity increases and the SFH reaching a peak and decreasing steeply towards low gas fractions. We note that even though the dust grain growth is slower in Model VI than Model V, it is still faster relative to the SFR and thus a more important term in the chemical evolution.

\begin{figure}
\center
  \includegraphics[trim=2mm 0mm 0mm
  0mm,clip=true,width=0.75\columnwidth]{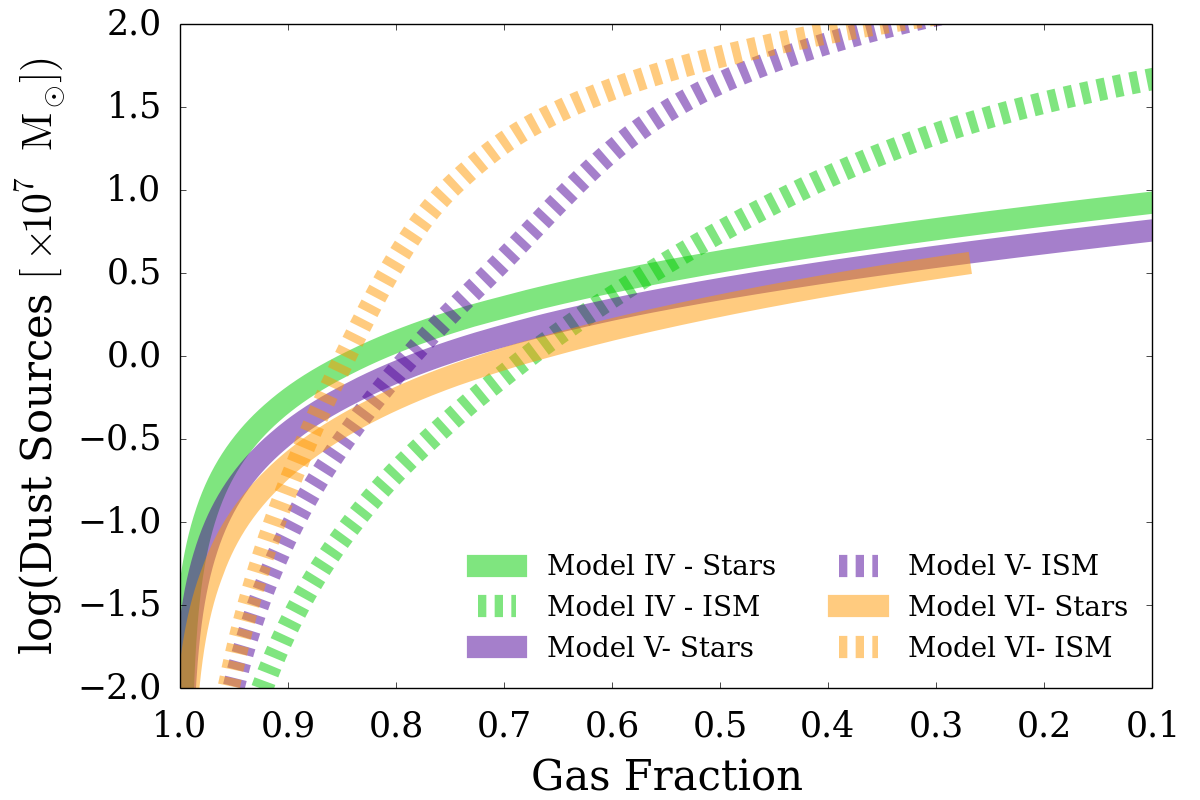}
  \includegraphics[trim=2mm 0mm 0mm
  0mm,clip=true,width=0.75\columnwidth] {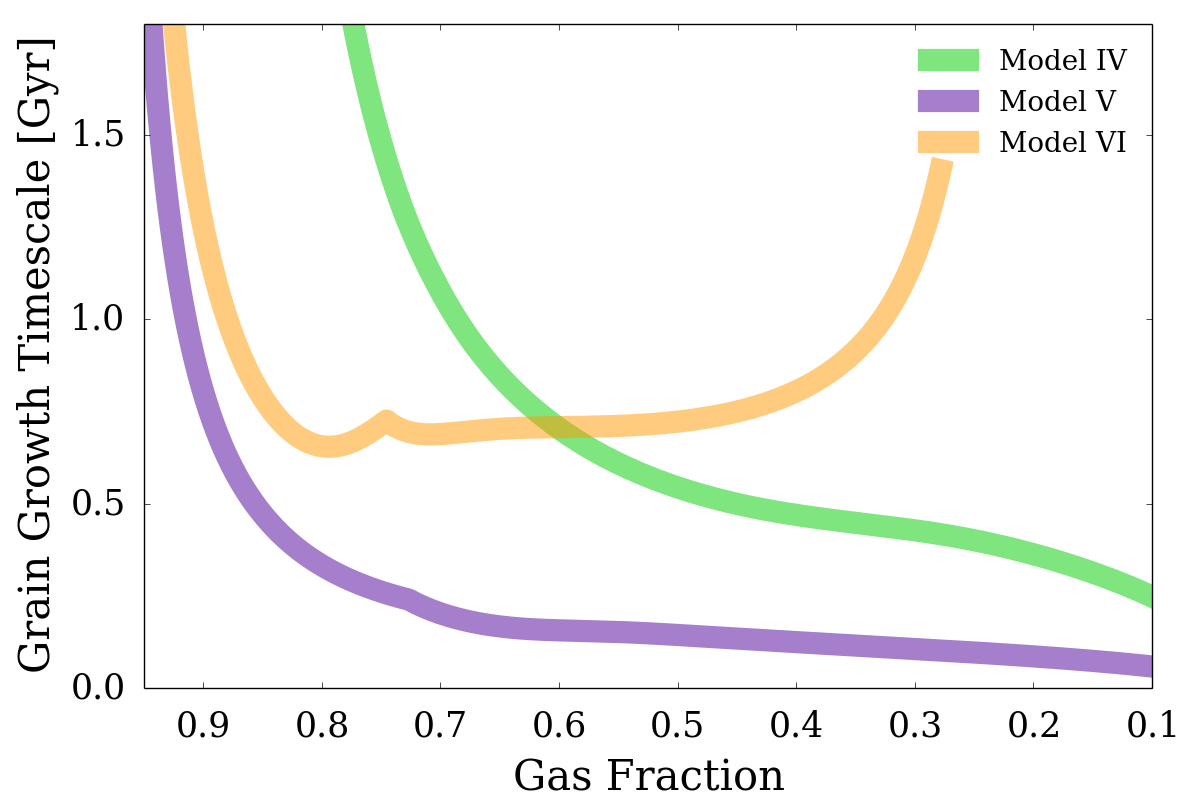}
\caption{({\it top:}) The dust mass produced by the various dust sources in Models IV, V and VI against gas fraction. Stellar dust sources dominate at the highest gas fractions and are overtaken by dust grain growth at lower gas fractions. ({\it bottom:}) The variation of the grain growth timescale $\tau_{\rm grow}$ (Equation~\ref{eq:tau_gg}) with gas fraction for Models IV, V and VI.  The growth timescale remains long until the critical metallicity is reached.}
  \label{fig:dust_sources}
\end{figure}
\subsection{Dust-to-metal ratio}
\label{sec:dust_metals}
\begin{figure}
\includegraphics[trim=3.5mm 0mm 0mm
  0mm,clip=true,width=1.0\columnwidth]{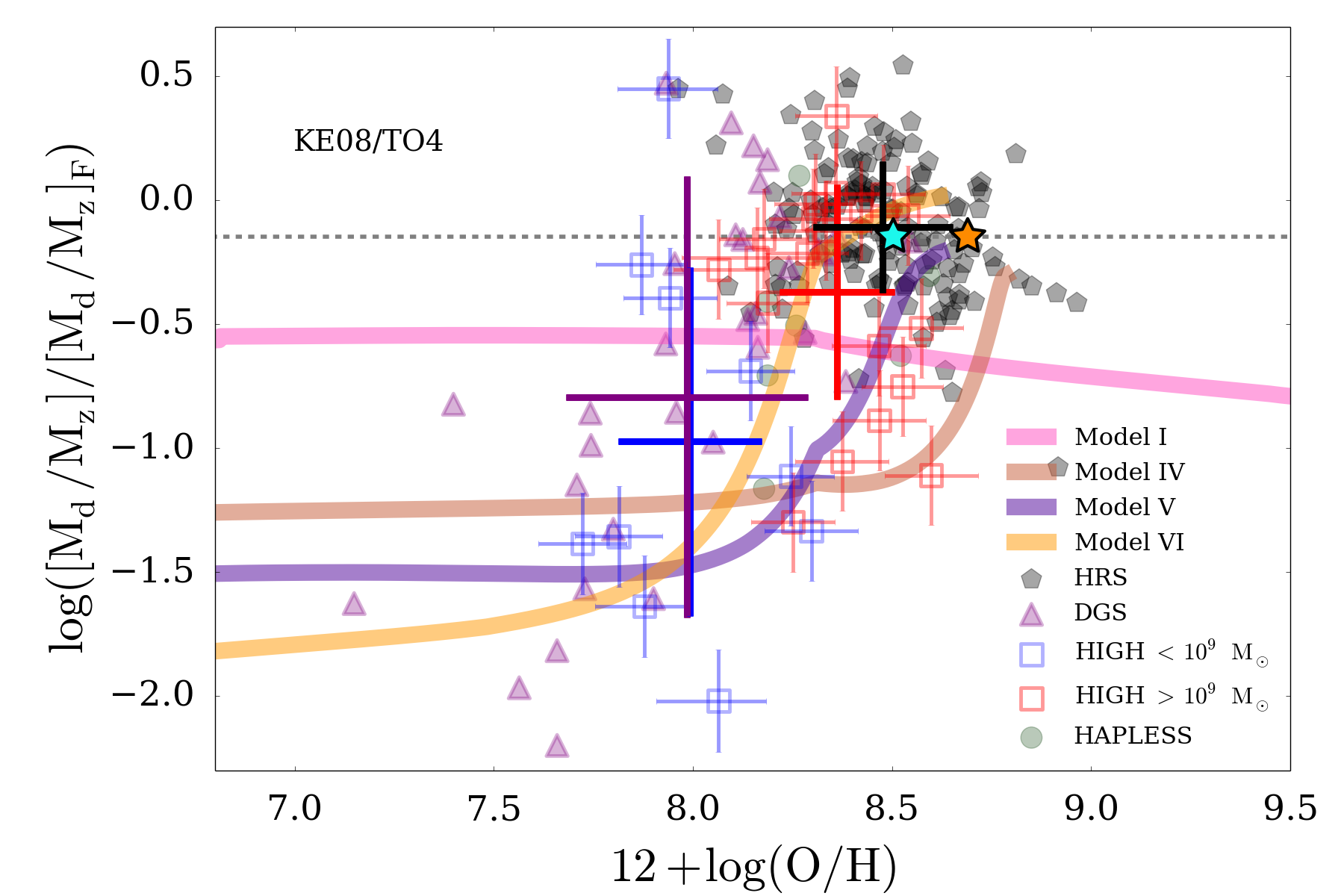}
\caption[$M_d/M_Z-Z$]{Dust-to-metal ratio versus metallicity (to allow comparison with \citet{Feldmann2015}, we use a normalisation\footnote{The normalisation in \citet{Feldmann2015} is given by their model $M_d/M_Z$ at solar metallicity.} of $[M_d/M_Z]_F = 0.7$) for \high, HRS, HAPLESS and DGS. The dust-to-metal ratio is significantly lower for galaxies in the low metallicity regime regardless of how actively star forming these galaxies are. The large crosses show the mean $\pm$ standard deviation of dust-to-gas within the samples. We also highlight the MW ($M_d/M_Z = 0.5$, orange star) and recent estimates for galaxies in the Virgo Cluster \citep[][cyan star]{Davies2014}.}
  \label{fig:DustMetals}
\end{figure}

\begin{table*}
\caption{Average dust-to-metal ratio for the different galaxy samples and metallicity calibrations quoted as mean $\pm$ standard deviation. }
\centering
\begin{tabular}{lcccccc} \hline\hline
Galaxy Sample & \multicolumn{3}{c}{Mean ${\rm 12+log(O/H)}$} &
\multicolumn{3}{c}{Mean $M_{\rm d}/M_Z$} \\
& N2 &KE08/T04 & PG16S &  N2 &KE08/T04 & PG16S  \\ \hline
DGS & $\rm 8.02 \pm 0.28 $ &$\rm 8.34 \pm 0.19$ &$\rm 7.98 \pm 0.30$ &
$\rm -0.72 \pm 0.71$ & $\rm -0.44 \pm 0.33$ &$\rm -0.69 \pm 0.66$\\
\high-low &  $\rm 8.17 \pm 0.12$ &$ \rm 8.37 \pm 0.16$ &$\rm 7.99 \pm
0.18$ & $\rm -1.15 \pm 0.70$ &$\rm -1.38 \pm 0.69$ &$\rm -0.97 \pm 0.70$
\\
\high-high  & $8.48 \pm 0.12$ &$8.81 \pm 0.16$ &$8.36 \pm 0.14$ & $\rm
-0.49 \pm 0.43$ &$\rm -0.82 \pm 0.42$ &$\rm -0.37 \pm 0.43$\\
HRS (LTGs) & $\rm 8.57 \pm 0.15$ &$ 8.75 \pm 0.17$ &$\rm 8.48 \pm 0.18$ &
$\rm -0.20 \pm 0.24$ &$\rm -0.38 \pm 0.25$ &$\rm -0.11 \pm 0.27 $ \\
\hline
\end{tabular}
\label{table:dust_metals}
\end{table*}

The variation in the dust-to-metals ratio is discussed in \citet{Mattsson2014}, where they show there is only a small change in $M_d/M_Z$ (but increased scatter) observed in low metallicity environments \citep{DeCia2013} and at high redshifts \citep{ZafarWatson2013} even down to $Z < 1$\,per\,cent of Solar. \footnotetext{The normalisation in \citet{Feldmann2015} is given by their model $M_d/M_Z$ at solar metallicity.}In Fig.~\ref{fig:DustMetals}, we find that for HRS, and a small fraction of the other sources, there is indeed not much variation in the $M_d/M_Z$\footnote{To estimate the total metal mass $M_Z$ from the observed oxygen abundance from Section~\ref{sec:metals}, we assume $12+{\rm log(O/H)}_\odot = 8.69$ and a Solar metal mass fraction $Z_\odot = 0.014$ following \citet{Asplund2009}.}. The HRS is in good agreement with the MW value \citep{Clark2016} and the recent survey by \citet{Davies2014} using {\it Herschel} observations of galaxies in the Virgo Cluster. Models with stardust only (Model I) predict an almost constant dust-to-metals ratio\footnote{The kink in Model I at $\rm{12+log(O/H)}\sim 8.4$ is due to an increasing metal mass from stars resulting from changing the input metal yield file \citep{Maeder1992}.}.

However, there are also $\sim25$ low metallicity (DGS and \high) sources for which $M_d/M_Z$ is significantly smaller. We note this result is independent on which metallicity calibration was used. Results for N2 and T04/KE08 are included in Table~\ref{table:dust_metals} for comparison with PG16S. We thus use our larger and more normal-star forming sample at low $Z$ to further support the \citet{Feldmann2015} result that the dust-to-metal ratio varies as a function of metallicity. The location of the low stellar mass samples (\high-low and DGS galaxies) is contrary to what we would expect if stellar sources were the dominant source of dust in the galaxies, which again shows we cannot model these sources without grain growth. \citet{Feldmann2015} attributes the rising dust-to-metal ratio to dust grain growth becoming more efficient as galaxies reach their critical metallicity. In contrast to the strong inflows and outflows, and the extremely efficient interstellar grain growth (timescale of $\sim 4$ Myr) from \citet{Feldmann2015}, we find we can also model these sources with more moderate inflows and outflows ($2.5\times$ SFR), and moderate grain growth (timescale of 1\,Gyr - 200\,Myrs) models (Model IV-VI). The dust growth timescales in our models are more similar to those quoted for the MW and local galaxies \citep{Draine2009,Asano2013,Mattsson2012,Mattsson2014}.

We note it is very difficult for a model with SFH consistent with the MW, with dust from LIMS combined with \emph{significant} dust production in supernovae and \emph{no dust destruction or grain growth} (Model I, peak $M_d/M_Z \sim 0.2$) to reach the observed MW dust-to-metals ratio ($\sim 0.5$; orange star in Fig.~\ref{fig:DustMetals}) as it evolves.
This issue demonstrates why significant interstellar grain growth is needed to supplement the dust mass and reduce the large offset in the predicted $M_d/M_Z$ in Model I compared to the observed values, even in our own galaxy.

\subsection{Statistical constraints}
We next attempt to check whether the comparison between the data and Models I-VI in earlier Sections provide strong constraints on the physical properties of these galaxies given the degeneracies in the model.  The scatter in the observed values between the different galaxy samples compiled in this work and others, and indeed within samples are often much larger than their error bars. This suggests there is an intrinsic source of scatter in the observed data, and in this case, one would not expect one model to provide a good fit to the whole sample or even to subsets based on simple flux selection criteria.  Here then, we focus on comparing whether one model, or class of models can provide a better description of the data than others, rather than derive the model that describes the data best in an absolute way. We aim to constrain the degeneracies and parameters of our chemical evolution models in a future paper (De Vis et al., \textit{in prep.}) using a Bayesian approach on a large grid of models.  In this section, we will test whether the ``eye-ball fits" used here to select ``preferred models" (and in similar other studies \citealp{Zhukovska2014,Remy-Ruyer2013,Feldmann2015}) are, in fact, statistically robust. 

We do this by calculating a statistic to measure the goodness of fit between the data and each of the models as plotted in Figs.~\ref{fig:baryonic}, \ref{fig:SFRMbary}, \ref{fig:HIDust}. For simplicity, we consider just two data samples: the full sample of all the galaxies together; and the HIGH-low subset. Using all of the galaxies equally weighted together is naturally dominated by the HRS sample which contains the most sources. The HIGH-low subset departs from the typical chemical trends seen in other, more evolved, galaxy samples, and so although we have to contend with small number statistics, this subset represents an unusual and interesting population. 

Since our measurements have significant uncertainties in both the $x$ and $y$ values, and our models are non-linear in $y$ as a function of $x$, it is not possible to simply use a standard $\chi^2$ approach to determine the goodness of fit.  We use a Bayesian approach to determine an effective $\chi^2$ which includes the  measurement uncertainties in 2 dimensions. We start from Bayes theorem:
\begin{equation}
P(\text{model} | \text{data}) = P(\text{data} | \text{model})\times P(\text{model})/P(\text{data}).
\end{equation}
For two measured parameters, $x$ and $y$, an observed data point $d_i$ is given by $(x_i,y_i)$ with Gaussian uncertainties ($\sigma_{x_i},\sigma_{y_i}$). If we have a model linking $x$ and $y$, $y=f(x)$, and we know that the true value for $x$ is $x_0$, then the true value for $y$ is given by $f(x_0)$. For this situation, the probability of observing $x_i$ and $y_i$ is given by:
\begin{align}
P( d_i | (x_0 \& f))  \propto \exp\left(-\frac{(x_i-x_0)^2}{2\sigma_{x_i}^2} - \frac{(y_i-f(x_0))^2}{2\sigma_{y_i}^2}\right).
\end{align}
Since the {\sc magphys} uncertainties used in this work are often asymmetric, we have used the lower x-errorbar if $x_0<x_i$ and the upper errorbar if $x_0>x_i$, and equivalently for the y-errorbars and $y_i$ compared to $f(x_0)$. 

To find the probability of observing the data point given the model $f$, we need to integrate over all possible values for the true value $x_0$. Assuming a uniform prior on $x_0$, this becomes, 
\begin{align}
P( d_i | f) \propto \displaystyle \int{ \exp\left(-\frac{(x_i-x_0)^2}{2\sigma_{x_i}^2} - \frac{(y_i-f(x_0))^2}{2\sigma_{y_i}^2} \right)} dx_0 . \label{eq:probdatapoint}
\end{align}
This integral can be solved numerically since the function $f(x_0)$ can be described numerically from the output of the chemical evolution models. The normalisation for Equation \ref{eq:probdatapoint} is found from determining the maximum probability that would be obtained for any y value for the same x-value as the real data point. We thus find the $y_{\text{max}}$ that maximises Equation~\ref{eq:probdatapoint} with respect to $y$,  and use this as the normalisation: 
\begin{align}
P( d_i | f) = \displaystyle \frac{\displaystyle \int{  \exp\left({-\frac{(x_i-x_0)^2}{2\sigma_{x_i}^2} - \frac{(y_i-f(x_0))^2}{2\sigma_{y_i}^2}} \right) dx_0 } }{\displaystyle\int{  \exp\left({-\frac{(x_i-x_0)^2}{2\sigma_{x_i}^2} - \frac{(y_{\text{max}}-f(x_0))^2}{2\sigma_{y_i}^2}}\right) dx_0}}. \label{eq:probdatapoint2}
\end{align}

\begin{table*}
\caption{The goodness of fit of each Model as expressed by $\langle \chi^2_{\text{eff}} \rangle$ for the combined HRS+HAPLESS+\high\ sample and for \high-low. Results for Figs.~\ref{fig:baryonic},~\ref{fig:SFRMbary}~and~\ref{fig:HIDust} are shown for both samples. $M_g/M_*$ was used instead of $f_g$ because for $f_g$ the approximation of a Gaussian error probability distribution is not valid.}
\centering
\begin{tabular}{lcccccc} \hline\hline
Galaxy Sample & \multicolumn{3}{c}{Combined} &
\multicolumn{3}{c}{\high-low} \\ \cmidrule(lr){2-4} \cmidrule(lr){5-7}
Fig. & \ref{fig:baryonic} & \ref{fig:SFRMbary} & \ref{fig:HIDust} & \ref{fig:baryonic} & \ref{fig:SFRMbary} & \ref{fig:HIDust} \\
y-axis &  $M_d/M_{\text{bary}}$ & SFR$/M_{\text{bary}}$ & $M_d/M_g$ & $M_d/M_{\text{bary}}$ & SFR$/M_{\text{bary}}$ & $M_d/M_g$ \\
x-axis &  $M_g/M_*$ & $M_g/M_*$ & $\rm{12+log(O/H)}$ & $M_g/M_*$ & $M_g/M_*$ & $\rm{12+log(O/H)}$ \\\hline
Model I & 1.90 & 7.55 & 2.73 & 1.28 & 20.45 & 1.65\\ 
Model II & 2.06 & 3.56 & 2.80 & 1.47 & 2.72 & 1.68\\ 
Model III & 0.84 & 10.45 & 2.79& 1.36 & 3.01 & 1.68\\ 
Model IV  & 1.31 & 3.55 & 8.64 & 0.94 & 2.72 & 2.38\\ 
Model V & 1.03 & 3.53 & 1.09 & 0.89 & 2.72 & 2.19\\ 
Model VI & 3.51 &  6.43 & 0.41 & 0.59 & 3.77 & 1.45\\ 
\hline
\end{tabular}
\label{table:chi2s}
\end{table*}

For a linear model, $y_{\text{max}}=f(x_0)$, so a data point that lies on the model will correspond to $P(d_i | f)=1$, which is equivalent to $\chi^2=0$, as expected. This normalization reproduces the standard form of $\chi^2$ for a linear function with uncertainties in both $x$ and $y$. However, our models are not linear and as a result $y_{\text{max}}$ is not necessarily equal to $f(x_0)$. By normalising $P(d_i |f)$ in this way, we ensure the estimated probability is never larger than 1. This normalisation is thus necessary to obtain sensible results. Formally we should normalise by the maximum of $P(d_i|f) $ in both $x$ and $y$, but this 2-d maximisation is time-consuming, and our approximation is valid so long as $\sigma_x \frac{ d^2y}{dx^2} < \frac{dy}{dx}$ and $\sigma_y \frac{d^2x}{dy^2} < \frac{dx}{dy}$, which is the case for our models and data. 

The likelihood for the entire sample ${\bf d}=\{d_i\}$ is then found from taking the product of each of the probabilities for the individual data points from Equation \ref{eq:probdatapoint2}. Equivalently, the logarithms of the probabilities can be added to give the log likelihood:
\begin{align}
\log({\cal L}( \mathbf{d} | f)) = \sum_i \log( P(d_i | f) ).
\end{align}
Additionally, in order to allow for more easily interpretable results, we convert $P(d_i | f)$ from Equation \ref{eq:probdatapoint2} to an effective $\chi^2$: 
\begin{align}
\chi^2_{\text{eff}}=  \sum_i -2\log( P(d_i | f) )
\end{align}
Finally, so we can easily compare samples with different number of galaxies, $n$, we calculate the average contribution to $\chi^2_{\text{eff}}$ per data point,
\begin{align}
\langle \chi^2_{\text{eff}} \rangle= \frac{\sum_i -2\log( P(d_i | f) )}{n}
\end{align}
This statistic is similar to reduced $\chi^2$, but does not allow for the number of degrees of freedom; the true number of degrees of freedom is hard to quantify because of correlations between the effects of parameters in our models.  Nevertheless, our models are essentially different parameter choices for one over-arching model, and so they all have the same effective degrees of freedom, and the statistic does allow a fair comparison between them. 

In this section we revisit three of the main figures in this work and determine how well Models I-VI fit the observations. Table~\ref{table:chi2s} shows the $\langle \chi^2_{\text{eff}} \rangle$ for the combined HRS+HAPLESS+\high\ sample and for \high-low for Figs.~\ref{fig:baryonic},~\ref{fig:SFRMbary}~and~\ref{fig:HIDust}. In these Figs., the models do not reach the lowest gas fractions and highest metallicities. Therefore in our determination of  $\langle \chi^2_{\text{eff}} \rangle$, we have discarded all data points with $f_g<0.2$ for Figs.~\ref{fig:baryonic}~and~\ref{fig:SFRMbary} and all data points with $12+\log(O/H)>8.5$ for Fig.~\ref{fig:HIDust}. Figs.~\ref{fig:baryonic}~and~\ref{fig:SFRMbary} use gas fraction $f_g$ on the x-axis. However, the uncertainties on $f_g$ become very non-linear near $f_g\sim 1$, so the approximation of a Gaussian error probability distribution is not valid, and the calculation of $\langle \chi^2_{\text{eff}} \rangle$ is unreliable if we use $f_g$ as the $x$ variable. Therefore we have used $M_g/M_*$ instead. This is entirely equivalent to $f_g$ ($M_g/M_*=f_g/(1-f_g)$), but has errors that are more close to Gaussian, and so our statistic is acceptable. It is worth noting that Models II and V provide a reasonable fit to each of the studied relations and for both sub-samples ($\langle \chi^2_{\text{eff}} \rangle <4$). In the rest of this section we will study which of the models fits which relation best and how we interpret these results.

In Table~\ref{table:chi2s} we find that the observations for $M_d/M_{\text{bary}}$ versus $M_g/M_*$ (or equivalently $f_g$) are best fitted by Model VI for \high-low. We thus find statistical confirmation for our result from Section~\ref{sec:baryon} that rapid dust grain growth is necessary to model the unevolved \high-low sources.
For the combined sample, Model III provides the best fit. Here the HRS dominates the $\langle \chi^2_{\text{eff}} \rangle$ because of its large sample size. As previously mentioned in Section~\ref{sec:baryon}, a reduced dust contribution from LIMS would reduce the $M_d/M_{\text{bary}}$ at low $f_g$ and thus provide a better match for Models I and II.

The large $\langle \chi^2_{\text{eff}} \rangle$ in Table~\ref{table:chi2s} for SFR$/M_{\text{bary}}$ against $M_g/M_*$ for all models and both sub-samples indicate that a single model cannot describe the spread in SFR$/M_{\text{bary}}$. For both the combined sample and \high-low, Model I provides a very poor fit. Delayed SFH thus provide a better description of the overall SFH of normal star-forming galaxies than exponentially declining SFR. For the combined sample, Model III also provides a poor match. The reduction in baryon mass due to the outflows is not matched by reduced star formation in the delayed SFH. We also find Model VI provides a poor match to the combined sample, and surprisingly also to \high-low. As can be seen in Fig.~\ref{fig:SFRMbary}, \high-low includes some very actively star-forming galaxies with small errorbars. These sources are more in line with the bursty sources in DGS and are different from the normal star-forming galaxies we tried to describe by the reduced delayed SFH in Model VI. Therefore, \high-low cannot successfully be described by a single Model in Fig.~\ref{fig:SFRMbary}.

For $M_d/M_g$ versus $\rm{12+log(O/H)}$, Table~\ref{table:chi2s} shows Model VI provides the best fit to both \high-low and the combined sample. This thus provides further corroboration of dust grain growth being necessary to model the build up of dust as galaxies evolve. For \high-low the scatter is rather large, and none of the models provides a particularly good fit.

\subsection{Discussion}
We have found that, in order to get satisfactory fits to the observed $M_d/M_{\rm bary}$, ${\rm SFR}/M_{\rm bary}$, ${\rm 12+log(O/H)}$, $M_d/M_{g}$ and $M_d/M_{Z}$ for the different nearby galaxy samples in this work, it is necessary to reduce the SNe dust contribution (by a factor of 6-100) and include moderate inflows and outflows, dust destruction and moderate grain growth in our models. 

The reverse shock in the remnants of SN likely reduces the produced dust \citep{Bianchi2007,Gall2011,DeLooze2016}. The \citet{Todini2001} SN dust prescription
used in our models does not include a correction for dust destruction by reverse shocks. The need to reduce the SN dust
contribution in Models IV-VI could thus, at least in part, be due to dust destruction by the reverse shock.

There is a growing number of studies that suggest significant amounts of dust grain growth are required to model observations in both high and low redshift studies \citep{Dwek2007,Matsuura2009,Michalowski2010,Mattsson2012,Asano2013,Grootes2013,Calura2014,Rowlands2014,Zhukovska2014,Nozawa2015,DeCia2016}.
On the other hand, \citet{Ferrara2016} point out the difficulties in obtaining high enough grain growth efficiencies to explain the observations. The subject of dust grain growth thus remains a debated subject.
For Models IV-VI, we find grain growth timescales ranging from 1 Gyr - 200 Myrs, similar to those quoted for
the Milky Way and local galaxies\citep[e.g.][]{Asano2013,Mattsson2012}). There is also evidence for shorter timescales \citep{Draine2009,Zhukovska2008,Feldmann2015}, which might be more appropriate for the more dust-rich sources at low-metallicity or higher metallicity sources. Variations in the dust growth timescales might also help to explain the differences between dust-rich and dust-poor sources at the same (high) gas fraction. If the reverse shock destroys the majority of the dust grains in SN for all galaxies (and not only the ones modelled well by Model IV-VI), then the higher dust mass sources (which are now fitted by Models I-III) could be explained by shorter dust grain growth timescales, and high dust mass can be reached in spite of a reduced SN contribution. In this scenario, all galaxies have a strongly reduced SN dust contribution compared to \citet{Todini2001}, and galaxies with short dust grain growth timescales result in a higher dust content (on the level of Model I), and galaxies with long dust grain growth timescales (such as in Model VI) will have a lower dust content at high gas fractions.

\subsection{Caveats }
\label{sec:caveats}
In the previous sections we have used a range of models to explain the dust properties in the dust-poor low-$Z$ sources, as well as dust-rich lower gas fraction sources. In this section we discuss potential caveats of our approach.
\begin{itemize}
\item {\bf Dust Emissivity} - If the dust emissivity is different across the samples, this could explain the reduced $M_d/\MHI$ seen in Fig.~\ref{fig:HIDust} and in $M_d/M_Z$ (Fig.~\ref{fig:DustMetals}).  For the dust poor \high-low sample to have a dust/metals ratio similar to the HRS and \high-high samples (i.e. $\sim 0.4-0.5$), $\kappa$ would have to be $\sim$4 times lower.
\item {\bf Missing molecular gas} - We lack sufficient molecular gas information for the HAPLESS and \high~samples. To affect our results, the molecular mass would have to be larger than the \hi\ mass. This does not agree with observed molecular gas masses for the HRS and DGS, nor with the galaxy gas-scaling laws from \citet{Saintonge2011} and \citet{Bothwell2014} for a wide range of stellar masses. These results suggest that $M_{\rm H_2}/M_{\rm bary}$ is small at all evolutionary phases (see \citetalias{DeVis2017} for more discussion). Using CO derived $\rm H_2$ masses for HRS from \citet{Boselli2014}, we find that including the molecular gas component does not change the conclusions of our work. At low gas fractions, $M_{\rm H_2}/\MHI$ is large for some sources, and subsequently these will shift to higher gas fractions and higher total gas masses when molecular gas is included. This shift only results in a better fit to the models at low $f_g$ (e.g.  Fig.~\ref{fig:HIDust}).

To study the effects of molecular gas at high gas fraction, we took $M_{\rm H_2}$ for DGS from \citet{Remy-Ruyer2014}. These were derived by converting CO fluxes using a constant conversion factor $X_\mathrm{CO,MW}$ \citep{Ackermann2011} or a metallicity-dependent conversion from CO $X_\mathrm{CO,Z}$ \citep{Schruba2012}. Using $X_\mathrm{CO,MW}$, we again find $M_{\rm H_2}/\MHI$ is small for all but the lowest gas fractions. The small shift at low gas fractions again results in a better fit with the models. However, if we use $M_{\rm H_2}$ derived using $X_\mathrm{CO,Z}$, we find significantly higher $M_{\rm H_2}/\MHI$ at high gas fractions and thus again a shift towards higher gas masses and gas fractions compared to not including molecular gas masses. For the high gas fraction sources this results in a poorer fit to the models assumed here, though the offset does not change our conclusions. Different $X_\mathrm{CO,Z}$ factors (see \citealt{Bolatto2013} for a review) lead typically to smaller $M_{\rm H_2}$ than for $X_\mathrm{CO,Z}$ from \citet{Schruba2012}, and would thus result in smaller offsets.

\item {\bf Increased dust destruction} - We have investigated whether it is possible to explain the observed dust-to-gas properties of the \high\ galaxies by increasing the amount of dust destruction as opposed to reducing the dust production from SNe. We can model increased dust destruction in two ways. Firstly we can increase the amount of dust which is destroyed per SNe (by adjusting the value of $m_{\rm ISM}$ in Eq.~\ref{eq:dustmasst}) and secondly by adjusting the value of $f_c$ (the fraction of the ISM in the cold phase). With a larger fraction of the ISM in the warm phase, the efficiency of the dust destruction in the galaxy will be increased. We find that changing dust destruction alone can not match the observed \mdmb\ and $M_d/M_g$ ratios, since an increased dust destruction does not reduce the dust produced at the high gas fractions ($f_g > 0.8$). Even an extreme model with $m_{\rm ISM} = 2500\,\rm{M_\odot}$ and $f_c = 0.01$ would still require significant SNe dust reduction to explain the observed dust-to-gas values. Alternatively, if we change $f_c$ to vary with $f_g$, this makes at most a factor of 2 difference to dust destruction, whereas we need a reduction of the dust mass by a factor of 10-100 at high $f_g$. Therefore the conclusion of needing a reduced dust yield from SNe first put forward by \citet{Zhukovska2014} is robust to changes in the values of $f_c$ and $m_{\rm ISM}$. We note that changes in $f_c$ and $m_{\rm ISM}$ could reduce the offset in \mdmb\ between the observations and some of the models at low gas fractions. 

Given our assumption that gas and dust are uniformly mixed, the dust destruction in our model is proportional to the global dust-to-gas ratio. However, in reality $M_d/M_g$ could be higher in star forming regions than the global average $M_d/M_g$. More of the SN dust will thus be destroyed before being mixed into the diffuse ISM than is currently the case in our model.  \citet{Hopkins2016} show that on molecular cloud scales, gas-grain decoupling can lead to fluctuations in the local dust-to-gas ratios.  For the highest dust-to-gas ratios they predict, dust destruction could remove a significant fraction of the dust (even at high gas fractions) compared to the current model. Conversely, dust grain growth would become more efficient.   This could thus provide an alternative interpretation for the scatter in $M_d/M_{\rm bary}$ at high gas fractions and the need to reduce the SN dust contributions.  A full treatment of this issue requires spatially resolved chemical evolution modeling, which is outside the scope of this work.

\item \textbf{Initial mass functions} - We have also tested how different IMFs change our results. Changing the model IMF to a more bottom-heavy IMF (e.g. \citeb{Salpeter1955} or \citeb{Cappellari2012}), reduces the dust and metals produced in the first generation of stars, which results in a better match of these models compared to the observations (i.e. smaller \mdmb\ at high gas fractions and smaller $Z$ at low gas fractions). Similarly at high gas fractions, a top-heavy IMF in the model could increase $Z$.  But to change the model IMF we must also scale the observational parameters which have been determined using the Chabrier function. For example, using a top-heavy IMF with slope $\alpha = -1.5$ \citep{Cappellari2012,Madau_2014} we would have to scale the stellar mass and SFRs by a factor of 0.32 \citep{Micha_2015}. This results in models that are nearly indistinguishable (in terms of a `good-fit') compared to the scatter in the relations.
\end{itemize}

\section{Conclusions}
\label{sec:conclusions}
In this paper, we have brought together the H{\sc i}-selected \high, dust-selected HAPLESS, stellar mass selected HRS and the metallicity-selected DGS sources.
Compared to the 126 sources from \citet{Remy-Ruyer2014}, we have increased the sample size to 382 sources (including 48 DGS sources in both samples).
Beyond the 37 DGS sources with $Z<1/5 \, Z_\odot$, we have added a further 67 sources with a metallicity smaller than $1/3 \, Z_\odot$, including 15 sources below $1/5 \, Z_\odot$.
Following \citet{Zhukovska2014} and \citet{Feldmann2015}, we have investigated the dust trends of these samples using a chemical evolution model (an updated version of \citeb{Rowlands2014b} and \citeb{Morgan2003}). We use the PG16S metallicity calibration, which was found to be the most reliable calibration for the low metallicity sources, and gas fraction (a proxy for the evolutionary state) to track and constrain the build-up of dust and metals as gas is converted into stars, from very high ($f_g= 0.97$) to very low ($f_g= 0.05$) gas fractions. We find that:
\begin{itemize}
\item DGS sources are selected to have low metallicities, which leads to a selection of very actively star forming galaxies. For a given gas fraction or stellar mass, we have found our low $M_*$ \high\ and HAPLESS samples to be more normal in terms of star formation properties and metallicity. These samples thus complement the DGS, and provide additional, new information on more normal star-forming galaxies in the nearby Universe.
\item Delayed star formation history models are necessary to match the evolution of ${\rm SFR}/M_{\rm bary}$ for our normal star-forming galaxies.
\item We find that low--moderate metallicity galaxies (a) can have dust properties that are consistent with dust production at early stages being dominated by SNe dust (as in \citetalias{Clark2015}), and thus with a linear $M_d/M_g - Z$ relationship and constant $M_d/M_Z$; or (b) have dust masses well below these trends, with a much smaller contribution from SNe dust. The lowest metallicity sources fall in the latter category and to model them we require a maximum of $0.01-0.16\,\rm M_\odot$ of dust per core collapse SN.
\item The dust properties and the observed correlation of $M_d/M_Z-Z$ for low metallicity sources are well matched when including: ({\sc i}) reduced stardust contribution by 6-100, particularly from core-collapse SNe as the reduced dust component has to act at very high gas fractions. ({\sc ii}) Moderate ($2.5\times$ SFR) enriched outflows and metal-poor inflows to keep the model metallicity from rising to higher than observed metallicties at low gas fractions. ({\sc iii}) Dust destruction and moderate grain growth (timescales ranging from 1\,Gyr - 200\,Myrs, similar to those quoted for the MW and local galaxies; \citealt{Draine2009,Asano2013,Mattsson2012,Mattsson2014}). The need for this moderate grain growth is corroborated by the good statistical match of Model VI to the $M_d/M_g-Z$ relation for both \high-low and the combined HRS+HAPLESS+\high\ sample.
\item As we show that neither the dust-to-metals nor the dust-to-gas ratio are constant during the evolution of a galaxy, we urge caution when using dust as a tracer of gas mass in galaxies \citep[e.g.][]{Eales2010,Scoville2014}. Assuming a universal value for either is unwise and unlikely to produce reliable results, particularly for low stellar mass systems.
\item In our best models, we find that grain growth produces, by mass, 70-93\,per\,cent of the total dust created over the lifetime of these galaxies, and the metallicity at which dust grain growth exceeds stellar dust sources in our model is reached between  $7.97 <12 + \rm{log(O/H)} < 8.63$ (or $0.88>f_g>0.53$). 
\end{itemize}

We show our Model VI (SN dust contribution reduced by factor 100, inflows and outflows of $2.5\times$ SFR, delayed SFH, dust grain growth and destruction) is consistent with all of the observed properties (except the SFR for some rather bursty sources) of the \high-low galaxies, the first normal star forming population of low stellar mass galaxies studied in this way. Comparing the data and models using a Bayesian approach confirms that Model VI provides the best statistical match to the \high-low data for the dust-to-baryon ratio against gas fraction and the dust-to-gas ratio against metallicity. For $\rm{SFR/M_{bary}}$, it is not possible to find one model that describes all the \high-low data since the intrinsic scatter within the sample is larger than the errorbars. When Model VI is combined with a bursty SFH (as shown originally in \citealt{Zhukovska2014}) and stronger outflows (\citeb{Feldmann2015}, Model VII), this scenario is also consistent with the DGS galaxies at similar $f_g$, $M_*$ and $Z$ without requiring extremely rapid grain growth timescales and extreme outflows for low metallicity galaxies.

\section*{Acknowledgments}
We gratefully acknowledge the anonymous referee for useful suggestions. We thank Suzanne Madden for providing the emission line tables for the Dwarf Galaxy Survey. PDV acknowledges funding from the Fund for Scientific Research Flanders. SPS acknowledges funding from the Science and Technology Facilities Council (STFC) Doctoral Training Grant scheme. HLG acknowledge support from the STFC Consolidated Grant scheme (ST/K000926/1). PDV, HLG, LD, PC and SJM acknowledge support from the European Research Council (ERC) in the form of Consolidator Grant {\sc CosmicDust}.  LD and SJM acknowledge support from the Advanced Investigator programme {\sc COSMICISM}. GAMA is a joint European-Australasian project based around a spectroscopic campaign using the Anglo-Australian Telescope. The GAMA input catalogue is based on data taken from the Sloan Digital Sky Survey and the UKIRT Infrared Deep Sky Survey. Complementary imaging of the GAMA regions is being obtained by a number of independent survey programmes including GALEX MIS, VST KiDS, VISTA VIKING, WISE, Herschel-ATLAS, GMRT and ASKAP providing UV to radio coverage. GAMA is funded by the STFC (UK), the ARC (Australia), the AAO, and the participating institutions. The GAMA website is http://www.gama-survey.org/ . This research has made use of GitHub, Astropy 21  \citep{Astropy2013}, NumPy 24 \citep{Walt2011}, SciPy 25, and MatPlotLib 26 \citep{Hunter2007A}.

\bibliographystyle{mnras}
\bibliography{Library}

\appendix
\section{Emission lines and galaxy properties}
The basic galaxy properties for the \high\ and HAPLESS galaxies are listed in Table~\ref{tab:properties} and their metallicities in Table~\ref{tab:HIGHmetals}. Star formation rates and stellar and dust masses were derived using {\sc magphys} (See \citetalias{DeVis2017} for details). Metallicities were derived for 4 different calibrations, using a weighted average of the metallicities from individual H{\sc ii} regions within the galaxy (Section~\ref{sec:metals}). The emission lines for each H{\sc ii} region in the \high\ and HAPLESS galaxies are listed in Table~\ref{tab:HIGHlines}. For DGS, emission lines from literature and derived metallicities are provided in Table~\ref{tab:DGSlines}.

\section{Dust mass equation}
\label{app:dusteq}
Here we reproduce the equation for the dust mass evolution, $M_d$, for the chemical evolution model used in this work:
\begin{align}
\frac{d(M_d)}{dt}=\int_{m_{\tau_m}}^{m_U}&\bigl(\left[m-m_{R}(m)\right]
Z(t-\tau_m)\delta_{\rm lims}+mp_z\delta_{\rm dust} \bigr) \nonumber \\
     & \mbox{} \times \psi(t-\tau_m)\phi(m)dm - (M_d/M_g)\psi(t) \nonumber \\
     & \mbox{} - \left(1-f_c\right) \frac{M_d}{\tau_{\rm dest}}+ f_c\left(1-\frac{M_d}{M_g} \right) \frac{M_d}{\tau_{\rm grow}}  \nonumber \\
     & \mbox{} + M_{d,i} + \left(\frac{M_d}{M_g}\right)_{I}I(t) - \left(\frac{M_d}{M_g}\right)_O {O}(t).
\label{eq:dustmasst}
\end{align}

$M_g$ is the gas mass, $\psi(t)$ is the star formation rate, $\phi(m)$ is the stellar IMF, $Z$ is the metal mass fraction defined as $M_Z/M_g$ and $m_R$ is the remnant mass of a star of mass $m$ \citep{Ferreras2000}.  The first term accounts for dust formed in stars and supernovae.  This includes metals re-released by stars after they die, and newly synthesised metals ejected in winds and supernovae.  The second term describes the removal of dust due to astration and the grain destruction and growth timescales are given in terms three and four. The fifth term allows us to include primordial dust in the galaxy for example associated with Pop III stars, we set this to zero \citep{Rowlands2014b}.  Finally, $I(t)$ and $O(t)$ are simple parameterisations of dust removed or contributed via inflows and outflows.   The lifetime $\tau_m$ of stars with initial mass $m$ is derived using the model in \citet{Schaller1992} and yields for LIMS and massive stars are taken from \citet{vandenHoek1997} and \citet{Maeder1992} respectively. A full discussion on the effect of using different yields can be seen in \citet{Rowlands2014b}, and \citet{Romano2010}.

\section{Comparison of models with Dwarf Galaxy Survey}
\label{app:dgs}

\begin{figure*}
\center
  \includegraphics[width=0.49\textwidth]{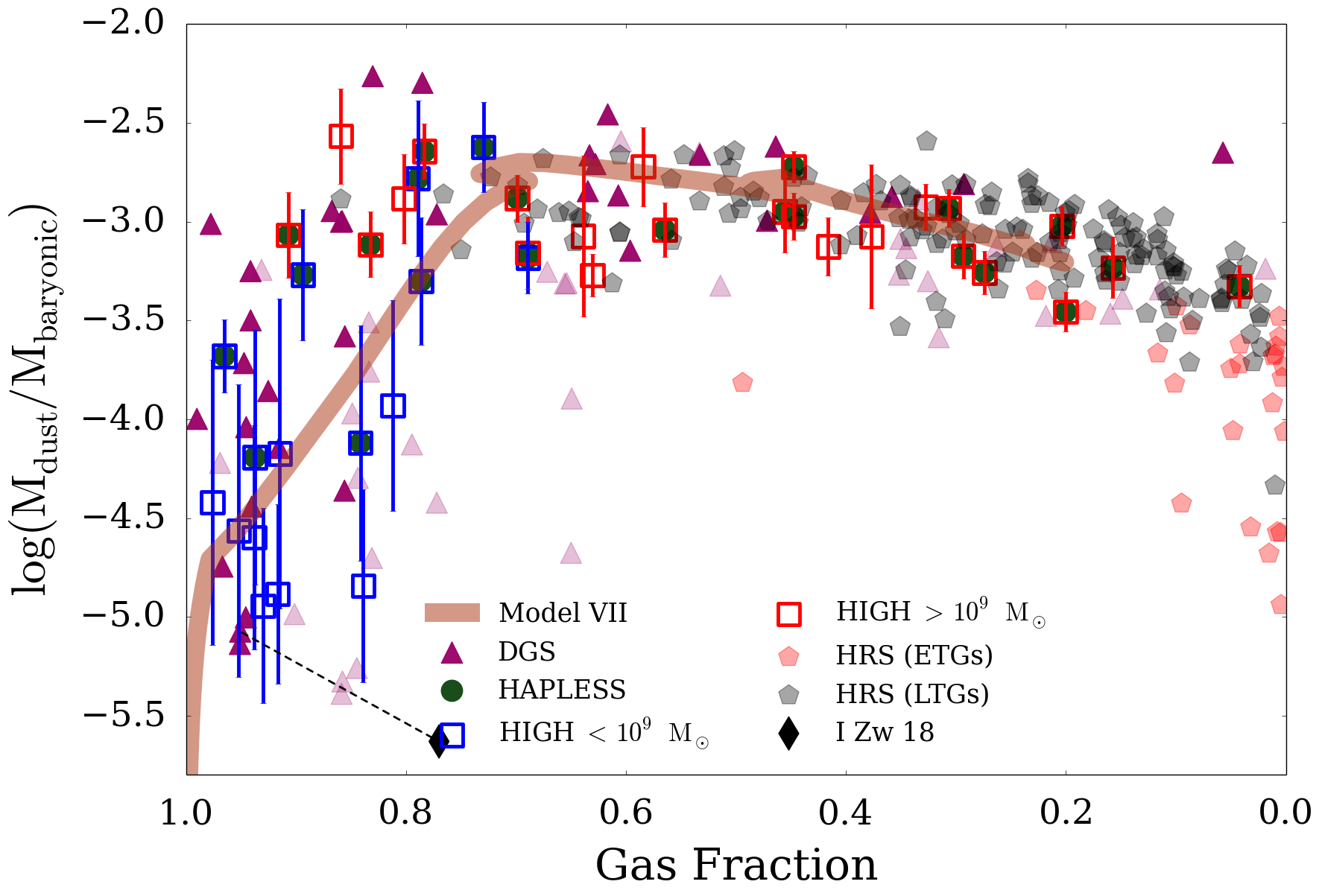}
  \includegraphics[width=0.49\textwidth]{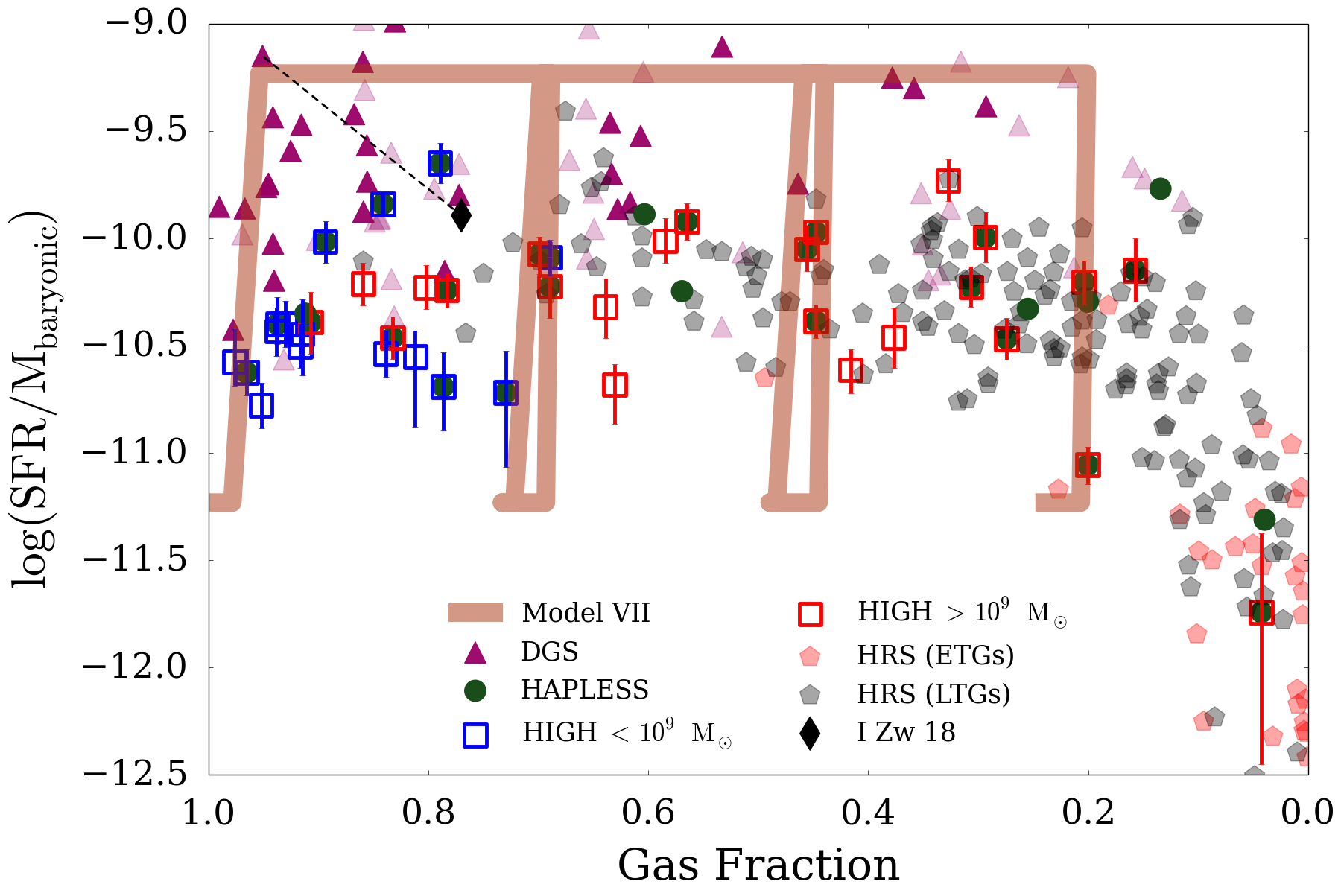}
  \includegraphics[width=0.49\textwidth]{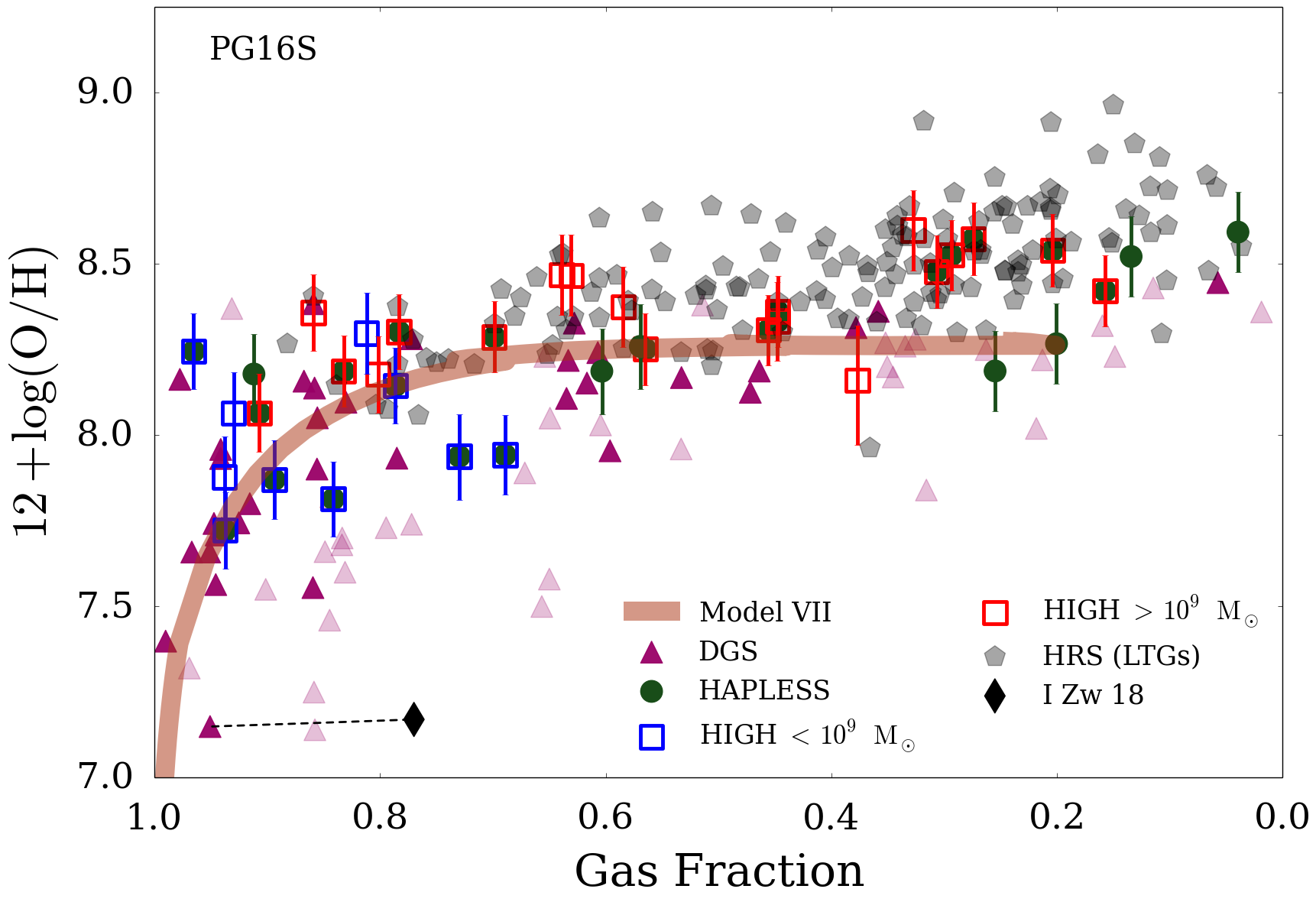}
  \includegraphics[width=0.50\textwidth]{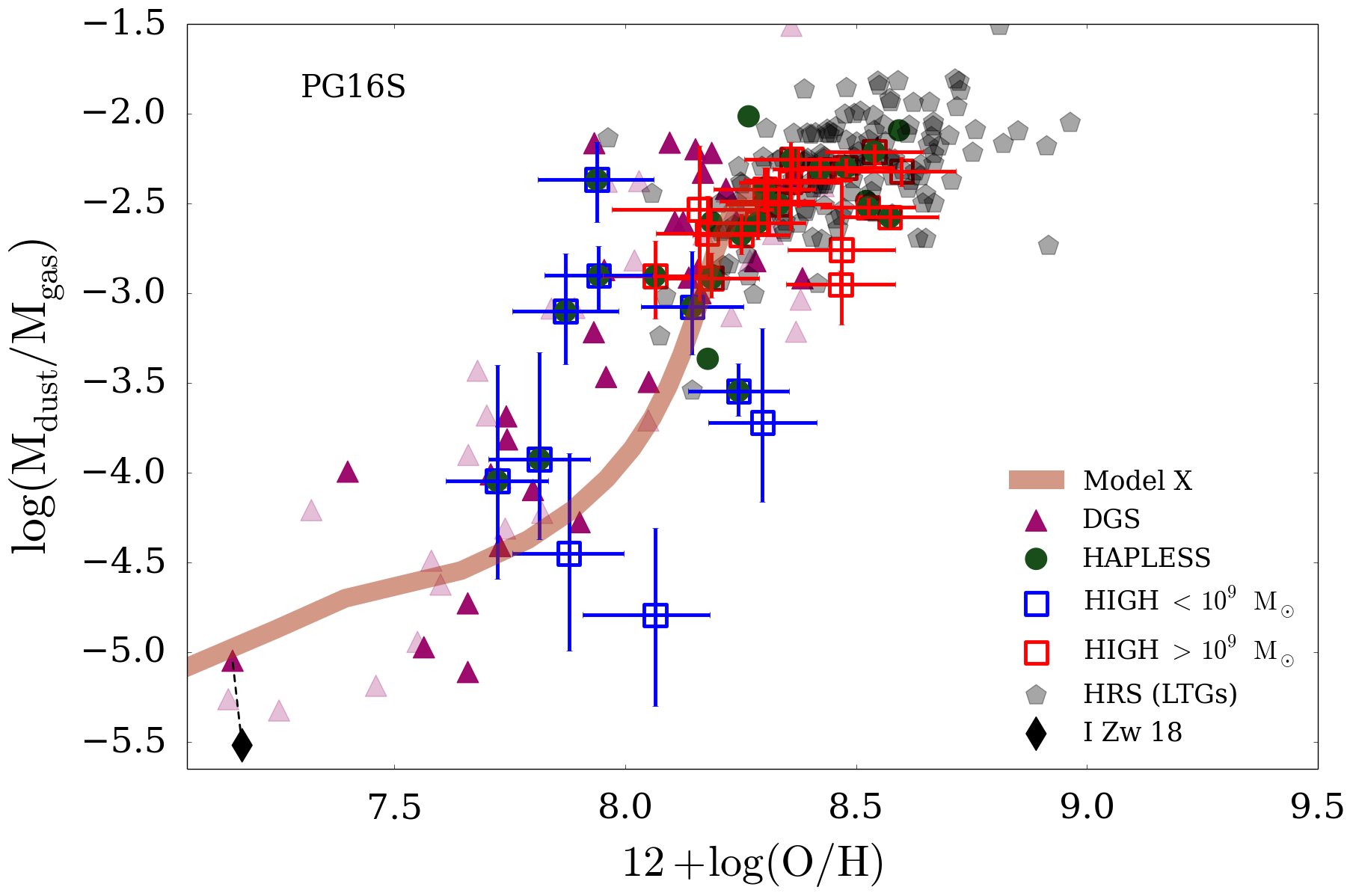}
  \caption{{\it Top:} \mdmb\ and \sfrmb\ evolution with gas fraction using a bursty SFH (Model VII). As shown in \citet{Zhukovska2014}, the DGS sample (purple triangles, transparent triangles are the same galaxies, yet with properties from \citet{Zhukovska2014}) can be explained with a model undergoing many bursts of star formation (brown line). {\it Bottom:} the metallicity variation is compared with gas fraction (\textit{left}) and $M_{d}/\MHI$ (\textit{right}). As shown in \citet{Feldmann2015}, the observed metallicity of the DGS galaxies can be explained by a chemical evolution model that incorporates strong inflows and outflows of gas (Model VII).}
  \label{fig:bursty_results}
\end{figure*}

We can model the properties of the DGS sources by including strong inflows and outflows \citep{Feldmann2015} and a bursty SFH \citep{Zhukovska2014} in the chemical evolution (Model VII). The results are shown in Fig.~\ref{fig:bursty_results} using the original DGS metallicities (transparent triangles), and the revised PG16S metallicities derived in this work. In the top-left panel, we compare the \mdmb\ of the DGS with Model VII (as we did with the HRS, \high\ and HAPLESS in Fig.~\ref{fig:baryonic}). Model VII matches the observed trend well. In the top-right panel of Fig.~\ref{fig:bursty_results}, we compare the predicted \sfrmb\ with gas fraction for Model VII. Here we see that the bursty model is required to explain the elevated \sfrmb\ of the DGS galaxies compared to the HAPLESS, HRS and \high~samples. In the bottom-left panel, we find the observed metallicities for DGS tend to be lower than for the other samples and are well matched by Model VII, due to including strong inflows and outflows at a rate of 4 times the SFR.

\FloatBarrier

\addtocounter{section}{-2}

\begin{table*}
            \caption{Basic properties for the H{\sc i}-selected \high\ sample (\citetalias{DeVis2017}) and dust-selected HAPLESS sample (\citetalias{Clark2015}). The check-marks indicate which sample the source is part of (24 sources are included in both samples).}
            \setlength{\tabcolsep}{6.5pt}
            \setlength{\extrarowheight}{4pt}
            \npdecimalsign{.}
            \scriptsize
            \nprounddigits{0}
            \begin{tabular}{llllrrrrrrcc}
            \hline\hline
            \# & Common name	& RA & DEC & Distance & log $ M_{_{\star}} $ & log $ \MHI $ & log $ M_{d} $ & log SFR & $f_g$ & \multirow{2}{*}{\rotatebox{90}{\high}} & \multirow{2}{*}{\rotatebox{90}{ HAPLESS}}\\
 & & (J2000 deg) & (J2000 deg) & (Mpc) & ($\rm M_{\odot}$) & ($\rm M_{\odot}$) & ($\rm M_{\odot}$) & ($\rm M_{\odot}\,yr^{-1}$) & & &   \\
 \\
  \hline
1 &   \setlength{\extrarowheight}{0pt} \begin{tabular}{@{}r@{}} 2MASXJ14265308 \\ +0057462 \end{tabular} & 216.72078 & 0.96285 & 120.35 & 9.60$_{-0.07}^{+0.13}$ & 9.62 & 7.26$_{-0.15}^{+0.18}$ & -0.03$_{-0.07}^{+0.07}$ & 0.58  & \checkmark  &  \\
2 & CGCG014-010 & 185.08868 & 0.36769 & 11.84 & 7.29$_{-0.12}^{+0.13}$ & 8.21 & 3.48$_{-0.40}^{+0.45}$ & -2.14$_{-0.04}^{+0.04}$ & 0.92  & \checkmark  &  \\
3 & CGCG014-014 & 185.27509 & 0.5519 & 41.91 & 8.54$_{-0.14}^{+0.17}$ &   & 6.13$_{-0.52}^{+0.48}$ & -1.18$_{-0.12}^{+0.12}$ &   &  & \checkmark  \\
4 & CGCG019-003 & 214.83333 & 1.16467 & 42.99 & 8.55$_{-0.13}^{+0.20}$ &   & 5.78$_{-0.57}^{+0.56}$ & -1.41$_{-0.25}^{+0.17}$ &  &  & \checkmark  \\
5 & CGCG019-084 & 220.62268 & 1.50173 & 34.63 & 9.06$_{-0.13}^{+0.13}$ &   & 6.39$_{-0.11}^{+0.12}$ & -0.83$_{-0.08}^{+0.07}$ &  &  & \checkmark  \\
6 & FGC1412 & 184.85783 & 0.21197 & 11.32 & 6.94$_{-0.10}^{+0.13}$ & 7.85 & 3.84$_{-0.53}^{+0.78}$ & -2.43$_{-0.17}^{+0.14}$ & 0.91  & \checkmark  &  \\
7 & IC1010 & 216.83483 & 1.02589 & 118.19 & 10.82$_{-0.25}^{+0.08}$ & 10.55 & 7.93$_{-0.13}^{+0.12}$ & 0.44$_{-0.06}^{+0.04}$ & 0.42  & \checkmark  &  \\
8 & IC1011 & 217.01885 & 1.00607 & 117.95 & 10.16$_{-0.07}^{+0.13}$ & 9.73 & 7.41$_{-0.09}^{+0.08}$ & 0.60$_{-0.04}^{+0.06}$ & 0.33  & \checkmark  &  \\
9 & LEDA1241857 & 222.59576 & 2.95833 & 28.64 & 8.2$_{-0.20}^{+0.14}$ &   & 4.93$_{-0.47}^{+0.68}$ & -1.90$_{-0.31}^{+0.12}$ &   &  & \checkmark  \\
10 & MGC0066574 & 219.99969 & -0.18714 & 33.37 & 7.18$_{-0.21}^{+0.20}$ &   & 5.22$_{-0.66}^{+0.55}$ & -2.28$_{-0.23}^{+0.14}$ &    &  & \checkmark  \\
11 & MGC0068525 & 221.3158 & -0.1602 & 29.19 & 8.68$_{-0.18}^{+0.12}$ &   & 4.50$_{-0.46}^{+0.51}$ & -1.90$_{-0.30}^{+0.13}$ &   &   & \checkmark  \\
12 & NGC4030 & 180.09841 & -1.10033 & 29.39 & 10.88$_{-0.09}^{+0.12}$ & 10.17 & 7.96$_{-0.08}^{+0.04}$ & 0.78$_{-0.05}^{+0.04}$ & 0.20  & \checkmark  & \checkmark  \\
13 & NGC4030b & 180.19873 & -0.02333 & 38.36 & 8.85$_{-0.14}^{+0.16}$ & 9.36 & 5.64$_{-0.44}^{+0.53}$ & -0.98$_{-0.32}^{+0.09}$ & 0.81  & \checkmark  &  \\
14 & NGC4202 & 184.53574 & -1.06413 & 93.20 & 10.30$_{-0.10}^{+0.11}$ & 10.41 & 7.46$_{-0.06}^{+0.07}$ & 0.05$_{-0.17}^{+0.06}$ & 0.63 & \checkmark  &  \\
15 & NGC5496 & 212.9082 & -1.15909 & 27.35 & 9.46$_{-0.05}^{+0.14}$ & 10.03 & 7.12$_{-0.11}^{+0.14}$ & -0.23$_{-0.04}^{+0.04}$ & 0.83  & \checkmark  & \checkmark  \\
16 & NGC5584 & 215.59857 & -0.3869 & 30.21 & 9.97$_{-0.16}^{+0.09}$ & 9.76 & 7.51$_{-0.10}^{+0.07}$ & 0.26$_{-0.04}^{+0.04}$ & 0.45  & \checkmark  & \checkmark  \\
17 & NGC5690 & 219.42 & 2.29162 & 32.10 & 10.38$_{-0.09}^{+0.11}$ & 9.90 & 7.61$_{-0.05}^{+0.05}$ & 0.31$_{-0.04}^{+0.05}$ & 0.31 & \checkmark  & \checkmark  \\
18 & NGC5691 & 219.47216 & -0.39846 & 33.35 & 10.01$_{-0.17}^{+0.10}$ & 9.16 & 6.85$_{-0.04}^{+0.07}$ & -0.06$_{-0.05}^{+0.06}$ & 0.16 & \checkmark  & \checkmark  \\
19 & NGC5705 & 219.95623 & -0.71874 & 31.35 & 9.33$_{-0.12}^{+0.08}$ & 9.77 & 7.35$_{-0.13}^{+0.12}$ & -0.24$_{-0.04}^{+0.04}$ & 0.78 & \checkmark  & \checkmark  \\
20 & NGC5713 & 220.04759 & -0.28933 & 33.60 & 10.56$_{-0.11}^{+0.14}$ & 10.06 & 7.54$_{-0.05}^{+0.05}$ & 0.72$_{-0.05}^{+0.06}$ & 0.29 & \checkmark  & \checkmark  \\
21 & NGC5719 & 220.23393 & -0.31856 & 30.72 & 10.79$_{-0.08}^{+0.09}$ & 10.07 & 7.43$_{-0.06}^{+0.07}$ & -0.17$_{-0.06}^{+0.04}$ & 0.20   & \checkmark  & \checkmark  \\
22 & NGC5725 & 220.24298 & 2.18655 & 29.42 & 9.13$_{-0.13}^{+0.08}$ & 8.93 & 6.45$_{-0.19}^{+0.19}$ & -0.65$_{-0.07}^{+0.07}$ & 0.46 & \checkmark  & \checkmark  \\
23 & NGC5733 & 220.69092 & -0.35132 & 30.13 & 8.89$_{-0.10}^{+0.12}$ &   & 6.47$_{-0.19}^{+0.23}$ & -0.68$_{-0.07}^{+0.04}$ &   &  & \checkmark  \\
24 & NGC5738 & 220.98402 & 1.60435 & 31.20 & 9.68$_{-0.12}^{+0.14}$ &   & 4.93$_{-0.34}^{+0.36}$ & -2.2$_{-0.26}^{+0.16}$ &   &   & \checkmark  \\
25 & NGC5740 & 221.10171 & 1.68019 & 28.08 & 10.28$_{-0.07}^{+0.11}$ & 9.74 & 7.16$_{-0.07}^{+0.07}$ & -0.05$_{-0.04}^{+0.04}$ & 0.27 & \checkmark  & \checkmark  \\
26 & NGC5746 & 221.23292 & 1.955 & 30.76 & 11.31$_{-0.10}^{+0.07}$ & 9.84 & 8.00$_{-0.07}^{+0.07}$ & -0.41$_{-0.70}^{+0.36}$ & 0.04 &  \checkmark  & \checkmark  \\
27 & NGC5750 & 221.54705 & -0.2236 & 31.12 & 10.59$_{-0.18}^{+0.06}$ & 9.08 & 6.99$_{-0.07}^{+0.07}$ & -0.70$_{-0.14}^{+0.08}$ & 0.04  &  & \checkmark  \\
28 & PGC037392 & 178.7696 & 1.71954 & 26.69 & 8.31$_{-0.12}^{+0.14}$ & 8.37 & 5.77$_{-0.53}^{+0.46}$ & -1.18$_{-0.12}^{+0.12}$ & 0.60 &  & \checkmark  \\
29 & PGC051719 & 217.15782 & 0.55312 & 29.01 & 8.82$_{-0.13}^{+0.11}$ &   & 6.42$_{-0.17}^{+0.20}$ & -1.1$_{-0.14}^{+0.07}$ &   &  & \checkmark  \\
30 & PGC052652 & 221.12776 & 1.52249 & 25.67 & 8.56$_{-0.17}^{+0.15}$ & 7.84 & 5.82$_{-0.40}^{+0.40}$ & -1.64$_{-0.18}^{+0.13}$ & 0.20 &  & \checkmark  \\
31 &  \setlength{\extrarowheight}{0pt} \begin{tabular}{@{}r@{}} SDSSJ084258.35 \\ +003838.5 \end{tabular} & 130.74318 & 0.64408 & 158.93 & 9.84$_{-0.13}^{+0.13}$ & 9.97 & 7.21$_{-0.41}^{+0.40}$ & -0.03$_{-0.13}^{+0.10}$ & 0.64 & \checkmark  &   \\
32 &  \setlength{\extrarowheight}{0pt}  \begin{tabular}{@{}r@{}} SDSSJ143353.30\\ +012905.6 \end{tabular} & 218.47167 & 1.48543 & 32.99 & 7.77$_{-0.18}^{+0.19}$ & 8.94 & 4.52$_{-0.63}^{+0.73}$ & -1.69$_{-0.04}^{+0.04}$ & 0.95  & \checkmark  &   \\
33 & UGC04673 & 133.967 & 2.52426 & 59.73 & 9.12$_{-0.17}^{+0.20}$ & 9.78 & 7.40$_{-0.27}^{+0.23}$ & -0.24$_{-0.05}^{+0.04}$ & 0.86  & \checkmark  &   \\
34 & UGC04684 & 134.17066 & 0.37591 & 40.56 & 9.35$_{-0.15}^{+0.14}$ & 9.58 & 6.70$_{-0.14}^{+0.16}$ & -0.36$_{-0.12}^{+0.10}$ & 0.69 & \checkmark  & \checkmark  \\
35 & UGC04996 & 140.81604 & -0.72945 & 57.25 & 9.36$_{-0.10}^{+0.12}$ & 9.84 & 7.18$_{-0.19}^{+0.21}$ & -0.17$_{-0.06}^{+0.06}$ & 0.80 & \checkmark  &  \\
36 & UGC06578 & 174.153 & 0.81678 & 20.42 & 8.02$_{-0.06}^{+0.10}$ & 8.82 & 5.72$_{-0.29}^{+0.32}$ & -1.02$_{-0.04}^{+0.03}$ & 0.89  & \checkmark  & \checkmark  \\
37 & UGC06780 & 177.20993 & -2.03249 & 34.16 & 9.00$_{-0.12}^{+0.18}$ & 9.87 & 6.97$_{-0.23}^{+0.20}$ & -0.36$_{-0.11}^{+0.11}$ & 0.91 & \checkmark  & \checkmark  \\
38 & UGC06877 & 178.55071 & 0.13681 & 18.30 & 8.94$_{-0.09}^{+0.13}$ & 8.01 & 5.53$_{-0.10}^{+0.10}$ & -0.76$_{-0.07}^{+0.06}$ & 0.13  &  & \checkmark  \\
39 & UGC06879 & 178.60538 & -2.3197 & 45.61 & 10.05$_{-0.15}^{+0.11}$ &   & 7.29$_{-0.12}^{+0.14}$ & -0.48$_{-0.09}^{+0.06}$ &   &   & \checkmark  \\
40 & UGC06903 & 178.9025 & 1.23817 & 37.66 & 9.89$_{-0.15}^{+0.09}$ & 9.68 & 7.17$_{-0.09}^{+0.10}$ & -0.24$_{-0.04}^{+0.04}$ & 0.45 & \checkmark  & \checkmark  \\
41 & UGC06970 & 179.69101 & -1.46169 & 30.31 & 9.39$_{-0.15}^{+0.12}$ & 9.05 & 6.52$_{-0.51}^{+0.35}$ & -0.86$_{-0.12}^{+0.10}$ & 0.38  & \checkmark  &  \\
42 & UGC07000 & 180.295 & -1.29751 & 30.76 & 9.11$_{-0.16}^{+0.08}$ & 9.10 & 6.43$_{-0.11}^{+0.12}$ & -0.45$_{-0.04}^{+0.04}$ & 0.56  & \checkmark  & \checkmark  \\
43 & UGC07053 & 181.0863 & -1.53071 & 30.13 & 8.19$_{-0.10}^{+0.18}$ & 9.25 & 4.80$_{-0.54}^{+0.56}$ & -1.03$_{-0.07}^{+0.06}$ & 0.94  & \checkmark  &  \\
44 & UGC07332 & 184.48653 & 0.43491 & 13.91 & 7.70$_{-0.13}^{+0.14}$ & 8.95 & 4.31$_{-0.40}^{+0.48}$ & -1.39$_{-0.04}^{+0.04}$ & 0.84   & \checkmark  &  \\
45 & UGC07394 & 185.11652 & 1.46789 & 32.65 & 8.93$_{-0.12}^{+0.14}$ & 9.24 & 6.87$_{-0.23}^{+0.21}$ & -1.22$_{-0.34}^{+0.17}$ & 0.73 & \checkmark  & \checkmark  \\
46 & UGC07396 & 185.14135 & 0.78863 & 41.30 & 9.11$_{-0.18}^{+0.14}$ & 9.11 & 6.49$_{-0.24}^{+0.28}$ & -0.77$_{-0.25}^{+0.16}$ & 0.57 &  & \checkmark  \\
47 & UGC07531 & 186.55054 & -1.30325 & 39.44 & 8.60$_{-0.08}^{+0.15}$ & 9.05 & 6.49$_{-0.39}^{+0.39}$ & -0.38$_{-0.04}^{+0.04}$ & 0.79 &  \checkmark  & \checkmark  \\
 \hline
    \end{tabular}
\end{table*}
\addtocounter{table}{-1}

\begin{table*}
            \caption{\textit{Continued}}
            \setlength{\tabcolsep}{6.5pt}
            \setlength{\extrarowheight}{4pt}
            \npdecimalsign{.}
            \nprounddigits{0}
            \scriptsize
            \begin{tabular}{llllrrrrrrcc}
              \hline\hline
            \# & Common name	& RA & DEC & Distance & log $ M_{_{\star}} $ & log $ \MHI $ & log $ M_{d} $ & log SFR & $f_g$ &  \multirow{2}{*}{\rotatebox{90}{\high}} & \multirow{2}{*}{\rotatebox{90}{ HAPLESS}}\\
 & & (J2000 deg) & (J2000 deg) & (Mpc) & ($\rm M_{\odot}$) & ($\rm M_{\odot}$) & ($\rm M_{\odot}$) & ($\rm M_{\odot}\,yr^{-1}$) &  & &   \\
 \\
  \hline
48 & UGC09215 & 215.86342 & 1.7243 & 25.65 & 9.31$_{-0.04}^{+0.14}$ & 9.56 & 6.95$_{-0.10}^{+0.09}$ & -0.24$_{-0.06}^{+0.02}$ & 0.70 & \checkmark  & \checkmark  \\
49 & UGC09299 & 217.39393 & -0.01906 & 28.29 & 8.61$_{-0.04}^{+0.19}$ & 9.94 & 6.39$_{-0.14}^{+0.15}$ & -0.55$_{-0.04}^{+0.04}$ & 0.97 & \checkmark  & \checkmark  \\
50 & UGC09348 & 218.11926 & 0.29425 & 30.36 & 9.41$_{-0.15}^{+0.14}$ &   & 6.66$_{-0.10}^{+0.10}$ & -0.89$_{-0.10}^{+0.07}$ &   &  & \checkmark  \\
51 & UGC09432 & 219.766 & 2.94708 & 28.53 & 8.19$_{-0.14}^{+0.02}$ & 9.19 & 4.4$_{-0.51}^{+0.48}$ & -1.06$_{-0.06}^{+0.05}$ & 0.93 & \checkmark  &  \\
52 & UGC09470 & 220.45274 & 0.68756 & 34.03 & 8.90$_{-0.13}^{+0.07}$ & 9.12 & 6.22$_{-0.19}^{+0.17}$ & -0.68$_{-0.04}^{+0.04}$ & 0.69 & \checkmark  & \checkmark  \\
53 & UGC09482 & 220.69539 & 0.66151 & 32.39 & 8.72$_{-0.14}^{+0.10}$ & 9.16 & 6.09$_{-0.27}^{+0.31}$ & -1.30$_{-0.18}^{+0.14}$ & 0.79 & \checkmark  & \checkmark  \\
54 & UM452 & 176.75239 & -0.29363 & 29.27 & 8.83$_{-0.17}^{+0.15}$ & 8.24 & 5.34$_{-0.34}^{+0.35}$ & -1.37$_{-0.17}^{+0.07}$ & 0.25 &  & \checkmark  \\
55 & UM456 & 177.65105 & -0.56613 & 33.73 & 8.28$_{-0.15}^{+0.15}$ & 8.89 & 4.96$_{-0.45}^{+0.59}$ & -0.76$_{-0.04}^{+0.04}$ & 0.84 & \checkmark  & \checkmark  \\
56 & UM456A & 177.6415 & -0.53795 & 35.53 & 7.88$_{-0.13}^{+0.11}$ & 8.93 & 4.89$_{-0.55}^{+0.65}$ & -1.32$_{-0.09}^{+0.12}$ & 0.94  & \checkmark  & \checkmark  \\
57 & UM491 & 184.97097 & 1.77326 & 39.71 & 8.46$_{-0.10}^{+0.09}$ & 9.36 & 5.99$_{-0.31}^{+0.34}$ & -0.83$_{-0.67}^{+0.05}$ & 0.91 &   & \checkmark  \\
58 & UM501 & 186.59463 & -1.2534 & 39.49 & 7.90$_{-0.10}^{+0.10}$ & 9.39 & 5.1$_{-0.54}^{+0.71}$ & -1.06$_{-0.04}^{+0.12}$ & 0.98 &  \checkmark  &  \\
 \hline
    \end{tabular}
    \label{tab:properties}
\end{table*}

\begin{table*}
            \caption{Metallicity measurements in the form $12+{\rm log (O/H)}$ for the H{\sc i}-selected \high\ sample (\citetalias{DeVis2017}) and dust-selected HAPLESS sample (\citetalias{Clark2015}). The check-marks indicate which sample the source is part of (24 sources are included in both samples).}
            \setlength{\tabcolsep}{10pt}
            \setlength{\extrarowheight}{5pt}
            \npdecimalsign{.}
            \nprounddigits{0}
            \begin{tabular}{llllrrrrrrcccccc}
            \hline\hline
            \# & Common name	& \multicolumn{4}{c}{12+log(O/H)} & \multirow{2}{*}{\rotatebox{90}{\high}} & \multirow{2}{*}{\rotatebox{90}{ HAPLESS}}\\ \cmidrule(lr){3-6}
 & & O3N2 & N2 & PG16S & KE08/T04 & &   \\
 \\
  \hline
1 & 2MASXJ14265308+0057462 & $8.53_{-0.12}^{+0.12}$ & $8.53_{-0.12}^{+0.12}$ & $8.37_{-0.12}^{+0.12}$ & $8.77_{-0.13}^{+0.13}$ & \checkmark & \\
2 & CGCG014-010 & $ _{ }^{ }$ & $ _{ }^{ }$ & $ _{ }^{ }$ & $ _{ }^{ }$ & \checkmark & \\
3 & CGCG014-014 & $8.25_{-0.14}^{+0.13}$ & $8.15_{-0.13}^{+0.12}$ & $7.75_{-0.26}^{+0.37}$ & $8.42_{-0.14}^{+0.13}$ &  & \checkmark  \\
4 & CGCG019-003 & $8.21_{-0.12}^{+0.12}$ & $8.22_{-0.12}^{+0.12}$ & $8.14_{-0.13}^{+0.12}$ & $8.36_{-0.12}^{+0.12}$ &  & \checkmark  \\
5 & CGCG019-084 & $8.59_{-0.13}^{+0.13}$ & $8.50_{-0.12}^{+0.12}$ & $ _{ }^{ }$ & $8.84_{-0.13}^{+0.13}$ &  & \checkmark  \\
6 & FGC1412 & $ _{ }^{ }$ & $ _{ }^{ }$ & $ _{ }^{ }$ & $ _{ }^{ }$ & \checkmark & \\
7 & IC1010 & $ _{ }^{ }$ & $ _{ }^{ }$ & $ _{ }^{ }$ & $ _{ }^{ }$ & \checkmark & \\
8 & IC1011 & $8.73_{-0.12}^{+0.12}$ & $8.77_{-0.20}^{+0.16}$ & $8.60_{-0.12}^{+0.12}$ & $9.01_{-0.13}^{+0.13}$ & \checkmark & \\
9 & LEDA1241857 & $8.34_{-0.12}^{+0.12}$ & $8.29_{-0.12}^{+0.12}$ & $8.19_{-0.12}^{+0.12}$ & $8.54_{-0.12}^{+0.12}$ &  & \checkmark  \\
10 & MGC0066574 & $ _{ }^{ }$ & $ _{ }^{ }$ & $ _{ }^{ }$ & $ _{ }^{ }$ &  & \checkmark  \\
11 & MGC0068525 & $ _{ }^{ }$ & $ _{ }^{ }$ & $ _{ }^{ }$ & $ _{ }^{ }$ &  & \checkmark  \\
12 & NGC4030 & $8.81_{-0.11}^{+0.11}$ & $8.55_{-0.11}^{+0.11}$ & $8.54_{-0.11}^{+0.11}$ & $9.11_{-0.11}^{+0.11}$ & \checkmark & \checkmark  \\
13 & NGC4030b & $8.36_{-0.12}^{+0.12}$ & $8.32_{-0.12}^{+0.12}$ & $8.30_{-0.12}^{+0.12}$ & $8.57_{-0.13}^{+0.13}$ & \checkmark & \\
14 & NGC4202 & $8.71_{-0.13}^{+0.16}$ & $8.44_{-0.12}^{+0.12}$ & $8.47_{-0.12}^{+0.12}$ & $8.98_{-0.15}^{+0.17}$ & \checkmark & \\
15 & NGC5496 & $8.34_{-0.10}^{+0.10}$ & $8.29_{-0.10}^{+0.10}$ & $8.19_{-0.10}^{+0.10}$ & $8.54_{-0.10}^{+0.10}$ & \checkmark & \checkmark  \\
16 & NGC5584 & $8.49_{-0.10}^{+0.10}$ & $8.43_{-0.10}^{+0.10}$ & $8.36_{-0.10}^{+0.10}$ & $8.72_{-0.10}^{+0.10}$ & \checkmark & \checkmark  \\
17 & NGC5690 & $8.76_{-0.11}^{+0.11}$ & $8.58_{-0.11}^{+0.11}$ & $8.48_{-0.11}^{+0.11}$ & $9.04_{-0.11}^{+0.11}$ & \checkmark & \checkmark  \\
18 & NGC5691 & $8.57_{-0.10}^{+0.10}$ & $8.54_{-0.10}^{+0.10}$ & $8.42_{-0.10}^{+0.10}$ & $8.82_{-0.10}^{+0.10}$ & \checkmark & \checkmark  \\
19 & NGC5705 & $8.56_{-0.11}^{+0.11}$ & $8.48_{-0.11}^{+0.11}$ & $8.30_{-0.11}^{+0.11}$ & $8.81_{-0.12}^{+0.12}$ & \checkmark & \checkmark  \\
20 & NGC5713 & $8.76_{-0.10}^{+0.10}$ & $8.63_{-0.10}^{+0.10}$ & $8.53_{-0.10}^{+0.10}$ & $9.05_{-0.10}^{+0.10}$ & \checkmark & \checkmark  \\
21 & NGC5719 & $ _{ }^{ }$ & $ _{ }^{ }$ & $ _{ }^{ }$ & $ _{ }^{ }$ & \checkmark & \checkmark  \\
22 & NGC5725 & $8.42_{-0.10}^{+0.10}$ & $8.41_{-0.10}^{+0.10}$ & $8.31_{-0.10}^{+0.10}$ & $8.65_{-0.10}^{+0.10}$ & \checkmark & \checkmark  \\
23 & NGC5733 & $8.28_{-0.11}^{+0.10}$ & $8.28_{-0.11}^{+0.11}$ & $8.18_{-0.11}^{+0.11}$ & $8.48_{-0.11}^{+0.11}$ &  & \checkmark  \\
24 & NGC5738 & $ _{ }^{ }$ & $ _{ }^{ }$ & $ _{ }^{ }$ & $ _{ }^{ }$ &  & \checkmark  \\
25 & NGC5740 & $8.68_{-0.11}^{+0.11}$ & $8.70_{-0.11}^{+0.11}$ & $8.57_{-0.11}^{+0.11}$ & $8.95_{-0.11}^{+0.11}$ & \checkmark & \checkmark  \\
 \hline
    \end{tabular}
    \label{tab:HIGHmetals}
\end{table*}
\addtocounter{table}{-1}

\begin{table*}
            \caption{\textit{Continued}}
            \setlength{\tabcolsep}{10pt}
            \setlength{\extrarowheight}{5pt}
            \npdecimalsign{.}
            \nprounddigits{0}
            \begin{tabular}{llllrrrrrrcccccc}
            \hline\hline
            \# & Common name	& \multicolumn{4}{c}{12+log(O/H)} & \multirow{2}{*}{\rotatebox{90}{\high}} & \multirow{2}{*}{\rotatebox{90}{ HAPLESS}}\\ \cmidrule(lr){3-6}
 & & O3N2 & N2 & PG16S & KE08/T04 & &   \\
 \\
  \hline
26 & NGC5746 & $ _{ }^{ }$ & $ _{ }^{ }$ & $ _{ }^{ }$ & $ _{ }^{ }$ & \checkmark & \checkmark  \\
27 & NGC5750 & $8.75_{-0.12}^{+0.12}$ & $8.79_{-0.12}^{+0.12}$ & $8.59_{-0.12}^{+0.12}$ & $9.03_{-0.12}^{+0.12}$ &  & \checkmark  \\
28 & PGC037392 & $8.33_{-0.12}^{+0.12}$ & $8.36_{-0.12}^{+0.12}$ & $8.19_{-0.12}^{+0.12}$ & $8.54_{-0.12}^{+0.12}$ &  & \checkmark  \\
29 & PGC051719 & $8.38_{-0.11}^{+0.11}$ & $8.33_{-0.11}^{+0.11}$ & $8.24_{-0.11}^{+0.11}$ & $8.59_{-0.11}^{+0.11}$ &  & \checkmark  \\
30 & PGC052652 & $8.35_{-0.12}^{+0.12}$ & $8.33_{-0.12}^{+0.12}$ & $8.27_{-0.12}^{+0.12}$ & $8.56_{-0.12}^{+0.12}$ &  & \checkmark  \\
31 & SDSSJ084258.35+003838.5 & $8.63_{-0.12}^{+0.12}$ & $8.58_{-0.12}^{+0.12}$ & $8.47_{-0.12}^{+0.12}$ & $8.89_{-0.13}^{+0.13}$ & \checkmark & \\
32 & SDSSJ143353.30+012905.6 & $ _{ }^{ }$ & $ _{ }^{ }$ & $ _{ }^{ }$ & $ _{ }^{ }$ & \checkmark & \\
33 & UGC04673 & $8.50_{-0.11}^{+0.11}$ & $8.45_{-0.11}^{+0.11}$ & $8.36_{-0.11}^{+0.11}$ & $8.75_{-0.12}^{+0.12}$ & \checkmark & \\
34 & UGC04684 & $8.58_{-0.12}^{+0.12}$ & $8.48_{-0.12}^{+0.12}$ & $ _{ }^{ }$ & $8.84_{-0.13}^{+0.13}$ & \checkmark & \checkmark  \\
35 & UGC04996 & $8.45_{-0.11}^{+0.11}$ & $8.36_{-0.11}^{+0.11}$ & $8.18_{-0.11}^{+0.11}$ & $8.68_{-0.12}^{+0.12}$ & \checkmark & \\
36 & UGC06578 & $8.04_{-0.11}^{+0.11}$ & $7.99_{-0.11}^{+0.11}$ & $7.87_{-0.12}^{+0.12}$ & $8.25_{-0.13}^{+0.13}$ & \checkmark & \checkmark  \\
37 & UGC06780 & $8.42_{-0.11}^{+0.11}$ & $8.32_{-0.11}^{+0.11}$ & $8.06_{-0.11}^{+0.11}$ & $8.64_{-0.12}^{+0.12}$ & \checkmark & \checkmark  \\
38 & UGC06877 & $8.58_{-0.12}^{+0.12}$ & $8.53_{-0.12}^{+0.12}$ & $8.52_{-0.12}^{+0.12}$ & $8.83_{-0.12}^{+0.12}$ &  & \checkmark  \\
39 & UGC06879 & $ _{ }^{ }$ & $ _{ }^{ }$ & $ _{ }^{ }$ & $ _{ }^{ }$ &  & \checkmark  \\
40 & UGC06903 & $8.59_{-0.11}^{+0.12}$ & $8.49_{-0.12}^{+0.12}$ & $8.33_{-0.12}^{+0.12}$ & $8.85_{-0.12}^{+0.12}$ & \checkmark & \checkmark  \\
41 & UGC06970 & $8.54_{-0.13}^{+0.13}$ & $8.45_{-0.16}^{+0.16}$ & $8.16_{-0.19}^{+0.16}$ & $8.79_{-0.15}^{+0.15}$ & \checkmark & \\
42 & UGC07000 & $8.39_{-0.10}^{+0.10}$ & $8.34_{-0.10}^{+0.10}$ & $8.25_{-0.10}^{+0.10}$ & $8.59_{-0.11}^{+0.11}$ & \checkmark & \checkmark  \\
43 & UGC07053 & $8.21_{-0.11}^{+0.11}$ & $8.15_{-0.11}^{+0.11}$ & $7.88_{-0.12}^{+0.12}$ & $8.36_{-0.12}^{+0.12}$ & \checkmark & \\
44 & UGC07332 & $ _{ }^{ }$ & $ _{ }^{ }$ & $ _{ }^{ }$ & $ _{ }^{ }$ & \checkmark & \\
45 & UGC07394 & $8.28_{-0.11}^{+0.11}$ & $8.20_{-0.11}^{+0.11}$ & $7.94_{-0.13}^{+0.12}$ & $8.47_{-0.12}^{+0.12}$ & \checkmark & \checkmark  \\
46 & UGC07396 & $8.44_{-0.12}^{+0.12}$ & $8.39_{-0.12}^{+0.12}$ & $8.26_{-0.12}^{+0.12}$ & $8.68_{-0.12}^{+0.12}$ &  & \checkmark  \\
47 & UGC07531 & $ _{ }^{ }$ & $ _{ }^{ }$ & $ _{ }^{ }$ & $ _{ }^{ }$ & \checkmark & \checkmark  \\
48 & UGC09215 & $8.40_{-0.10}^{+0.10}$ & $8.37_{-0.10}^{+0.10}$ & $8.29_{-0.10}^{+0.10}$ & $8.62_{-0.11}^{+0.11}$ & \checkmark & \checkmark  \\
49 & UGC09299 & $8.39_{-0.11}^{+0.11}$ & $8.36_{-0.11}^{+0.11}$ & $8.25_{-0.11}^{+0.11}$ & $8.60_{-0.12}^{+0.12}$ & \checkmark & \checkmark  \\
50 & UGC09348 & $8.49_{-0.12}^{+0.12}$ & $8.49_{-0.12}^{+0.12}$ & $8.33_{-0.12}^{+0.12}$ & $8.73_{-0.12}^{+0.12}$ &  & \checkmark  \\
51 & UGC09432 & $8.25_{-0.12}^{+0.12}$ & $8.21_{-0.12}^{+0.12}$ & $8.06_{-0.16}^{+0.12}$ & $8.42_{-0.13}^{+0.13}$ & \checkmark & \\
52 & UGC09470 & $8.13_{-0.12}^{+0.12}$ & $8.15_{-0.12}^{+0.12}$ & $7.94_{-0.12}^{+0.12}$ & $8.22_{-0.13}^{+0.13}$ & \checkmark & \checkmark  \\
53 & UGC09482 & $8.33_{-0.11}^{+0.11}$ & $8.27_{-0.11}^{+0.11}$ & $8.14_{-0.11}^{+0.11}$ & $8.54_{-0.12}^{+0.12}$ & \checkmark & \checkmark  \\
54 & UM452 & $8.27_{-0.12}^{+0.12}$ & $8.26_{-0.12}^{+0.12}$ & $8.19_{-0.12}^{+0.12}$ & $8.45_{-0.12}^{+0.12}$ &  & \checkmark  \\
55 & UM456 & $8.04_{-0.11}^{+0.11}$ & $8.05_{-0.11}^{+0.11}$ & $7.81_{-0.11}^{+0.11}$ & $8.20_{-0.13}^{+0.13}$ & \checkmark & \checkmark  \\
56 & UM456A & $8.06_{-0.11}^{+0.11}$ & $8.04_{-0.11}^{+0.11}$ & $7.72_{-0.11}^{+0.11}$ & $8.10_{-0.13}^{+0.13}$ & \checkmark & \checkmark  \\
57 & UM491 & $8.25_{-0.12}^{+0.12}$ & $8.24_{-0.12}^{+0.12}$ & $8.18_{-0.12}^{+0.12}$ & $8.42_{-0.12}^{+0.12}$ &  & \checkmark  \\
58 & UM501 & $ _{ }^{ }$ & $ _{ }^{ }$ & $ _{ }^{ }$ & $ _{ }^{ }$ & \checkmark & \\
 \hline
    \end{tabular}
\end{table*}

\begin{landscape}
\begin{table}
\caption{Emission line measurements for the H{\sc i}GH and HAPLESS samples. The first three columns give the ID from Table \ref{tab:properties}, the common name, and the GAMA cataID respectively. All emission lines have been corrected for reddening using the Balmer decrement $C(H\beta)$ and the \citet{Cardelli1989} dust obscuration curve. The Origin column specifies whether the stellar absorption corrected fluxes were extracted using GANDALF or the GaussFitComplexv05 (GFC) catalogue. Measurements from multiple regions of the galaxies are presented where appropriate. }
		\label{tab:HIGHlines}
		\scriptsize
		\setlength{\tabcolsep}{6.5pt}
    		\begin{tabular}{lccccccccccclcc}
    		\hline\hline
\multirow{4}{*}{ID} & \multirow{4}{*}{name} & \multirow{4}{*}{cataID} & \multirow{4}{*}{$C(H\beta)$}  &  &  &  &  &  & & & & \multirow{4}{*}{Origin} & \multirow{4}{*}{\rotatebox{90}{H{\sc i}GH}} & \multirow{4}{*}{\rotatebox{90}{HAPLESS}}    \\
& & & &  \multicolumn{8}{c}{$I_{\lambda}/I_{H\beta}$} & & & \\ \cmidrule(lr){5-12}
 &  &  &  & [O{\sc ii}] & $H\beta$  & [O{\sc iii},4959] & [O{\sc iii},5007] & $H\alpha$ & [N{\sc ii}] & [S{\sc ii},6713] & [S{\sc ii},6731] & & &   \\
   &  &  &  &  &  &  &  &  & & & & & &   \\     		\hline
1& 2MASXJ142...& 106616& 0.53& $2.36 \pm 0.10$& $1.00 \pm 0.02$& $0.37 \pm 0.01$& $1.04 \pm 0.02$& $2.86 \pm 0.03$& $0.703 \pm 0.010$& $0.674 \pm 0.013$& $0.539 \pm 0.011$& GANDALF& \checkmark & \\
3& CGCG014-014& 86115& 0.69& $5.78 \pm 0.53$& $1.00 \pm 0.31$& $0.59 \pm 0.05$& $1.50 \pm 0.09$& $2.86 \pm 0.09$& $0.139 \pm 0.043$& $0.357 \pm 0.035$& $0.237 \pm 0.029$& GFC&   & \checkmark \\
4& CGCG019-003& 227753& 0.0& $3.32 \pm 0.11$& $1.00 \pm 0.01$& $1.05 \pm 0.01$& $3.00 \pm 0.02$& $2.60 \pm 0.02$& $0.190 \pm 0.008$& $0.309 \pm 0.009$& $0.206 \pm 0.008$& GANDALF&   & \checkmark \\
5& CGCG019-084& 240108& 1.2& & $1.00 \pm 0.22$& $0.23 \pm 0.07$& $0.63 \pm 0.19$& $2.86 \pm 0.10$& $0.658 \pm 0.066$& $0.884 \pm 0.066$& & GANDALF&   & \checkmark \\
6& FGC1412& 611445& 0.24& & $1.00 \pm 1.27$& $0.17 \pm 2.52$& $0.49 \pm 7.17$& $2.86 \pm 0.40$& $0.224 \pm 1.491$& $0.154 \pm 0.794$& & GANDALF& \checkmark & \\
6& FGC1412& 611446& 0.18& $2.61 \pm 20.76$& $1.00 \pm 1.32$& $0.37 \pm 0.63$& $1.05 \pm 1.80$& $2.86 \pm 0.91$& $0.476 \pm 0.676$& $0.927 \pm 0.587$& $0.620 \pm 0.757$& GANDALF& \checkmark & \\
8& IC1011& 106717& 0.92& & $1.00 \pm 0.05$& $0.14 \pm 0.02$& $0.41 \pm 0.03$& $2.87 \pm 0.68$& $1.211 \pm 0.028$& $0.458 \pm 0.019$& $0.359 \pm 0.017$& GFC& \checkmark & \\
9& LEDA1241857& 367540& 0.1& & $1.00 \pm 0.14$& $0.63 \pm 0.03$& $1.78 \pm 0.05$& $2.86 \pm 0.06$& $0.305 \pm 0.022$& $0.452 \pm 0.025$& $0.325 \pm 0.016$& GFC&   & \checkmark \\
10& MGC0066574& 594420& 0.0& $3.79 \pm 4.23$& $1.00 \pm 0.55$& $0.20 \pm 0.65$& $0.58 \pm 1.87$& $0.77 \pm 0.28$& $0.226 \pm 0.382$& $0.257 \pm 0.296$& $0.399 \pm 0.331$& GANDALF&   & \checkmark \\
12& NGC4030& 31521& 0.0& $0.35 \pm 0.15$& $1.00 \pm 0.03$& $0.05 \pm 0.01$& $0.13 \pm 0.03$& $2.09 \pm 0.02$& $0.554 \pm 0.020$& $0.149 \pm 0.017$& $0.108 \pm 0.016$& GANDALF& \checkmark & \checkmark \\
12& NGC4030& 31523& 0.55& $2.55 \pm 0.22$& $1.00 \pm 0.03$& $0.03 \pm 0.01$& $0.09 \pm 0.02$& $2.86 \pm 0.05$& $0.630 \pm 0.015$& $0.281 \pm 0.012$& $0.240 \pm 0.009$& GANDALF& \checkmark & \checkmark \\
12& NGC4030& 690077& 0.0& $14.76 \pm 2.09$& $1.00 \pm 0.05$& $0.08 \pm 0.02$& $0.24 \pm 0.04$& $2.31 \pm 0.04$& $0.689 \pm 0.031$& $0.230 \pm 0.036$& $0.159 \pm 0.036$& GANDALF& \checkmark & \checkmark \\
13& NGC4030b& 584731& 0.0& $3.94 \pm 0.37$& $1.00 \pm 0.04$& $0.60 \pm 0.01$& $1.73 \pm 0.04$& $2.53 \pm 0.03$& $0.298 \pm 0.017$& $0.265 \pm 0.015$& $0.168 \pm 0.015$& GANDALF& \checkmark & \\
14& NGC4202& 32362& 2.57& $4.25 \pm 25.06$& $1.00 \pm 0.18$& $0.44 \pm 0.16$& $0.21 \pm 0.12$& $2.88 \pm 0.08$& $0.538 \pm 0.027$& $0.256 \pm 0.019$& $0.164 \pm 0.015$& GFC& \checkmark & \\
15& NGC5496& 496980& 0.0& $2.20 \pm 5.24$& $1.00 \pm 0.42$& $0.33 \pm 0.03$& $0.89 \pm 0.05$& $1.65 \pm 0.05$& $0.170 \pm 0.031$& $0.378 \pm 0.028$& $0.262 \pm 0.022$& GFC& \checkmark & \checkmark \\
15& NGC5496& 463393& 0.25& $2.35 \pm 0.30$& $1.00 \pm 0.03$& $0.48 \pm 0.01$& $1.36 \pm 0.03$& $2.86 \pm 0.02$& $0.342 \pm 0.015$& $0.513 \pm 0.016$& $0.367 \pm 0.015$& GANDALF& \checkmark & \checkmark \\
15& NGC5496& 463394& 1.25& $2.46 \pm 0.25$& $1.00 \pm 0.04$& $0.83 \pm 0.03$& $2.30 \pm 0.08$& $2.86 \pm 0.07$& $0.272 \pm 0.009$& $0.265 \pm 0.009$& $0.185 \pm 0.007$& GANDALF& \checkmark & \checkmark \\
15& NGC5496& 496979& 0.91& $5.29 \pm 0.86$& $1.00 \pm 0.13$& $0.74 \pm 0.08$& $2.05 \pm 0.23$& $2.86 \pm 0.24$& $0.234 \pm 0.022$& $0.289 \pm 0.024$& $0.209 \pm 0.018$& GANDALF& \checkmark & \checkmark \\
15& NGC5496& 496981& 1.21& $10.32 \pm 0.71$& $1.00 \pm 0.03$& $0.81 \pm 0.01$& $2.24 \pm 0.02$& $2.86 \pm 0.02$& $0.344 \pm 0.008$& $0.423 \pm 0.008$& $0.295 \pm 0.008$& GANDALF& \checkmark & \checkmark \\
15& NGC5496& 496982& 0.0& $6.06 \pm 0.59$& $1.00 \pm 0.04$& $0.89 \pm 0.01$& $2.54 \pm 0.04$& $2.05 \pm 0.02$& $0.147 \pm 0.013$& $0.273 \pm 0.012$& $0.199 \pm 0.013$& GANDALF& \checkmark & \checkmark \\
15& NGC5496& 496985& 0.35& $4.85 \pm 0.72$& $1.00 \pm 0.06$& $0.43 \pm 0.02$& $1.22 \pm 0.05$& $2.86 \pm 0.06$& $0.454 \pm 0.032$& $0.664 \pm 0.031$& $0.462 \pm 0.034$& GANDALF& \checkmark & \checkmark \\
15& NGC5496& 496986& 0.0& $11.33 \pm 1.72$& $1.00 \pm 0.08$& $0.46 \pm 0.02$& $1.32 \pm 0.06$& $2.57 \pm 0.06$& $0.238 \pm 0.044$& $0.436 \pm 0.041$& $0.299 \pm 0.045$& GANDALF& \checkmark & \checkmark \\
16& NGC5584& 693091& 0.29& $1.38 \pm 7.39$& $1.00 \pm 0.05$& $0.43 \pm 0.02$& $1.26 \pm 0.04$& $2.86 \pm 0.04$& $0.448 \pm 0.018$& $0.532 \pm 0.031$& $0.383 \pm 0.023$& GFC& \checkmark & \checkmark \\
16& NGC5584& 63349& 0.0& $187.48 \pm 0.84$& $1.00 \pm 0.12$& $0.18 \pm 0.04$& $0.51 \pm 0.12$& $2.56 \pm 0.10$& $0.365 \pm 0.089$& $0.149 \pm 0.124$& $0.179 \pm 0.116$& GANDALF& \checkmark & \checkmark \\
16& NGC5584& 63351& 0.0& $1.91 \pm 0.27$& $1.00 \pm 0.09$& $0.35 \pm 0.02$& $1.00 \pm 0.07$& $2.86 \pm 0.17$& $0.509 \pm 0.028$& $0.448 \pm 0.030$& $0.320 \pm 0.022$& GANDALF& \checkmark & \checkmark \\
16& NGC5584& 63353& 0.86& $2.54 \pm 0.32$& $1.00 \pm 0.04$& $0.33 \pm 0.01$& $0.92 \pm 0.04$& $2.86 \pm 0.05$& $0.577 \pm 0.014$& $0.428 \pm 0.012$& $0.300 \pm 0.013$& GANDALF& \checkmark & \checkmark \\
16& NGC5584& 63354& 0.66& $2.84 \pm 0.70$& $1.00 \pm 0.14$& $0.31 \pm 0.04$& $0.89 \pm 0.12$& $2.86 \pm 0.23$& $0.535 \pm 0.054$& $0.569 \pm 0.059$& $0.397 \pm 0.042$& GANDALF& \checkmark & \checkmark \\
16& NGC5584& 693086& 0.01& $3.23 \pm 0.77$& $1.00 \pm 0.18$& $1.14 \pm 0.20$& $3.25 \pm 0.56$& $2.86 \pm 0.43$& $0.239 \pm 0.032$& $0.173 \pm 0.023$& $0.138 \pm 0.018$& GANDALF& \checkmark & \checkmark \\
16& NGC5584& 693088& 0.27& $2.16 \pm 0.22$& $1.00 \pm 0.02$& $0.25 \pm 0.01$& $0.72 \pm 0.02$& $2.86 \pm 0.02$& $0.882 \pm 0.013$& $0.555 \pm 0.011$& $0.445 \pm 0.010$& GANDALF& \checkmark & \checkmark \\
17& NGC5690& 262444& 1.63& $4.54 \pm 25.79$& $1.00 \pm 0.07$& $0.08 \pm 0.03$& $0.22 \pm 0.05$& $2.87 \pm 0.05$& $0.984 \pm 0.029$& $0.499 \pm 0.037$& $0.352 \pm 0.031$& GFC& \checkmark & \checkmark \\
17& NGC5690& 262445& 2.13& $11.63 \pm 25.93$& $1.00 \pm 0.05$& $0.05 \pm 0.02$& $0.16 \pm 0.04$& $2.87 \pm 0.03$& $0.625 \pm 0.010$& $0.356 \pm 0.008$& $0.251 \pm 0.007$& GFC& \checkmark & \checkmark \\
17& NGC5690& 716324& 0.73& $2.00 \pm 0.88$& $1.00 \pm 0.07$& $0.10 \pm 0.02$& $0.28 \pm 0.04$& $2.86 \pm 0.07$& $0.805 \pm 0.032$& $0.542 \pm 0.027$& $0.340 \pm 0.023$& GANDALF& \checkmark & \checkmark \\
18& NGC5691& 64554& 0.4& $3.71 \pm 0.34$& $1.00 \pm 0.03$& $0.25 \pm 0.01$& $0.70 \pm 0.03$& $2.86 \pm 0.02$& $0.773 \pm 0.015$& $0.698 \pm 0.015$& $0.488 \pm 0.015$& GANDALF& \checkmark & \checkmark \\
18& NGC5691& 693423& 0.0& $6.78 \pm 0.25$& $1.00 \pm 0.01$& $0.39 \pm 0.00$& $1.11 \pm 0.01$& $2.75 \pm 0.01$& $0.804 \pm 0.005$& $0.529 \pm 0.005$& $0.442 \pm 0.006$& GANDALF& \checkmark & \checkmark \\
18& NGC5691& 693424& 0.0& $1.67 \pm 0.10$& $1.00 \pm 0.02$& $0.23 \pm 0.01$& $0.64 \pm 0.01$& $2.50 \pm 0.02$& $0.688 \pm 0.011$& $0.510 \pm 0.011$& $0.365 \pm 0.010$& GANDALF& \checkmark & \checkmark \\
18& NGC5691& 693425& 0.75& $1.90 \pm 0.12$& $1.00 \pm 0.01$& $0.33 \pm 0.00$& $0.92 \pm 0.01$& $2.86 \pm 0.01$& $0.888 \pm 0.006$& $0.669 \pm 0.008$& $0.527 \pm 0.007$& GANDALF& \checkmark & \checkmark \\
18& NGC5691& 693426& 0.62& $3.13 \pm 0.16$& $1.00 \pm 0.02$& $0.29 \pm 0.00$& $0.81 \pm 0.01$& $2.86 \pm 0.01$& $0.611 \pm 0.007$& $0.397 \pm 0.007$& $0.283 \pm 0.007$& GANDALF& \checkmark & \checkmark \\
18& NGC5691& 756783& 0.58& & $1.00 \pm 0.02$& $0.24 \pm 0.00$& $0.66 \pm 0.01$& $2.86 \pm 0.03$& $0.491 \pm 0.008$& $0.370 \pm 0.008$& $0.248 \pm 0.007$& GANDALF& \checkmark & \checkmark \\
19& NGC5705& 49167& 0.0& $2.18 \pm 0.46$& $1.00 \pm 0.07$& $0.25 \pm 0.02$& $0.71 \pm 0.07$& $2.29 \pm 0.09$& $0.590 \pm 0.054$& $0.772 \pm 0.059$& $0.532 \pm 0.050$& GANDALF& \checkmark & \checkmark \\
19& NGC5705& 49169& 0.21& $4.11 \pm 0.19$& $1.00 \pm 0.03$& $0.28 \pm 0.01$& $0.81 \pm 0.02$& $2.86 \pm 0.07$& $0.530 \pm 0.018$& $0.567 \pm 0.018$& $0.431 \pm 0.015$& GANDALF& \checkmark & \checkmark \\
19& NGC5705& 49171& 0.0& $4.53 \pm 1.82$& $1.00 \pm 0.33$& $0.20 \pm 0.05$& $0.57 \pm 0.15$& $0.98 \pm 0.14$& $0.233 \pm 0.134$& $0.336 \pm 0.103$& $0.211 \pm 0.148$& GANDALF& \checkmark & \checkmark \\
\hline\hline
\end{tabular}
\end{table}
\end{landscape}
\addtocounter{table}{-1}

\begin{landscape}
\begin{table}
\caption{\textit{Continued}}
		\scriptsize
		\setlength{\tabcolsep}{6.5pt}
    		\begin{tabular}{lccccccccccclcc}
    		\hline\hline
\multirow{4}{*}{ID} & \multirow{4}{*}{name} & \multirow{4}{*}{cataID} & \multirow{4}{*}{$C(H\beta)$}  &  &  &  &  &  & & & & \multirow{4}{*}{Origin} & \multirow{4}{*}{\rotatebox{90}{H{\sc i}GH}} & \multirow{4}{*}{\rotatebox{90}{HAPLESS}}    \\
& & & &  \multicolumn{8}{c}{$I_{\lambda}/I_{H\beta}$} & & & \\ \cmidrule(lr){5-12}
 &  &  &  & [O{\sc ii}] & $H\beta$  & [O{\sc iii},4959] & [O{\sc iii},5007] & $H\alpha$ & [N{\sc ii}] & [S{\sc ii},6713] & [S{\sc ii},6731] & & &   \\
   &  &  &  &  &  &  &  &  & & & & & &   \\     		\hline
20& NGC5713& 693034& 1.38& $1.11 \pm 9.33$& $1.00 \pm 0.02$& $0.06 \pm 0.01$& $0.19 \pm 0.01$& $2.87 \pm 0.01$& $1.001 \pm 0.007$& $0.443 \pm 0.005$& $0.328 \pm 0.004$& GFC& \checkmark & \checkmark \\
20& NGC5713& 64772& 1.52& $7.49 \pm 0.43$& $1.00 \pm 0.02$& $0.16 \pm 0.00$& $0.45 \pm 0.01$& $2.86 \pm 0.04$& $0.855 \pm 0.012$& $0.386 \pm 0.006$& $0.279 \pm 0.006$& GANDALF& \checkmark & \checkmark \\
20& NGC5713& 64773& 0.53& $1.18 \pm 0.06$& $1.00 \pm 0.01$& $0.10 \pm 0.00$& $0.29 \pm 0.01$& $2.86 \pm 0.01$& $0.912 \pm 0.005$& $0.454 \pm 0.005$& $0.334 \pm 0.004$& GANDALF& \checkmark & \checkmark \\
20& NGC5713& 693033& 0.51& $2.10 \pm 0.11$& $1.00 \pm 0.01$& $0.11 \pm 0.00$& $0.30 \pm 0.01$& $2.86 \pm 0.01$& $0.871 \pm 0.005$& $0.387 \pm 0.004$& $0.285 \pm 0.004$& GANDALF& \checkmark & \checkmark \\
20& NGC5713& 693035& 0.94& & $1.00 \pm 0.01$& $0.09 \pm 0.00$& $0.24 \pm 0.00$& $2.86 \pm 0.02$& $0.779 \pm 0.006$& $0.326 \pm 0.004$& $0.227 \pm 0.004$& GANDALF& \checkmark & \checkmark \\
20& NGC5713& 693036& 0.19& $1.24 \pm 0.07$& $1.00 \pm 0.01$& $0.06 \pm 0.00$& $0.16 \pm 0.01$& $2.86 \pm 0.01$& $1.001 \pm 0.008$& $0.483 \pm 0.007$& $0.348 \pm 0.006$& GANDALF& \checkmark & \checkmark \\
20& NGC5713& 693037& 0.62& $1.29 \pm 0.16$& $1.00 \pm 0.02$& $0.07 \pm 0.01$& $0.21 \pm 0.02$& $2.86 \pm 0.02$& $0.895 \pm 0.013$& $0.452 \pm 0.011$& $0.330 \pm 0.010$& GANDALF& \checkmark & \checkmark \\
20& NGC5713& 693038& 0.63& $0.76 \pm 0.29$& $1.00 \pm 0.04$& $0.09 \pm 0.01$& $0.24 \pm 0.03$& $2.86 \pm 0.02$& $1.005 \pm 0.017$& $0.501 \pm 0.016$& $0.359 \pm 0.015$& GANDALF& \checkmark & \checkmark \\
22& NGC5725& 343415& 0.24& $5.46 \pm 25.00$& $1.00 \pm 0.28$& $0.31 \pm 0.13$& $0.92 \pm 0.21$& $2.86 \pm 0.33$& $0.550 \pm 0.151$& $0.942 \pm 0.161$& $0.622 \pm 0.123$& GFC& \checkmark & \\
22& NGC5725& 343407& 1.02& $9.35 \pm 69.01$& $1.00 \pm 0.04$& $0.57 \pm 0.02$& $1.75 \pm 0.04$& $2.87 \pm 0.02$& $0.435 \pm 0.012$& $0.482 \pm 0.011$& $0.339 \pm 0.009$& GFC& \checkmark & \\
22& NGC5725& 343410& 0.34& $3.74 \pm 15.39$& $1.00 \pm 0.14$& $0.34 \pm 0.02$& $0.96 \pm 0.04$& $2.86 \pm 0.05$& $0.461 \pm 0.023$& $0.539 \pm 0.021$& $0.354 \pm 0.016$& GFC& \checkmark & \\
22& NGC5725& 343414& 0.69& $3.97 \pm 11.92$& $1.00 \pm 0.10$& $0.44 \pm 0.05$& $1.32 \pm 0.09$& $2.86 \pm 0.06$& $0.491 \pm 0.037$& $0.766 \pm 0.070$& $0.481 \pm 0.065$& GFC& \checkmark & \\
22& NGC5725& 343405& 0.0& & $1.00 \pm 0.01$& $0.55 \pm 0.01$& $1.59 \pm 0.02$& $2.72 \pm 0.04$& $0.471 \pm 0.008$& $0.443 \pm 0.009$& $0.333 \pm 0.006$& GANDALF& \checkmark & \\
22& NGC5725& 343409& 0.31& $1.34 \pm 0.29$& $1.00 \pm 0.05$& $0.68 \pm 0.02$& $1.94 \pm 0.06$& $2.86 \pm 0.04$& $0.369 \pm 0.022$& $0.418 \pm 0.023$& $0.306 \pm 0.022$& GANDALF& \checkmark & \\
22& NGC5725& 343411& 0.0& $2.80 \pm 0.27$& $1.00 \pm 0.03$& $0.54 \pm 0.01$& $1.53 \pm 0.03$& $2.63 \pm 0.02$& $0.453 \pm 0.016$& $0.419 \pm 0.018$& $0.306 \pm 0.017$& GANDALF& \checkmark & \\
22& NGC5725& 343413& 0.15& $1.22 \pm 0.50$& $1.00 \pm 0.14$& $0.80 \pm 0.10$& $2.27 \pm 0.28$& $2.86 \pm 0.28$& $0.265 \pm 0.038$& $0.232 \pm 0.036$& $0.139 \pm 0.028$& GANDALF& \checkmark & \\
22& NGC5725& 722438& 0.0& $5.81 \pm 0.42$& $1.00 \pm 0.04$& $0.47 \pm 0.02$& $1.35 \pm 0.05$& $2.66 \pm 0.05$& $0.824 \pm 0.033$& $0.721 \pm 0.032$& $0.635 \pm 0.039$& GANDALF& \checkmark & \\
23& NGC5733& 64893& 0.43& & $1.00 \pm 0.04$& $0.84 \pm 0.02$& $2.60 \pm 0.05$& $2.86 \pm 0.05$& $0.361 \pm 0.020$& $0.542 \pm 0.019$& $0.378 \pm 0.015$& GFC&   & \checkmark \\
23& NGC5733& 64894& 0.0& $2.85 \pm 0.12$& $1.00 \pm 0.02$& $0.81 \pm 0.01$& $2.30 \pm 0.02$& $2.69 \pm 0.02$& $0.237 \pm 0.007$& $0.329 \pm 0.007$& $0.232 \pm 0.007$& GANDALF&   & \checkmark \\
23& NGC5733& 64895& 0.0& $0.98 \pm 0.11$& $1.00 \pm 0.03$& $0.85 \pm 0.01$& $2.42 \pm 0.04$& $2.66 \pm 0.02$& $0.225 \pm 0.014$& $0.412 \pm 0.015$& $0.278 \pm 0.012$& GANDALF&   & \checkmark \\
25& NGC5740& 321075& 1.87& & $1.00 \pm 0.04$& $0.20 \pm 0.01$& $0.58 \pm 0.03$& $2.87 \pm 0.04$& $1.116 \pm 0.019$& $0.324 \pm 0.013$& $0.268 \pm 0.013$& GFC& \checkmark & \checkmark \\
25& NGC5740& 321076& 0.11& $2.46 \pm 0.57$& $1.00 \pm 0.06$& $0.10 \pm 0.02$& $0.30 \pm 0.05$& $2.86 \pm 0.08$& $1.078 \pm 0.039$& $0.452 \pm 0.028$& $0.318 \pm 0.027$& GANDALF& \checkmark & \checkmark \\
25& NGC5740& 321077& 0.0& $2.64 \pm 0.29$& $1.00 \pm 0.03$& $0.26 \pm 0.01$& $0.74 \pm 0.02$& $2.72 \pm 0.02$& $0.930 \pm 0.015$& $0.415 \pm 0.013$& $0.296 \pm 0.015$& GANDALF& \checkmark & \checkmark \\
27& NGC5750& 65076& 0.0& $2.42 \pm 0.20$& $1.00 \pm 0.02$& $0.13 \pm 0.01$& $0.38 \pm 0.02$& $2.68 \pm 0.02$& $1.177 \pm 0.012$& $0.456 \pm 0.012$& $0.336 \pm 0.011$& GANDALF&   & \checkmark \\
28& PGC037392& 288461& 0.06& & $1.00 \pm 0.05$& $0.85 \pm 0.02$& $2.44 \pm 0.06$& $2.86 \pm 0.06$& $0.406 \pm 0.046$& $0.815 \pm 0.046$& $0.492 \pm 0.046$& GANDALF&   & \checkmark \\
29& PGC051719& 92677& 0.32& $2.66 \pm 0.20$& $1.00 \pm 0.05$& $0.85 \pm 0.04$& $2.41 \pm 0.13$& $2.86 \pm 0.11$& $0.244 \pm 0.011$& $0.331 \pm 0.014$& $0.236 \pm 0.011$& GANDALF&   & \checkmark \\
29& PGC051719& 92676& 0.22& $4.08 \pm 0.32$& $1.00 \pm 0.04$& $0.35 \pm 0.02$& $1.00 \pm 0.04$& $2.86 \pm 0.06$& $0.511 \pm 0.022$& $0.656 \pm 0.025$& $0.427 \pm 0.024$& GANDALF&   & \checkmark \\
30& PGC052652& 240202& 0.46& $3.20 \pm 0.25$& $1.00 \pm 0.02$& $0.65 \pm 0.01$& $1.83 \pm 0.02$& $2.86 \pm 0.01$& $0.352 \pm 0.006$& $0.389 \pm 0.007$& $0.274 \pm 0.006$& GANDALF&   & \checkmark \\
31& SDSSJ084...& 622084& 0.19& $1.75 \pm 0.09$& $1.00 \pm 0.02$& $0.20 \pm 0.01$& $0.56 \pm 0.02$& $2.86 \pm 0.02$& $0.796 \pm 0.011$& $0.480 \pm 0.013$& $0.356 \pm 0.011$& GANDALF& \checkmark & \\
33& UGC04673& 517868& 0.65& & $1.00 \pm 0.17$& $0.21 \pm 0.07$& $0.63 \pm 0.12$& $2.86 \pm 0.24$& $0.776 \pm 0.060$& $0.672 \pm 0.071$& $0.568 \pm 0.056$& GFC& \checkmark & \\
33& UGC04673& 517869& 0.53& $2.89 \pm 0.33$& $1.00 \pm 0.07$& $0.51 \pm 0.03$& $1.42 \pm 0.09$& $2.86 \pm 0.12$& $0.426 \pm 0.025$& $0.372 \pm 0.021$& $0.257 \pm 0.017$& GANDALF& \checkmark & \\
34& UGC04684& 600168& 0.31& & $1.00 \pm 0.05$& $0.22 \pm 0.01$& $0.62 \pm 0.04$& $2.86 \pm 0.05$& $0.615 \pm 0.027$& $0.629 \pm 0.025$& & GANDALF& \checkmark & \checkmark \\
35& UGC04996& 198771& 0.0& $1.66 \pm 0.20$& $1.00 \pm 0.07$& $0.27 \pm 0.02$& $0.78 \pm 0.06$& $1.91 \pm 0.05$& $0.346 \pm 0.033$& $0.531 \pm 0.040$& $0.396 \pm 0.036$& GANDALF& \checkmark & \\
35& UGC04996& 198772& 0.22& $6.01 \pm 0.89$& $1.00 \pm 0.07$& $0.51 \pm 0.02$& $1.44 \pm 0.06$& $2.86 \pm 0.05$& $0.301 \pm 0.037$& $0.466 \pm 0.035$& $0.405 \pm 0.032$& GANDALF& \checkmark & \\
36& UGC06578& 6821& 0.0& $0.62 \pm 14.85$& $1.00 \pm 0.21$& $1.39 \pm 0.00$& $4.13 \pm 0.01$& $1.19 \pm 0.00$& $0.013 \pm 0.000$& $0.031 \pm 0.000$& $0.023 \pm 0.000$& GFC& \checkmark & \checkmark \\
36& UGC06578& 6822& 0.25& & $1.00 \pm 0.02$& $1.01 \pm 0.01$& $2.88 \pm 0.03$& $2.86 \pm 0.03$& $0.126 \pm 0.005$& $0.292 \pm 0.005$& $0.200 \pm 0.007$& GANDALF& \checkmark & \checkmark \\
37& UGC06780& 177588& 0.13& $1.76 \pm 0.39$& $1.00 \pm 0.05$& $0.33 \pm 0.02$& $0.95 \pm 0.05$& $2.86 \pm 0.06$& $0.500 \pm 0.043$& $0.640 \pm 0.044$& $0.459 \pm 0.045$& GANDALF& \checkmark & \checkmark \\
37& UGC06780& 177591& 0.0& $3.09 \pm 0.28$& $1.00 \pm 0.05$& $0.43 \pm 0.01$& $1.24 \pm 0.04$& $2.28 \pm 0.03$& $0.169 \pm 0.016$& $0.317 \pm 0.021$& $0.207 \pm 0.018$& GANDALF& \checkmark & \checkmark \\
38& UGC06877& 70114& 1.02& $2.48 \pm 0.06$& $1.00 \pm 0.02$& $0.24 \pm 0.01$& $0.72 \pm 0.01$& $2.87 \pm 0.02$& $0.712 \pm 0.005$& $0.251 \pm 0.007$& $0.193 \pm 0.003$& GFC&   & \checkmark \\
40& UGC06903& 22742& 0.0& $4.11 \pm 3.47$& $1.00 \pm 0.25$& $0.17 \pm 0.08$& $0.49 \pm 0.24$& $1.93 \pm 0.29$& $0.524 \pm 0.126$& $0.445 \pm 0.091$& $0.323 \pm 0.134$& GANDALF& \checkmark & \checkmark \\
40& UGC06903& 272331& 0.33& $2.04 \pm 0.44$& $1.00 \pm 0.06$& $0.23 \pm 0.02$& $0.65 \pm 0.07$& $2.86 \pm 0.07$& $0.592 \pm 0.043$& $0.678 \pm 0.050$& $0.429 \pm 0.043$& GANDALF& \checkmark & \checkmark \\
41& UGC06970& 185266& 0.0& $1.42 \pm 0.90$& $1.00 \pm 0.19$& $0.27 \pm 0.07$& $0.76 \pm 0.21$& $1.88 \pm 0.12$& $0.370 \pm 0.125$& $0.687 \pm 0.121$& $0.463 \pm 0.127$& GANDALF& \checkmark & \\

\hline\hline
\end{tabular}
\end{table}
\end{landscape}
\addtocounter{table}{-1}

\begin{landscape}
\begin{table}
\caption{\textit{Continued}}
		\scriptsize
		\setlength{\tabcolsep}{6.5pt}
    		\begin{tabular}{lccccccccccclcc}
    		\hline\hline
\multirow{4}{*}{ID} & \multirow{4}{*}{name} & \multirow{4}{*}{cataID} & \multirow{4}{*}{$C(H\beta)$}  &  &  &  &  &  & & & & \multirow{4}{*}{Origin} & \multirow{4}{*}{\rotatebox{90}{H{\sc i}GH}} & \multirow{4}{*}{\rotatebox{90}{HAPLESS}}    \\
& & & &  \multicolumn{8}{c}{$I_{\lambda}/I_{H\beta}$} & & & \\ \cmidrule(lr){5-12}
 &  &  &  & [O{\sc ii}] & $H\beta$  & [O{\sc iii},4959] & [O{\sc iii},5007] & $H\alpha$ & [N{\sc ii}] & [S{\sc ii},6713] & [S{\sc ii},6731] & & &   \\
   &  &  &  &  &  &  &  &  & & & & & &   \\     		\hline
42& UGC07000& 144491& 0.56& $3.52 \pm 13.57$& $1.00 \pm 0.15$& $0.44 \pm 0.06$& $1.64 \pm 0.15$& $2.86 \pm 0.08$& $0.450 \pm 0.051$& $0.442 \pm 0.041$& $0.362 \pm 0.036$& GFC& \checkmark & \checkmark \\
42& UGC07000& 144493& 0.0& $2.97 \pm 0.14$& $1.00 \pm 0.02$& $0.74 \pm 0.01$& $2.11 \pm 0.02$& $2.11 \pm 0.02$& $0.238 \pm 0.010$& $0.301 \pm 0.011$& $0.206 \pm 0.010$& GANDALF& \checkmark & \checkmark \\
42& UGC07000& 144494& 0.23& $3.24 \pm 0.37$& $1.00 \pm 0.04$& $0.32 \pm 0.01$& $0.91 \pm 0.03$& $2.86 \pm 0.04$& $0.453 \pm 0.023$& $0.544 \pm 0.023$& $0.395 \pm 0.025$& GANDALF& \checkmark & \checkmark \\
42& UGC07000& 144495& 0.0& $5.38 \pm 1.69$& $1.00 \pm 0.20$& $0.62 \pm 0.12$& $1.77 \pm 0.34$& $2.69 \pm 0.34$& $0.332 \pm 0.057$& $0.430 \pm 0.062$& $0.286 \pm 0.054$& GANDALF& \checkmark & \checkmark \\
42& UGC07000& 700775& 0.27& $1.02 \pm 0.50$& $1.00 \pm 0.36$& $1.00 \pm 0.34$& $2.83 \pm 0.97$& $2.86 \pm 0.71$& $0.279 \pm 0.085$& $0.395 \pm 0.097$& $0.328 \pm 0.075$& GANDALF& \checkmark & \checkmark \\
43& UGC07053& 185622& 0.21& & $1.00 \pm 0.05$& $1.04 \pm 0.03$& $3.09 \pm 0.08$& $2.86 \pm 0.06$& $0.145 \pm 0.021$& $0.233 \pm 0.037$& $0.178 \pm 0.027$& GFC& \checkmark & \\
43& UGC07053& 185623& 0.64& $0.74 \pm 0.53$& $1.00 \pm 0.20$& $0.15 \pm 0.15$& $0.67 \pm 0.17$& $2.86 \pm 0.27$& $0.328 \pm 0.213$& $0.614 \pm 0.142$& $0.561 \pm 0.134$& GFC& \checkmark & \\
43& UGC07053& 791635& 0.0& $4.10 \pm 0.89$& $1.00 \pm 0.05$& $0.60 \pm 0.02$& $1.72 \pm 0.06$& $2.35 \pm 0.05$& $0.085 \pm 0.025$& $0.258 \pm 0.028$& $0.185 \pm 0.027$& GANDALF& \checkmark & \\
44& UGC07332& 85881& 0.0& $86.23 \pm 52.80$& $1.00 \pm 1.30$& $0.39 \pm 0.48$& $1.12 \pm 1.37$& $1.62 \pm 0.57$& $0.155 \pm 1.762$& $1.524 \pm 0.572$& $1.116 \pm 0.439$& GANDALF& \checkmark & \\
45& UGC07394& 221194& 0.52& $2.97 \pm 6.90$& $1.00 \pm 0.19$& $0.37 \pm 0.18$& $0.69 \pm 0.24$& $2.86 \pm 0.24$& $0.302 \pm 0.101$& $0.563 \pm 0.133$& $0.555 \pm 0.134$& GFC& \checkmark & \checkmark \\
45& UGC07394& 221195& 0.5& $0.72 \pm 0.12$& $1.00 \pm 0.07$& $0.82 \pm 0.05$& $2.31 \pm 0.14$& $2.86 \pm 0.11$& $0.143 \pm 0.012$& $0.272 \pm 0.015$& $0.177 \pm 0.013$& GANDALF& \checkmark & \checkmark \\
46& UGC07396& 9163& 0.02& $5.82 \pm 1.15$& $1.00 \pm 0.10$& $0.43 \pm 0.02$& $1.24 \pm 0.07$& $2.86 \pm 0.06$& $0.455 \pm 0.050$& $0.590 \pm 0.059$& $0.447 \pm 0.064$& GANDALF&   & \checkmark \\
48& UGC09215& 238952& 0.7& $2.50 \pm 56.85$& $1.00 \pm 0.16$& $0.52 \pm 0.02$& $1.38 \pm 0.03$& $2.86 \pm 0.03$& $0.503 \pm 0.018$& $0.461 \pm 0.014$& $0.318 \pm 0.011$& GFC& \checkmark & \checkmark \\
48& UGC09215& 714924& 0.0& $43.90 \pm 191.72$& $1.00 \pm 0.04$& $0.41 \pm 0.01$& $1.09 \pm 0.03$& $1.97 \pm 0.02$& $0.321 \pm 0.007$& $0.326 \pm 0.006$& $0.234 \pm 0.005$& GFC& \checkmark & \checkmark \\
48& UGC09215& 714923& 0.43& $2.82 \pm 39.15$& $1.00 \pm 0.02$& $0.60 \pm 0.01$& $1.82 \pm 0.02$& $2.86 \pm 0.01$& $0.478 \pm 0.006$& $0.365 \pm 0.008$& $0.267 \pm 0.004$& GFC& \checkmark & \checkmark \\
48& UGC09215& 319800& 0.13& $1.91 \pm 0.07$& $1.00 \pm 0.01$& $0.97 \pm 0.00$& $2.76 \pm 0.01$& $2.86 \pm 0.01$& $0.263 \pm 0.005$& $0.309 \pm 0.005$& $0.223 \pm 0.005$& GANDALF& \checkmark & \checkmark \\
48& UGC09215& 319801& 0.0& $1.95 \pm 0.09$& $1.00 \pm 0.02$& $0.40 \pm 0.01$& $1.13 \pm 0.02$& $1.91 \pm 0.01$& $0.272 \pm 0.010$& $0.329 \pm 0.012$& $0.214 \pm 0.010$& GANDALF& \checkmark & \checkmark \\
49& UGC09299& 593645& 0.29& & $1.00 \pm 0.07$& $0.60 \pm 0.04$& $1.80 \pm 0.07$& $2.86 \pm 0.06$& $0.483 \pm 0.035$& $0.684 \pm 0.033$& $0.474 \pm 0.026$& GFC& \checkmark & \checkmark \\
49& UGC09299& 593646& 0.0& $2.95 \pm 0.24$& $1.00 \pm 0.03$& $0.55 \pm 0.01$& $1.56 \pm 0.03$& $2.82 \pm 0.03$& $0.330 \pm 0.017$& $0.430 \pm 0.018$& $0.308 \pm 0.016$& GANDALF& \checkmark & \checkmark \\
50& UGC09348& 619104& 0.39& $5.13 \pm 0.62$& $1.00 \pm 0.07$& $0.43 \pm 0.02$& $1.22 \pm 0.06$& $2.86 \pm 0.04$& $0.637 \pm 0.032$& $0.734 \pm 0.028$& $0.554 \pm 0.030$& GANDALF&   & \checkmark \\
51& UGC09432& 367146& 0.1&& $1.00 \pm 0.03$& $0.74 \pm 0.01$& $2.15 \pm 0.03$& $2.86 \pm 0.05$& $0.195 \pm 0.014$& $0.450 \pm 0.018$& $0.313 \pm 0.014$& GANDALF& \checkmark & \\
52& UGC09470& 16827& 0.0& $5.06 \pm 0.13$& $1.00 \pm 0.01$& $1.21 \pm 0.00$& $3.44 \pm 0.01$& $2.22 \pm 0.01$& $0.103 \pm 0.001$& $0.173 \pm 0.002$& $0.121 \pm 0.002$& GANDALF& \checkmark & \checkmark \\
53& UGC09482& 16863& 0.0& $5.18 \pm 0.38$& $1.00 \pm 0.02$& $0.86 \pm 0.01$& $2.45 \pm 0.03$& $2.65 \pm 0.02$& $0.252 \pm 0.011$& $0.404 \pm 0.013$& $0.265 \pm 0.011$& GANDALF& \checkmark & \\
53& UGC09482& 16899& 0.0& $6.51 \pm 0.69$& $1.00 \pm 0.05$& $0.38 \pm 0.02$& $1.08 \pm 0.05$& $2.49 \pm 0.05$& $0.243 \pm 0.034$& $0.492 \pm 0.039$& $0.342 \pm 0.038$& GANDALF& \checkmark & \\
54& UM452& 54103& 0.21& $2.46 \pm 0.07$& $1.00 \pm 0.01$& $0.85 \pm 0.00$& $2.42 \pm 0.01$& $2.86 \pm 0.01$& $0.258 \pm 0.004$& $0.366 \pm 0.005$& $0.266 \pm 0.004$& GANDALF&   & \checkmark \\
55& UM456& 559583& 0.0& $1.61 \pm 0.12$& $1.00 \pm 0.01$& $1.80 \pm 0.01$& $5.15 \pm 0.02$& $1.93 \pm 0.01$& $0.037 \pm 0.003$& $0.078 \pm 0.003$& $0.053 \pm 0.003$& GANDALF& \checkmark & \checkmark \\
55& UM456& 559584& 0.0& $4.76 \pm 0.16$& $1.00 \pm 0.02$& $1.05 \pm 0.01$& $3.00 \pm 0.02$& $2.06 \pm 0.01$& $0.078 \pm 0.005$& $0.197 \pm 0.006$& $0.138 \pm 0.006$& GANDALF& \checkmark & \checkmark \\
56& UM456A& 559608& 0.0& $3.65 \pm 0.13$& $1.00 \pm 0.02$& $1.15 \pm 0.01$& $3.30 \pm 0.02$& $1.73 \pm 0.01$& $0.049 \pm 0.006$& $0.125 \pm 0.006$& $0.083 \pm 0.006$& GANDALF& \checkmark & \checkmark \\
56& UM456A& 559610& 0.0& $2.27 \pm 0.13$& $1.00 \pm 0.02$& $1.10 \pm 0.01$& $3.15 \pm 0.02$& $2.67 \pm 0.02$& $0.060 \pm 0.006$& $0.166 \pm 0.006$& $0.110 \pm 0.006$& GANDALF& \checkmark & \checkmark \\
57& UM491& 290172& 0.0& $4.52 \pm 0.10$& $1.00 \pm 0.01$& $0.87 \pm 0.00$& $2.47 \pm 0.01$& $2.53 \pm 0.01$& $0.201 \pm 0.005$& $0.277 \pm 0.005$& $0.191 \pm 0.004$& GANDALF& & \checkmark \\
\hline\hline
\end{tabular}
\end{table}
\end{landscape}

\begin{landscape}
\begin{table}
\caption{Emission line measurements and derived metallicities for the Dwarf Galaxy Survey using the O3N2, N2, PG16S and KE08/T04 methods. The literature emission lines have been corrected for stellar absorption and reddening using methods in the listed references.    }
		\label{tab:DGSlines}
		\setlength{\tabcolsep}{12pt}
    		\begin{tabular}{lrrrrrrrccccr}
    		\hline\hline
  		\multirow{2}{*}{name} & \multicolumn{7}{c}{$I_{\lambda}/I_{H\beta}$} & \multicolumn{4}{c}{12+log(O/H)} & \multirow{2}{*}{Ref}\\ \cmidrule(lr){2-8} \cmidrule(lr){9-12}
	 & $H\alpha$& $[O\textsc{ii}]$& $[O{\textsc{iii},4959}]$ & $[O{\textsc{iii},5007}]$& $[N\textsc{ii}]$& $[S{\textsc{ii},6713}]$& $[S{\textsc{ii},6731}]$ & N2 & O3N2 & PG16S & KE08/T04 & \\ 	\hline
Haro II & 2.91 & 1.08 & 1.19 & 3.72 & 0.488 & 0.156 & 0.102 & 8.40 & 8.30 & 8.45 & 8.49 & 1 \\
Haro 2 & 2.86 & 3.41 & 0.55 & 1.45 & 0.461 & 0.297 & 0.298 & 8.39 & 8.42 & 8.36 & 8.65 & 2 \\
Haro 3 & 2.86 & 2.71 & 1.08 & 3.22 & 0.236 & 0.223 & 0.180 & 8.24 & 8.22 & 8.24 & 8.37 & 1 \\
He 2-10 & 3.00 & 2.29 & 0.36 & 1.06 & 0.878 & 0.301 & 0.266 & 8.60 & 8.55 & 8.55 & 8.8 & 3 \\
HS0822+3542 & 2.75 & 0.31 & 1.19 & 3.58 & 0.010 & 0.026 & 0.018 & 7.29 & 7.78 & 7.40 &   & 4 \\
HS1304+3259 & 2.86 & 1.85 & 1.52 & 4.51 & 0.100 & \multicolumn{2}{c}{0.330$^a$} & 8.10 & 8.05 & 8.03 & 8.07 & 5 \\
HS1319+3224 & 2.86 & 1.24 & 1.63 & 5.03 &   & \multicolumn{2}{c}{0.140$^a$}  &   &   &   &   & 4 \\
HS1330+3651 & 2.86 & 0.48 & 1.64 & 4.9 & 0.090 & \multicolumn{2}{c}{0.350$^a$}  & 8.08& 8.03 & 8.03 &   & 4 \\
HS1442+4250 & 2.76 & 0.54 & 1.75 & 4.98 & 0.023 & 0.042 & 0.029 & 7.72& 7.84 & 7.73 &   & 6 \\
I Zw 18 & 2.74 & 0.408 & 0.636 & 1.906&  0.012 & 0.029 & 0.022 &  7.39 & 7.88 & 7.15 & & 7,13 \\
II Zw 40 & 2.87 & 0.84 & 2.46 & 7.41 & 0.063 & 0.067 & 0.054 & 8.09 & 7.92 & 8.14 &   & 8 \\
IC10 & 2.85 & 1.01 & 1.35 & 3.97 & 0.241 & 0.118 & 0.094 &8.25& 8.20 & 8.33 & 8.33 & 9 \\
Mrk 1089 & 2.86 & 1.50 & 0.74 & 2.24 & 0.315 & 0.141 & 0.101 & 8.30 & 8.31 & 8.38 & 8.51 & 10 \\
Mrk 1450 & 2.83 & 1.35 & 1.79 & 4.76 & 0.067 & 0.117 & 0.083  & 8.02& 7.99 & 7.95 &   & 11 \\
Mrk 153 & 2.81 & 0.00 & 1.51 &   & 0.059 & \multicolumn{2}{c}{0.247$^a$} & 7.99 &   &   &   & 12 \\
Mrk 209 & 2.78 & 0.72 & 1.96 & 5.54 & 0.029 & 0.061 & 0.045 & 7.80 & 7.86 & 7.80 &   & 13 \\
Mrk 930 & 2.85 & 2.37 & 1.39 & 4.17 & 0.143 & 0.269 & 0.198  & 8.15 & 8.12 & 8.11 & 8.19 & 14 \\
NGC 1140 & 2.88 & 2.32 & 0.97 & 0.29 & 0.256 & 0.231 & 0.175 & 8.25 & 8.57 & 8.28 & 8.81 & 15 \\
NGC 1569 & 2.84 & 0.99 & 1.50 & 4.51 & 0.137 & 0.205 & 0.147 & 8.15 & 8.10 & 8.13 & 8.16 & 16 \\
NGC 1705 & 2.86 & 3.74 & 1.05 & 3.00 & 0.105 & \multicolumn{2}{c}{0.060$^a$} & 8.10  & 8.12 & 7.99 & 8.2 & 17 \\
NGC 1705 & 2.86 & 3.37 & 1.66 & 4.87 &   & \multicolumn{2}{c}{0.026$^a$} &  &   &   &   & 16 \\
NGC 1705 & 2.91 & 2.54 & 1.47 & 4.25 & 0.111 & \multicolumn{2}{c}{0.034$^a$} & 8.11 & 8.07 & 8.24 & 8.11 & 16 \\
NGC 1705 & 2.76 & 2.43 & 1.52 & 3.83 &   &  &  &   &   &   &   & 16 \\
NGC 1705 & 2.85 & 4.00 & 1.03 & 3.15 & 0.039 & \multicolumn{2}{c}{0.049$^a$} & 7.89 & 7.97 & 7.73 &   & 16 \\
NGC 1705 & 2.86 & 4.75 & 1.32 & 3.67 & 0.122 & \multicolumn{2}{c}{0.016$^a$} & 8.13 & 8.07 & 8.33 & 8.18 & 16 \\
NGC 1705 & 2.81 & 2.74 & 1.67 & 4.86 & 0.098 & \multicolumn{2}{c}{0.037$^a$} & 8.09  & 8.04 & 8.25 &   & 16 \\
NGC 1705 & 2.86 & 3.12 & 1.71 & 4.94 & 0.069 & \multicolumn{2}{c}{0.036$^a$} & 8.03  & 7.99 & 8.15 &   & 16 \\
NGC 1705 & 2.84 & 3.49 & 1.69 & 4.87 & 0.095 & \multicolumn{2}{c}{0.038$^a$} & 8.09  & 8.04 & 8.24 &   & 16 \\
NGC 1705 & 2.84 & 4.36 & 1.45 & 4.27 & 0.148 & \multicolumn{2}{c}{0.047$^a$} & 8.16 & 8.12 & 8.28 & 8.2 & 16 \\
NGC 1705 & 2.83 & 2.64 & 1.13 & 3.40 &   & \multicolumn{2}{c}{0.033$^a$}   &  &   &   &   & 16 \\
NGC 1705 & 2.86 & 5.13 & 0.59 & 2.04 &   &   &   &  &   &   &   & 16 \\
NGC 1705 & 2.86 & 4.01 & 1.09 & 3.19 & 0.097 & \multicolumn{2}{c}{0.043$^a$} & 8.09  & 8.10 & 8.03 & 8.16 & 16 \\
NGC 1705 & 2.86 & 4.02 & 1.26 & 3.59 & 0.154 & \multicolumn{2}{c}{0.043$^a$} & 8.17  & 8.15 & 8.23 & 8.25 & 16 \\
NGC 1705 & 2.86 & 4.08 & 1.57 & 4.39 & 0.122 & \multicolumn{2}{c}{0.046$^a$} & 8.13  & 8.09 & 8.24 & 8.14 & 16 \\
NGC 1705 & 2.86 & 3.52 & 1.63 & 4.64 & 0.016 & \multicolumn{2}{c}{0.038$^a$} & 7.42  & 7.80 & 7.67 &   & 16 \\
NGC 1705 & 2.79 & 3.02 & 1.37 & 3.93 & 0.062 & \multicolumn{2}{c}{0.036$^a$} & 8.01  & 8.01 & 8.02 &   & 16 \\
NGC 2366 & 2.86 & 0.84 & 1.34 & 3.93 & 0.139 & 0.234 & 0.167 & 8.15 & 8.12 & 8.08 & 8.2 & 18 \\
NGC 2366 & 2.86 & 1.75 & 1.93 & 5.73 & 0.037 & 0.099 & 0.074 & 7.87 & 7.88 & 7.84 &   & 17 \\
NGC 4214 & 2.85 & 3.04 & 1.00 & 3.03 & 0.241 & 0.267 & 0.194 & 8.24 & 8.23 & 8.23 & 8.39 & 15 \\
NGC 4214 & 2.80 & 3.08 & 2.47 & 7.52 & 0.041 & 0.035 & 0.033 & 7.91 & 7.86 & 8.08 &   & 17 \\
NGC 4449 & 2.87 & 3.89 & 0.69 & 2.07 & 0.338 & 0.476 & 0.334 & 8.32 & 8.33 & 8.22 & 8.53 & 15 \\
NGC 4861 & 2.86 & 1.41 & 1.26 & 3.76 & 0.204 & 0.346 & 0.173 & 8.22 & 8.18 & 8.16 & 8.3 & 17 \\
 \hline\hline
\end{tabular}

$^a$ The [S{\rm II}] $\lambda 6717$ and [S{\rm II}] $\lambda 6731$ lines are blended.
\end{table}
\end{landscape}
\addtocounter{table}{-1}

\begin{landscape}
\begin{table}
\caption{Continued}
		\setlength{\tabcolsep}{12pt}
    		\begin{tabular}{lrrrrrrrccccr}
    		\hline\hline
  		\multirow{2}{*}{name} & \multicolumn{7}{c}{$I_{\lambda}/I_{H\beta}$} & \multicolumn{4}{c}{12+log(O/H)} & \multirow{2}{*}{Ref}\\ \cmidrule(lr){2-8} \cmidrule(lr){9-12}
	 & $H\alpha$& $[O\textsc{ii}]$& $[O{\textsc{iii},4959}]$ & $[O{\textsc{iii},5007}]$& $[N\textsc{ii}]$& $[S{\textsc{ii},6713}]$& $[S{\textsc{ii},6731}]$ & O3N2 & N2 & PG16S & KE08/T04 & \\ 	\hline
NGC 5253 & 2.83 & 2.60 & 1.42 & 4.22 & 0.200 & 0.270 & 0.205 & 8.21 & 8.16 & 8.17 & 8.28 & 15 \\
NGC 625 & 2.86 & 1.76 & 1.52 & 4.53 & 0.123 & 0.136 & 0.101 & 8.13 & 8.08 & 8.10 & 8.13 & 19 \\
NGC 625 & 2.84 & 2.31 & 0.99 & 2.95 & 0.204 & 0.197 & 0.146 & 8.21 & 8.21 & 8.24 & 8.36 & 18 \\
NGC 625 & 2.82 & 3.38 & 0.87 & 2.55 & 0.262 & 0.341 & 0.240 & 8.27 & 8.27 & 8.21 & 8.45 & 18 \\
NGC 625 & 2.86 & 4.86 & 0.33 & 1.03 & 0.422 & 0.676 & 0.520 & 8.37 & 8.46 & 8.2 & 8.69 & 18 \\
NGC 6822 & 2.85 & 0.94 & 1.77 & 5.35 & 0.051 & 0.067 & 0.050 & 7.96 & 7.94 & 7.94 &   & 20 \\
NGC 6822 & 2.85 & 1.47 & 1.44 & 4.26 & 0.071 & 0.103 & 0.072 & 8.03 & 8.02 & 7.92 &   & 19 \\
Pox 186 & 2.86 & 0.35 & 2.09 & 6.22 &   &   &   &   &   &  &   & 21 \\
SBS 0335-052 & 2.86 & 0.30 & 1.09 & 3.24 & 0.009 & 0.021 & 0.020 & 7.17 & 7.77 & 7.33 &   & 17 \\
SBS 0335-052 & 2.86 & 0.25 & 1.1 & 3.29 & 0.009 & 0.020 & 0.017 & 7.17 & 7.76 & 7.35 &   & 17 \\
SBS 0335-052 & 2.86 & 0.26 & 1.42 & 4.27 & 0.061 & 0.124 & 0.084 & 8.00 & 7.99 & 7.87 &   & 17 \\
SBS 0335-052 & 2.86 & 0.23 & 1.50 & 4.49 & 0.124 & 0.221 & 0.215 & 8.13 & 8.09 & 8.09 & 8.13 & 17 \\
SBS 1159+545 & 2.76 & 0.65 & 1.29 & 3.80 & 0.085 & 0.187 & 0.133 & 8.07 & 8.06 & 7.92 & 8.08 & 17 \\
SBS 1211+540 & 2.71 & 0.80 & 2.04 & 6.07 & 0.041 & 0.093 & 0.069 & 7.92 & 7.90 & 7.90 &   & 17 \\
SBS 1249+493 & 2.86 & 1.24 & 2.01 & 5.88 & 0.047 & 0.097 & 0.072 & 7.94 & 7.91 & 7.93 &   & 17 \\
SBS 1415+437 & 2.86 & 1.18 & 1.19 & 3.54 & 0.037 & 0.090 & 0.067 & 7.87 & 7.95 & 7.66 &   & 12 \\
SBS 1533+574 & 2.81 & 2.46 & 1.30 & 3.80 & 0.122 & 0.234 & 0.170 & 8.13 & 8.11 & 8.03 & 8.18 & 12 \\
SBS 1533+574 & 2.85 & 2.04 & 1.79 & 5.33 & 0.087 & 0.167 & 0.117 & 8.07 & 8.01 & 8.05 &   & 12 \\
Tol 0618-402 & 2.86 & 2.11 & 1.62 & 4.95 &   &   &   &   &   &   &  & 22 \\
Tol 0618-402 & 2.86 & 2.35 & 1.60 & 4.92 &   &   &   &   &   &   &  & 21 \\
Tol 1214-277 & 2.74 & 0.36 & 1.76 & 5.28 & 0.009 & 0.019 & 0.016 & 7.23 & 7.71 & 7.58 &   & 14 \\
UGC 4483 & 2.86 & 1.32 & 0.90 & 2.73 & 0.037 & 0.070 & 0.052 & 7.88 & 7.99 & 7.56 &   & 17 \\
UGCA 20 & 2.76 & 0.92 & 0.89 & 2.60 & 0.035 & 0.073 & 0.050 & 7.87 & 7.99 & 7.53 &   & 23 \\
UGCA 20 & 2.76 & 1.35 & 0.85 & 2.58 & 0.044 & 0.088 & 0.064 & 7.93 & 8.02 & 7.58 &   & 22 \\
UM 133 & 2.86 & 1.81 & 1.25 & 3.74 & 0.043 & 0.118 & 0.083 & 7.91 & 7.96 & 7.71 &   & 17 \\
UM 311 & 2.89 & 1.80 & 1.32 & 3.98 & 0.180 & 0.167 & 0.124 & 8.19 & 8.15 & 8.16 & 8.26 & 17 \\
UM 448 & 2.85 & 2.78 & 0.87 & 2.60 & 0.409 & 0.366 & 0.285 & 8.36 & 8.33 & 8.31 & 8.53 & 13 \\
UM 461 & 2.78 & 0.53 & 2.04 & 6.02 & 0.021 & 0.052 & 0.042 & 7.68 & 7.80 & 7.74 &   & 13 \\
VII Zw 403 & 2.83 & 1.36 & 1.17 & 3.52 & 0.051 & 0.105 & 0.077 & 7.96 & 8.00 & 7.74 &   & 17 \\ \hline\hline
\end{tabular}

\textbf{References:}
(1) \citet{Guseva2012}, (2) \citet{Kong2002}, (3) \citet{Kobulnicky1999}, (4) \citet{Pustilnik2003}, (5) \citet{Popescu2000}, (6) \citet{Guseva2003}, (7) \citet{Skillman1993}, (8) \citet{Guseva2000}, (9) \citet{Magrini2009}, (10) \citet{Lopez2004}, (11) \citet{Izotov1994}, (12) \citet{Izotov2006}, (13) \citet{Izotov1997}, (14) \citet{Izotov1998}, (15) \citet{Izotov2004}, (16) \citet{Kobulnicky1997}, (17) \citet{Lee2004}, (18) \citet{Izotov2007}, (19) \citet{Skillman2003}, (20) \citet{Peimbert2005}, (21) \citet{Guseva2007}, (22) \citet{Masegosa1994}, (23) \citet{VanZee1996}.
\end{table}
\end{landscape}

\end{document}